\documentclass[a4paper,11pt]{article}
\pdfoutput=1 
\usepackage{jcappub} 

\usepackage{amssymb}
\usepackage{mathtools}
\usepackage{amsmath}
\usepackage{amstext}
\usepackage{graphicx,epsfig}
\usepackage{epsfig}
\usepackage{verbatim} 	
\usepackage{caption}
\usepackage{fancybox}
\usepackage{slashed}
\usepackage{afterpage}
\usepackage{color}
\usepackage{ulem}
\usepackage{enumitem}
\usepackage{subfigure}
\usepackage{parskip}
\usepackage{dsfont}
\usepackage{tabu}
\usepackage{tikz}
\usepackage{booktabs}
\usepackage[numbers,sort&compress]{natbib}
\usepackage{cancel}

\usepackage{bbding}
\usepackage{wasysym}
\usepackage{multirow,hhline,array}
\usepackage[utf8]{inputenc}

\definecolor{navyblue}{rgb}{0.0, 0.0, 0.5}
\definecolor{royalblue}{rgb}{0.25, 0.41, 0.88}
\definecolor{cadmiumgreen}{rgb}{0.0, 0.42, 0.24}
\definecolor{blue-violet}{rgb}{0.54, 0.17, 0.89}
\definecolor{darkviolet}{rgb}{0.58, 0.0, 0.83}
\definecolor{orange(colorwheel)}{rgb}{1.0, 0.5, 0.0}

\usepackage{hyperref}
\hypersetup{colorlinks=true,linkcolor=royalblue,citecolor=magenta,pdfencoding=auto}
\usepackage{pifont}
\usepackage{dcolumn}
\usepackage{colortbl}

\newcommand{\D}{{\rm d}}

\newcommand{\overbar}[1]{\mkern 1.5mu\overline{\mkern-1.5mu#1\mkern-1.5mu}\mkern 1.5mu}

\newcommand{\Mpl}{M_{\text{Pl}}}

\definecolor{magenta(process)}{rgb}{1.0, 0.0, 0.56}

\definecolor{darkspringgreen}{rgb}{0.09, 0.45, 0.27}

\definecolor{royalblue(web)}{rgb}{0.25, 0.41, 0.88}

\title{\centering Effective Field Theory \\ of Black Hole Quasinormal Modes \\ in Scalar-Tensor Theories}

\author[a]{Gabriele Franciolini,}
\author[b]{Lam Hui,}
\author[c]{Riccardo Penco,}
\author[d]{Luca Santoni,}
\author[e]{Enrico Trincherini}

\affiliation[a]{Department of Theoretical Physics and Center for Astroparticle Physics (CAP)
24 quai E. Ansermet, CH-1211 Geneva 4, Switzerland}
\affiliation[b]{Department of Physics, Center for Theoretical Physics, Columbia University, New York, NY 10027, USA}
\affiliation[c]{Department of Physics, Carnegie Mellon University, Pittsburgh, PA 15213, USA}
\affiliation[d]{Institute for Theoretical Physics and Center for Extreme Matter and Emergent Phenomena,
	Utrecht University, 
	Princetonplein 5, 3584 CC Utrecht, The Netherlands}
\affiliation[e]{Scuola Normale Superiore, Piazza dei Cavalieri 7, 56126, Pisa, Italy and INFN – Sezione di Pisa, 56200, Pisa, Italy}

\emailAdd{gabriele.franciolini@unige.ch}
\emailAdd{lh399@columbia.edu}
\emailAdd{rpenco@cmu.edu}
\emailAdd{l.santoni@uu.nl}
\emailAdd{enrico.trincherini@sns.it}

\abstract{\noindent
The final ringdown phase in a coalescence process is a valuable laboratory to test General Relativity and potentially constrain additional degrees of freedom in the gravitational sector. We introduce here an effective description for perturbations around spherically symmetric spacetimes in the context of scalar-tensor theories, which we apply to study quasi-normal modes for black holes with scalar hair. 
We derive the equations of motion governing the dynamics of both the polar and the axial modes in terms of the coefficients of the effective theory. Assuming the deviation of the background
from Schwarzschild is small, we use the WKB method to introduce the notion of ``light ring expansion''. This approximation is
analogous to the slow-roll expansion used for inflation, and it allows us to express the quasinormal mode spectrum in terms of a small number of parameters. This work is a first step in describing, in a model independent way, how the scalar hair can affect the ringdown stage and leave signatures on the emitted gravitational wave signal. Potential signatures include the shifting of the quasi-normal spectrum, the breaking of isospectrality between polar and axial modes, and the existence of scalar radiation.}

\begin{document}
\maketitle
\flushbottom

\section{Introduction}

The detection of gravitational waves by LIGO and Virgo~\cite{Abbott:2016blz,Abbott:2016nmj,TheLIGOScientific:2017qsa} has opened up a new window into the strong gravity regime. Up until now, observations appear to be very well described by General Relativity (GR). Nevertheless, as more and more events are observed, it becomes important to determine quantitatively the extent to which alternative theories of gravity are ruled out. Effective field theories (EFTs) provide a framework to carry out this program in a model-independent way. They also provide a general foil against which GR can be tested. The only inputs required are the number and type of ``light'' degrees of freedom in the gravity sector, and the symmetries that constrain their interactions. 

If the only relevant degrees of freedom in the gravity sector are the two graviton polarizations, the only modification of GR that can (and should) be considered is the addition of higher powers of curvature invariants to the Einstein-Hilbert Lagrangian~\cite{Endlich:2017tqa}. These higher-derivative operators modify (among other things) the phase, amplitude and polarization of gravitational waves. Such corrections are suppressed schematically by powers of $\omega/\Lambda$, where $\omega$ is the frequency of gravitational waves and $\Lambda$ is the scale at which ``new physics'' kicks in. Observations can then be used to place a lower bound on the magnitude of $\Lambda$, providing a model-independent constraint on new ``heavy'' physics in the gravitational sector.\footnote{Such  constraint can then be mapped onto specific models with a procedure known as matching~\cite{Rothstein:2003mp}.} The exact same strategy is used at the LHC to place model-independent bounds on physics beyond the Standard Model. 

In this paper, we will focus on less minimal modifications of GR, in which the light degrees of freedom include one additional light scalar besides the graviton---{\it i.e.}, we will consider scalar-tensor theories.
Many examples of scalar-tensor theories exist of course, but our goal is to be general: what is the most general dynamics of fluctuations around a black hole with scalar hair? 
 We are particularly interested in the way in which this extra scalar mode affects the ringdown that takes place at the end of the merger process. Of course, testing  gravity using the ringdown is not a new idea (see {\it e.g.}~\cite{Berti:2005ys,Berti:2018vdi}). However, until now these tests have been carried out on a model-by-model basis ({\it e.g.} \cite{Bhattacharyya:2017tyc,Bhattacharyya:2018qbe}). In this paper, we propose instead a different approach---based on EFT techniques---which can be used to place {\it model-independent} constraints on scalar tensor theories. More precisely, we introduce an EFT framework to describe quasi-normal modes (QNMs) of static, isolated\footnote{A more realistic program would require to estimate to what extent environmental effects are negligible or affect instead the QNM spectrum. Disentangling the impact of the astrophysical environment from the observations is a key ingredient in order to really constrain additional degrees of freedom in the gravitational sector. For a study in this direction see \textit{e.g.} \cite{Barausse:2014tra}.} black hole solutions with a scalar hair.   

It is well known that such solutions are hard to come by if we demand asymptotic flatness.\footnote{In the presence of a negative cosmological constant, instead, a non-minimal coupling of the form $\Phi^2 R$ is sufficient to give rise to a stable scalar hair in a certain range of parameter space~\cite{Winstanley:2005fu}.} This fact is encoded in a variety of ``no-hair theorems'' which, under fairly general assumptions, forbid the existence of non-trivial scalar profiles surrounding black hole solutions (see {\it e.g.}~\cite{Bekenstein:1971hc,Bekenstein:1995un,Hui:2012qt}). These assumptions can nevertheless be violated, and as a result several solutions with non-trivial scalar hair can be found in the literature (see {\it e.g.}~\cite{Dennhardt:1996cz,Sotiriou:2014pfa,Antoniou:2017hxj}). Depending on the circumstances, such hairy solutions can even be dynamically preferred over solutions with a vanishing scalar profile~\cite{Doneva:2017bvd,Silva:2017uqg}. 
Indeed, a fairly simple way to endow black holes with hair does not even involve sophisticated dynamics; all one needs is non-trivial boundary conditions. Take the example of a minimally coupled, free scalar with a standard kinetic term: Jacobson \cite{Jacobson:1999vr} showed that a black hole has scalar hair if the scalar approaches a pure function of time at spatial infinity. The amount of hair, quantified by the scalar charge to mass ratio for instance, is generally small if the time scale of variation (for instance, Hubble time scale) is long compared to the black hole horizon size \cite{Berti:2013gfa}. However, if the time variation is due to close-by objects (e.g. stars which can well have scalar hair), the induced black hole scalar hair could be non-negligible~\cite{Horbatsch:2011ye}.\footnote{A spatial gradient in the boundary condition is expected to have a similar effect.}
For a fairly exhaustive review of no-hair theorems as well as asymptotically flat solutions with scalar hair, we refer the reader to~\cite{Herdeiro:2015waa}.

While allowed from a technical viewpoint, one may still worry that hairy black holes could already be ruled out by existing observations. For instance, it was recently argued~\cite{Tattersall:2018map} that a non-negligible amount of scalar hair would be incompatible with present cosmological constraints when combined with current bounds on the speed of propagation of gravitational waves~\cite{Monitor:2017mdv}. Such a claim however is based on the assumption that the scalar field remains weakly coupled from cosmological scales all the way to the scales relevant for black hole mergers. This does not necessarily have to be the case, as was pointed out for instance in the context of Horndeski theories~\cite{deRham:2018red}. 
Moreover, the speed of propagation constraint can be viewed as putting a lower bound on the cut-off scale associated with certain higher dimension operators. The bound is significant for cosmological applications, but sufficiently weak to allow non-trivial effects on the horizon scale of astrophysically interesting black holes.

In fact, one could even argue that a scalar hair is a natural feature to consider if we are after observable departures from GR. The reason is that, if the scalar field profile around two well-separated black holes is initially constant, scalar perturbations can be excited at linear level by the merging process only if the scalar couples to the Riemann tensor. For instance, one can consider a  coupling between $\Phi$ and the Gauss-Bonnet invariant $\mathcal{G}_\text{GB} = R_{\mu\nu\lambda\rho} R^{\mu\nu\lambda\rho}-4R_{\mu\nu}R^{\mu\nu}+R^2$ of the form $f(\Phi) \mathcal{G}_\text{GB}$, which is known to generate scalar hair~\cite{Sotiriou:2015pka,Blazquez-Salcedo:2017txk}. This argument is admittedly more suggestive than it is rigorous, as it discounts for instance the possibility that a small hair of cosmological origin~\cite{Jacobson:1999vr} could get amplified by non-linearities during the merger process, and in turn lead to a sizable emission of scalar modes. Regardless, we are finally in a position where the absence of scalar hair is a feature that can be tested experimentally rather than ruled out by no-go theorems based on a set of assumptions. As such, one can also view our formalism as a pragmatic attempt to study the possible observational consequence of a scalar hair during ringdown.

Another important constraint that scalar-tensor theories need to contend with is the lack of evidence for additional polarizations in the gravitational wave spectra observed so far~\cite{Abbott:2018utx,OBeirne:2018slh}. This, however, should be mostly viewed as a constraint on the strength of their coupling to baryons rather than on their actual existence \cite{Damour:1998jk,Barausse:2012da}. To illustrate this point, let's consider scalar-Gauss-Bonnet gravity~\cite{Sotiriou:2014pfa} without a tree-level coupling between the scalar and baryons. The couplings with baryons induced by radiative corrections will be suppressed by derivatives due to the shift invariance of this model, and thus can be easily rendered weak enough to be undetectable.\footnote{Strictly speaking this is true only in the asymptotic region close to the detector, where the background value for the scalar hair vanishes and the metric is nearly Minkowski. Indeed, in general, a non-negligible mixing with the graviton degrees of freedom induces shift-symmetry breaking corrections suppressed by powers of $\Mpl$ and proportional to background quantities, once the non-dynamical metric components are integrated out. The non-invariance under shifts of the Hamiltonian constraints, responsible for this fact, is at the core of the construction of shift-symmetric adiabatic modes on FLRW spacetimes \cite{Finelli:2017fml,Finelli:2018upr}.}
At the same time, one could have a significant departure from the Schwarzschild metric for sufficiently large values of the scalar charge. In this example, the  polarizations of QNMs observed would be the same as in GR, although their spectrum could be significantly different. As a matter of fact, the QNM spectrum could differ significantly from the GR one even if the metric remained exactly Schwarzschild, as is the case for the so-called stealth solutions~\cite{Babichev:2017guv,Minamitsuji:2018vuw}. This is because metric perturbations can mix with the scalar perturbation on a spherically symmetric background.

The spectrum of QNMs predicted by GR in the static case is comprised of two isospectral towers of modes that are respectively even and odd under parity~\cite{Chandrasekhar:1985kt}. Setting aside additional polarizations for a moment, modifications of GR can then be classified into three distinct groups, depending on whether they (1) modify the spectrum of even and odd modes while preserving isospectrality; (2) break isospectrality; or (3) mix the even and odd modes, so that there is no longer a distinction between the two. The latter possibility is realized if the scalar field that acquires a non-trivial profile is odd under parity. In fact, a spherically symmetric profile for a pseudo-scalar spontaneously breaks parity, and therefore perturbations around it are not parity eigenstates. The approach we develop in this paper will make it explicit that this third option is always a higher derivative effect.

In this paper, we will make a first step towards a systematic exploration of the three possibilities mentioned above by introducing an EFT for {\it perturbations} around static black holes with scalar hair (Sec. \ref{Sec:EFT}). Our approach will follow blueprints that were first developed in the context of inflation~\cite{Cheung:2007st}. The main idea is that, if the scalar field has a non-trivial radial profile, one can always choose to work in ``unitary gauge'' and set to zero the scalar perturbation. This can be achieved by performing an appropriate radial diffeomorphism. When this is done at the level of the action, one is left with an effective theory that is invariant under time- and angular-diffeomorphisms, but not under radial ones. 

We will show how to appropriately reorganize this action in such a way that only a finite number of terms contribute to the Lagrangian at any given order in perturbations and derivatives. As a result, we will see that at quadratic order in perturbations (which is all we need to study QNMs) and lowest order in derivatives, there are at most three operators that we can add to the Einstein-Hilbert Lagrangian. Moreover, it turns out that these three operators only affect the even sector (meaning the odd sector is completely determined by the background metric, which could still be different from Schwarzschild in general).\footnote{In other words, the relevant potential for the odd modes depends exclusively on the background metric components, with exactly the same functional dependence as in GR. This still leaves open the possibility of a modification to the odd QNM spectrum if the metric is different from Schwarzschild.}
The power of our approach lies in the fact that these three most relevant operators, whose effect will be studied in detail in Secs. \ref{Sec:odd}--\ref{sec:declim}, could in principle arise from an infinite number of diff-invariant scalar-tensor theories.
 
One obvious drawback of working with a theory that is not invariant under radial diffeomorphisms is that its action will contain arbitrary functions of the radial coordinate. These functions will in turn appear in the potential for the QNMs and, because of this, calculating their  spectrum could at first sight seem a hopeless task. In order to make it more tractable, we can exploit the fact that current gravitational wave observations appear to be consistent with GR. This suggests that, barring (un)fortunate coincidences, we can assume the background to be ``quasi-Schwarzschild''. This is certainly a natural assumption to make from an EFT viewpoint, and when coupled with the WKB approximation~\cite{Schutz:1985zz,Iyer:1986np} it allows us to express the QNM spectrum (Sec. \ref{QNMgeneral}) in terms of a small set of parameters---namely the values and derivatives of the EFT coefficients and the background metric components evaluated at the light ring. We call this procedure {\it light ring expansion}.
This state of affairs should be reminiscent of what happens in inflation, where a WKB approximation can also be employed to calculate the spectrum of primordial perturbations~\cite{Martin:2002vn}, and departures from exact de Sitter is also encoded by a small number of parameters---the first few derivatives of the inflaton potential. Our light-ring expansion is the analog of the usual slow-roll expansion in inflation.

We should emphasize however that our EFT remains a useful tool even in situations where the WKB approximation is not needed. For instance, this is the case if the EFT coefficients are known functions of the radius. Then, our approach still provides a particularly convenient way of organizing the calculation of the QNM spectrum in scalar-tensor theories.  

This is not the first time that the idea of writing down an EFT for perturbations around spherically symmetric backgrounds is put forward in the literature. For instance, an approach very similar to the one we are proposing here was first explored in~\cite{Kase:2014baa}. The authors of that paper also focused their attention on static black hole solutions with a scalar hair, and chose to work in unitary gauge.  However, they performed a 2+1+1 ADM decomposition and considered the most general action that is manifestly invariant under angular diffeomorphisms, with the additional requirement that its coefficients depend only on the radial coordinate. \footnote{Based on invariance under angular diffeomorphisms, these coefficients could in principle also depend on time. However, the requirement that they are time-independent is technically natural, because it is protected by invariance under time-translations, which is an isometry of the background we are considering.} This construction is however more general than is necessary:
it gives rise to an effective action that is not invariant under time-diffeomorphisms, and involves more free functions. Keep in mind that
a single scalar acquiring a static radial profile can only define a single preferred radial foliation, defined by the condition $\Phi = $ constant. As we discussed earlier, by working in unitary gauge one should obtain a theory that only breaks radial diffeomorphisms. For this reason, we expect that the effective action put forward in~\cite{Kase:2014baa} will generically propagate two scalar degrees of freedom (besides the graviton, of course) already at lowest order in the derivative expansion. In other words, tunings are necessary to ensure that the low energy spectrum contains a single scalar mode in the construction
of \cite{Kase:2014baa}. In our formalism, these tunings are already enforced by symmetries. 

More recently, a different approach to perturbations of spherically symmetric gravitational solutions was proposed in~\cite{Tattersall:2017erk}. The advantage of this approach is that it applies to any number and type of light degrees of freedom and that, unlike ours, none of these are required to have a non-trivial background configuration---{\it i.e.}, there is no need to consider ``hairy'' solutions. The downside is that an in principle straightforward but in practice quite lengthy procedure is needed to ensure invariance under diffeomorphisms. This procedure was carried out explicitly in~\cite{Tattersall:2017erk,Tattersall:2018nve} for a certain class of scalar tensor-theories assuming that the background metric is exactly Schwarzschild and that the black hole has no scalar hair. This procedure would need to be redone for more general backgrounds.

Invariance under diffeomorphisms is built into our formalism from the very beginning.\footnote{Note, however, that invariance under temporal and angular diffeomorphisms
is not maintained on a term by term basis in the action---see discussion around Eq. (\ref{EFT-ss0}).
Also, invariance under radial diffeomorphism is manifest only after introducing the Goldstone $\pi$---see
discussion in Sec. \ref{sec:declim}.} Moreover, the derivative counting in unitary gauge is such that one is naturally led to consider ``beyond Horndeski'' operators~\cite{Achour:2016rkg}. These operators do not lead to additional propagating degrees of freedom, but would naively appear to be of higher order in the formalism of~\cite{Tattersall:2017erk}. Finally, we should point out that our approach could be easily extended to include additional light degrees of freedom, as was already done for the EFT of inflation in~\cite{Senatore:2010wk}. 

{\it A road map:} let us give a road map of the main results of this paper,
especially for readers who are not interested in the technical details.

\begin{itemize}

\item The most general action for perturbations around a spherically symmetric
background at second order in derivatives is given in Eq. (\ref{EFT-ss0}). Note that the background needs not be that of
a black hole in general. The background metric is given by Eq. (\ref{GMN-2}) while
the background scalar profile is some general function of radius $\bar\Phi(r)$. 

\item  The quadratic action for perturbations in Eq. \eqref{EFT-ss0} does not involve any epsilon tensor. This means that, at quadratic order in derivatives, there is no way to tell whether the underlying covariant scalar-tensor theory involves a pseudoscalar or a scalar. This implies that mixing between even and odd modes cannot occur at this order in the derivative expansion. In other words, mixing is always a higher derivative effect.

\item The EFT for perturbations involves free functions of the radius $r$. For practical applications, the freedom can be much reduced by assuming small departures from GR. 
In that case, the position of the light ring is slightly displaced from its GR value. Adopting the WKB approximation, we introduce the light-ring expansion to express the quasi-normal spectrum in terms of the potential and its derivatives at the GR light ring. This is discussed in Sec. \ref{sec:WKB} and a concrete example is provided in \ref{sec:the inverse problem}.

\item The quadratic action simplifies greatly if one restricts
to the leading order in derivatives, given in Eq. (\ref{appEFT-ex1}) for odd perturbations,
and Eq. (\ref{action even lowest order}) for even perturbations.
There are only 2 free radial functions $\Lambda(r)$ and $f(r)$ in the odd perturbation action
(3 in the even perturbation action, with the addition $M_2^4 (r)$), assuming
a conformal transformation to the Einstein frame has been performed.
Working out the experimental signatures would require specification
of the coupling of matter to both the metric and scalar perturbations (see discussion in
Sec.  \ref{sec:outlook}). 

\item For this lowest-order-in-derivatives action,  there is a simple
(tadpole) constraint on the background metric given by Eq. (\ref{consrEe}).
In this case, the functions $\Lambda(r)$ and $f(r)$ are completely determined by the
background metric (Eqs. (\ref{fm}) and (\ref{lm}), setting $\alpha = 0$ and 
$M_1 = \Mpl$ assuming conformal transformation to Einstein frame). 

\item The general metric perturbations are labeled in a manner following Regge and Wheeler
\cite{Regge:1957td} (Eqs. (\ref{odd metric perturbations version 0}) and (\ref{even metric perturbations version 0}), or
more explicitly Eqs. (\ref{odd metric perturbations})  and (\ref{even metric perturbations})).
We adopt the Regge-Wheeler-unitary gauge whenever a gauge choice is made.
This means the scalar field fluctuation $\delta\Phi = 0$,  the odd sector metric perturbation
${\rm h_2} = 0$ (Eq. (\ref{odd metric perturbations})) and the even sector metric perturbations
$\mathcal{H}_0 = G = 0$ (Eq. (\ref{unitary gauge metric pert})). Note that Regge and Wheeler
also chose $\mathcal{H}_1 = 0$ which we can no longer do because of having set $\delta\Phi = 0$.

\item For the lowest-order-in-derivatives action, the odd sector perturbations obey a simple equation of motion (see Eqs. (\ref{sch-odd-e1})
and (\ref{rwpot})), which takes the same form as in GR i.e. it has the same dependence on the background metric (\ref{GMN-2}) and its derivatives, though the background metric itself can in general be different from that of GR. The corresponding action can be read off from Eqs. (\ref{japa1}) 
and (\ref{u and v}) by setting $\alpha =  M^{2}_{10} = M_{12} = 0$.
The dynamics of the even sector is significantly more complicated than that of the odd sector---the same is true in GR, but we also have an additional scalar mode----and the corresponding action is given
by Eq. (\ref{action even lowest order}). Nonetheless, the speeds of propagation of the even modes
are simply expressed by Eq. (\ref{eigencs}).

\item Our action is invariant under time and angle diffeomorphisms. Restoring radial diffeomorphism invariance can be achieved by introducing a Goldstone mode $\pi$. This, and the decoupling limit, is discussed in Sec. \ref{sec:declim}.

\item For readers interesting in going beyond the lowest order in derivatives, for instance
necessary to describe theories such as the galileon or a scalar coupled to the Gauss-Bonnet term, 
a discussion can be found in Sec. \ref{sec:nlorder}.
\end{itemize}

\vspace{0.5cm}

{\it Conventions:} throughout this paper we will work in units such that $c=\hbar=1$ and adopt a ``mostly plus'' metric signature. Unless otherwise specified, we will work in spherical coordinates with Greek indices $\mu, \nu, \lambda, ...$ running over the $(t,\theta, \phi, r)$ components, lowercase Latin indices $a, b, c, ...$ from the beginning of the alphabet running over the $(t, \theta, \phi)$ components, and lowercase Latin indices $i,j,k, ...$ from the middle of the alphabet running over the $(\theta, \phi)$. Finally, we will denote the scalar field with $\Phi$, to avoid any potential confusion with the angular variable $\phi$.

{\it A note regarding our notation on the transformed quantities.}
By transform we mean spherical harmonic transform and/or Fourier transform.
Take the example of the metric fluctuation variable ${\rm h_0}$ (Eq. (\ref{odd metric perturbations version 0}) or Eq. (\ref{odd metric perturbations})). We use the same symbol ${\rm h_0}$ to denote
(1) the fluctuation in configuration space i.e. a function of time and space ${\rm h_0} (t, r, \theta,\phi)$,
or (2) the fluctuation in spherical harmonic space i.e. ${\rm h_0} (t, r, \ell, m)$ (where the
spherical harmonic $Y_{\ell m} (\theta,\phi)$ has been used),
or (3) the fluctuation in Fourier/spherical harmonic space i.e. ${\rm h_0} (\omega, r, \ell, m)$
(where the spherical harmonic $Y_{\ell m} (\theta, \phi)$ and the Fourier wave $e^{-i\omega t}$ have been used). We often omit the arguments for ${\rm h_0}$ altogether and rely on the context 
to differentiate between these different meanings. See the discussion around Eqs. (\ref{H0 transform 1}) and (\ref{H0 transform 2}) for more details.

\section{Effective theory in unitary gauge}
\label{Sec:EFT}

In this Section we construct an effective theory for perturbations around static and spherically symmetric backgrounds.
We highlight here the main steps, mostly focusing on the results, and refer to App. \ref{Sec:constructionEFT} for all the details.
The procedure closely follows the logic underlying the construction of the EFT of inflation \cite{Creminelli:2006xe,Cheung:2007st}, but with respect to the case of Friedmann-Lema\^itre-Robertson-Walker (FLRW) spacetimes some important differences arise at the level of perturbations, as we shall discuss in details.

We assume that the theory consists of the metric $g_{\mu\nu}$ and a single scalar degree of freedom $\Phi$, which takes on an $r$-dependent profile $\bar \Phi(r)$ that sources the background metric $\bar g_{\mu\nu}$, defined by
\begin{equation}
\D s^2 = -a^2(r)\D t^2 + \frac{\D r^2}{b^2(r)} + c^2(r)\left(\D\theta^2+\sin^2\theta\D \phi^2 \right) \, .
\label{GMN-2}
\end{equation} 
Notice that, without loss of generality, one is always free to rescale the radial coordinate in such a way to get rid of one of the three functions in \eqref{GMN-2}. Nevertheless, in this Section, we will keep the metric in the redundant  form \eqref{GMN-2}: this will make the comparison with known models in the current literature more transparent. 

A convenient way of describing the low energy physics for the tensor and scalar excitations is to work in the unitary gauge, defined by $\delta \Phi\equiv0$. This condition is equivalent to using the radial diffeomorphism invariance to fix a specific hypersurface in the radial foliation of the spacetime manifold.\footnote{Notice that this requires a nontrivial background scalar profile, i.e. $\bar{\Phi}'(r)\neq0$.}
After gauge fixing, the residual symmetries of the action are the temporal and angular diffeomorphisms. Therefore, besides the Riemann tensor, the full metric $g_{\mu\nu}$, covariant derivatives $\nabla_\mu$ and the epsilon-tensor $\epsilon^{\mu\nu\lambda\rho}$, the most general unitary gauge action contains as additional ingredients the contravariant component $g^{rr}$, the extrinsic curvature $K_{\mu\nu}$ associated with equal-$r$ hypersurfaces and arbitrary functions of $r$. Explicitly, it takes on the form
\begin{equation}
S = \int\D^4 x\sqrt{-g} \, \mathcal{L}\left(g_{\mu\nu}, \epsilon^{\mu\nu\lambda\rho}, 
R_{\mu\nu\alpha\beta}, g^{rr} , K_{\mu\nu} ,\nabla_\mu , r 
\right) \, .
\label{Tintro}
\end{equation}
Now, any bona fide effective theory for perturbations around the time-independent, spherically symmetric background metric \eqref{GMN-2} can be obtained by expanding each operator in the action \eqref{Tintro} in fluctuations up to some order in the number of fields and derivatives. In this respect, it turns out that the symmetries of the background play a crucial role in dictating the structure of the building blocks entering the final action for perturbations. In order to make this fact manifest, it is worth reviewing briefly how the construction is implemented in the context of FLRW backgrounds \cite{Creminelli:2006xe,Cheung:2007st}. This will also help clarifying the differences arising in systems with spherical symmetry.
Let us denote with $\mathcal{O}=\lbrace R_{\mu\nu\alpha\beta}, g^{00} , K_{\mu\nu} \rbrace$ the building blocks of the EFT on time-foliated FLRW spacetimes \cite{Cheung:2007st}. Here $K_{\mu\nu}$ is now the extrinsic curvature associated with the constant-time hypersurface, not to be confused with the analogue in \eqref{Tintro}.
Since the hypersurface is maximally symmetric, a generic background quantity $\bar{\mathcal{O}}$ can always be written just in terms of the background metric, the unit normal vector $n_\mu$ and functions of time \cite{Cheung:2007st}. For instance, $\bar K_{\mu\nu}=H(t)\bar  h_{\mu\nu}$, where $h_{\mu\nu}$ is the induced metric and $H(t)$ is the Hubble function. This remarkable fact allows to define the perturbation $\delta\mathcal{O}$ associated with a generic operator $\mathcal{O}$ as follows: $\mathcal{O}=\mathcal{O}^{(0)}+\delta\mathcal{O}$, where \textit{(i)} $\mathcal{O}^{(0)}$ contains the background value, \textit{i.e.} $\overbar{{\mathcal{O}}^{(0)}}\equiv\bar{\mathcal{O}}$ (where `` $\bar{}$ '' denotes setting the perturbations to zero), while $\delta\mathcal{O}$ starts linearly in perturbations,  {\it and} \textit{(ii)} both $\mathcal{O}^{(0)}$ and $\delta\mathcal{O}$ transform covariantly. For instance, one can split $K_{\mu\nu}=H(t)h_{\mu\nu}+\delta K_{\mu\nu}$, where both $K_{\mu\nu}^{(0)}\equiv H(t) h_{\mu\nu}$ and $\delta K_{\mu\nu}$ are covariant quantities \cite{Cheung:2007st}. As a result, all the operators in the Lagrangian for perturbations of \cite{Cheung:2007st} are separately invariant under residual (spatial) diffeomorphisms, and counting
powers of $\delta K_{\mu\nu}$ is the same as counting the order of perturbations.
In the class of theories that non-linearly realize time translation invariance, this turns out to be a distinctive feature of the FLRW subclass and crucially relies on the high degree of symmetry of the background (i.e. homogeneity and isotropy) \cite{Cheung:2007st}. 
By contrast, in the case of non-maximally symmetric backgrounds of the type in \eqref{GMN-2}, one cannot define for all the operators in \eqref{Tintro} perturbations that transform covariantly under residual (temporal and angular) diffeomorphisms. As an example, consider the extrinsic curvature $K_{\mu\nu}$ in \eqref{Tintro}. On the background, $\bar K_{ab}=\frac{1}{2}\partial_r\bar  h_{ab}$ (see App. \ref{Sec:constructionEFT}) and it is clear from \eqref{GMN-2} that $\bar K_{ab} \, \cancel{\propto} \, \bar h_{ab}$. As a byproduct, there is no way to define a $K_{\mu\nu}^{(0)}$ in terms of the metric only in such a way that it transforms covariantly.\footnote{For instance, one might be tempted to define $K_{\mu\nu}^{(0)} = \frac{1}{2}\partial_r h_{\mu\nu}$, but
this is not a good tensor under $r$ dependent ($t,\theta,\phi$) diffeomorphisms.}
Thus, we define $\delta K_{\mu\nu}$ by $K_{\mu\nu} - \bar K_{\mu\nu}$, and it is not a covariant tensor.
The two main consequences of this fact are: \textit{i)} new independent operators are in principle allowed at any order in perturbations, including as we will see one additional tadpole; \textit{ii)} invariance under residual (temporal and angular) diffeomorphisms in general will not be manifest in the Lagrangian at a given order in perturbations
i.e. an object like $\delta K_{\mu\nu} \delta K^{\mu\nu}$ is not invariant because 
$\delta K_{\mu\nu}$ is not a covariant tensor.

In the case of spherically symmetric backgrounds, the perturbation of a given operator $\mathcal{O}_i$ that belongs to the building blocks $\lbrace R_{\mu\nu\alpha\beta}, g^{rr} , K_{\mu\nu} \rbrace$ or their derivatives can be defined by subtracting the background value of the operator, $\delta\mathcal{O}_i \equiv \mathcal{O}_i-\bar{\mathcal{O}}_i$. Even if the $\delta\mathcal{O}_i $ so defined does not transform covariantly, at a given order $n$ in the number of perturbations, the most general action will be of the form: 
\begin{equation}
S^{(n)} = \int\D^4 x\sqrt{-g} \,  \sum_{i_1, \dots, i_n} C^{(n)}_{i_1, \dots, i_n}(r) \delta\mathcal{O}_{i_1} \dots \delta\mathcal{O}_{i_n} \; ,
\label{actionperturb}
\end{equation} 
where the indices $i_m$ run on the operators up a to given order in derivatives. Now, for every choice of the functions of the radial coordinate $C^{(n)}_{i_1, \dots, i_n}(r)$ there is clearly a gauge invariant Lagrangian (with respect to temporal and angular diffs.) of the form (\ref{Tintro}) such that its expansion in perturbations up to order $n$ gives the desired coefficients. 
On the other hand, in general, at fixed order $n$ a specific Lagrangian (\ref{Tintro}) in two different gauges will give rise to an action for perturbations with different values for the coefficients $C^{(n)}_{i_1, \dots, i_n}(r)$, so a Lagrangian of the form (\ref{actionperturb}) is well defined only once the gauge choice for perturbations is made.    
This aspect, as we will see in the next sections, does not turn out to be a limitation in any practical application of the EFT for perturbations that we are constructing.

As we prove in App. \ref{Sec:constructionEFT}, the most general action for perturbations in unitary gauge up to quadratic order and with no more than two derivatives can be written as 
\begin{equation}
\begin{split}
& S =  \int\D^4x \, \sqrt{-g} \bigg[
\frac{1}{2}M^2_1(r) R -\Lambda(r) - f(r)g^{rr} - \alpha(r)\bar K_{\mu\nu} K^{\mu\nu}
\\
&	+ M_2^4(r)(\delta g^{rr})^2
	+M_3^3(r) \delta g^{rr }\delta K  + M_4^2(r) \bar K_{\mu\nu} \delta g^{rr }\delta K^{\mu\nu}
\\
&	+ M_5^2(r)(\partial_r\delta g^{rr})^2
	+M_6^2(r) (\partial_r\delta g^{rr})\delta K  + M_7(r)\bar  K_{\mu\nu} (\partial_r\delta g^{rr})\delta K^{\mu\nu}
	+ M_8^2(r)(\partial_a\delta g^{rr})^2
\\
&	+M_9^2(r)(\delta K)^2 + M_{10}^2(r)\delta K_{\mu\nu}\delta K^{\mu\nu} 
+ M_{11}(r) \bar K_{\mu\nu}\delta K \delta K^{\mu\nu} 
+ M_{12}(r) \bar  K_{\mu\nu}\delta K^{\mu\rho} \delta {K^{\nu}}_\rho 
\\
&	+ \lambda(r)\bar  K_{\mu\rho}{{\bar{K}}^{\rho}}_\nu \delta K \delta K^{\mu\nu}
+ M_{13}^2(r) \delta g^{rr } \delta\hat{R} 
+ M_{14}(r) \bar K_{\mu\nu} \delta g^{rr }\delta\hat{R}^{\mu\nu} + \ldots
\bigg] \, ,
\label{EFT-ss0}
\end{split}
\end{equation}
where $\hat R_{\mu\nu}$ is the Ricci tensor built out of the induced metric $h_{\mu\nu}$. 

A few comments are in order at this point. As anticipated above, one has formally a larger number of operators for perturbations with respect to \cite{Cheung:2007st}. In particular, there is in principle an additional tadpole parametrized by the function $\alpha(r)$, which, together with $\Lambda(r)$ and $f(r)$, will be determined by the Einstein equations.\footnote{The analog of the additional tadpole term in the case of the EFT for inflation
would be ${K}_{\mu\nu}^{(0)} K^{\mu\nu}$, and this can be shown to be rewritable in
terms of the other tadpole terms, for an FLRW background (see Appendix of \cite{Cheung:2007st}).}

Furthermore, notice that in general the $r$-dependence of the coefficient $M^2_1(r)$ can be re-absorbed by a conformal transformation that brings us back to the Einstein frame. This will generically affect the coupling to additional matter fields, which have been left unexpressed in \eqref{EFT-ss0}. We will have more to say about this in Sec. \ref{sec:outlook}.

A remarkable feature of the quadratic action above is that it does not depend on the epsilon tensor. In the covariant action for the underlying scalar-tensor theory, epsilon tensors will appear if either (1) the action is not invariant under parity, or (2) the scalar field is odd under parity, {\it i.e.} it is a pseudo-scalar. In the first case parity is explicitly broken, in the second case it is spontaneously broken by the scalar hair. Both types of breaking would lead to a mixing between even and odd perturbations. However, the fact that an epsilon tensor cannot appear at second order in derivatives in perturbations means that mixing is always a higher derivative effect. In other words, at lowest order parity is an accidental symmetry of our action for perturbations.\footnote{A concrete example is provided by 
Chern-Simons gravity \cite{Alexander:2007zg,Molina:2010fb}. In this case, the non-minimal coupling $\Phi R_{\mu\nu\lambda\rho} \tilde{R}^{\mu\nu\lambda\rho}$ requires $\Phi$ to be a pseudoscalar. The expansion of this term up to quadratic order in perturbations contains several terms. Some of these terms contain two derivatives acting on perturbations and therefore yield a contribution to the action in Eq. \eqref{EFT-ss0}. However, the only term in which derivatives and metric perturbations are actually contracted with each other through an epsilon tensor is $\bar{\Phi} \delta R_{\mu\nu\lambda\rho} \delta \tilde{R}^{\mu\nu\lambda\rho}$. This term contains two derivatives acting {\it on each} metric perturbation, and thus it is a higher derivative correction to the action \eqref{EFT-ss0}.}

It is worth also commenting on the fact that some of the combinations among the operators in \eqref{EFT-ss0} may secretly propagate an extra unwanted ghost-like degree of freedom.\footnote{This well-known fact has already been discussed in the unitary gauge language in the context of  the EFT for FLRW spacetimes in \cite{Langlois:2017mxy}.}  The absence of this kind of pathology can be guaranteed by enforcing degeneracy conditions, which determine specific relations among some of the coefficients in \eqref{EFT-ss0} therefore reducing the number of independent operators in the EFT. This study requires in general a detailed classification of the Hamiltonian constraints associated with \eqref{EFT-ss0}. However, since in the rest of the paper we will mainly focus on the leading order terms in the derivative expansion, these complications will never affect our discussion and hence can be safely disregarded in the following. 

Finally, we conclude this section stressing again that the form of the unitary gauge action \eqref{EFT-ss0} is dictated only by the spontaneous breaking pattern of the Poincaré group down to spatial rotation and time translation invariance.
In other words, there is no input from additional internal or spacetime symmetries, which would further constrain the couplings in the effective action \eqref{EFT-ss0}. We will not discuss this possibility here, leaving it for future work.\footnote{For a discussion on how to impose additional internal symmetries, {\it e.g.} a shift symmetry of the type $\Phi\rightarrow\Phi+c$, in the unitary gauge action, we refer the interested reader to \cite{Finelli:2018upr}, where this is explained in details in the context of FLRW cosmologies.}

\subsection{Tadpole conditions} \label{sec:tadpoles}

In this Section we focus on the tadpole operators in the EFT \eqref{EFT-ss0},
\begin{equation}
S_{\text{tadpole}} = \int\D^4 x\sqrt{-g}\left[ 
\frac{1}{2}M^2_1(r) R - f(r)g^{rr}-\Lambda(r) - \alpha(r)\bar K_{\mu\nu} K^{\mu\nu} \right] \, .
\label{bSm}
\end{equation}
In particular, we shall see that the Einstein equations can be used to fix $\Lambda$ and $f$ in terms of the background metric \eqref{GMN-2}, $M^2_1$ and $\alpha$. In addition, they provide a first order differential equation which can be used to relate $M^2_1$ and $\alpha$. In this respect, we start writing the most general energy momentum tensor compatible with the symmetries of the system as ${T^\mu}_\nu= \text{diag}(-\rho, p_r,p_\Omega,p_\Omega)$, being $p_\Omega$ and $p_r$ the tangential and radial pressures respectively. Plugging into the Einstein equations
\begin{equation}
\left( R_{\mu\nu} -\frac{1}{2}Rg_{\mu\nu} - \nabla_\nu  \nabla_\mu + g_{\mu\nu} \square \right) M^2_1 -
T_{\mu\nu} = 0 \, , 
\end{equation}
one can solve for the fluid variables and find
\begin{align}
{T^0}_0 &=-\rho =  b^2 \left(  \frac{2c''}{c}+\frac{c'^2}{c^2}+\frac{2b'c'}{bc}-\frac{1}{b^2c^2}\right)M^2_1 + b^2\left(\frac{b'}{b}+ \frac{2c'}{c} \right)(M^2_1)' + b^2(M^2_1)'' \, ,
\label{Ee0}\\
{T^r}_r&=p_r = b^2\left( \frac{2a'c'}{ac}+\frac{c'^2}{c^2}-\frac{1}{b^2c^2} \right)M^2_1 + b^2\left(\frac{a'}{a}+\frac{2c'}{c}\right)(M^2_1)'  \, ,
\label{Eer}\\
{T^i}_j&= \delta^i_j \, p_\Omega = \delta^i_j \, b^2\left[\left( \frac{a''}{a}+\frac{c''}{c}+\frac{a'b'}{ab}+\frac{a'c'}{ac}+\frac{b'c'}{bc} \right)M^2_1 + \left(\frac{a'}{a}+\frac{b'}{b}+\frac{c'}{c}\right)(M^2_1)' + (M^2_1)'' \right]  .
\label{Eet}
\end{align}
For our purpose, $T_{\mu\nu}$ comes from the terms in Eq. (\ref{bSm}) beyond the Einstein Hilbert term.
Using Eq. \eqref{dKMNv} for the variation of the extrinsic curvature, the background energy momentum tensor $T_{\mu\nu}$ associated with the tadpole action \eqref{bSm} reads
\begin{multline}
T_{\mu\nu}  = -(f g^{rr}+\Lambda+\alpha K_{\alpha\beta}K^{\alpha\beta}) g_{\mu\nu} + 2 f \delta_\mu^r\delta_\nu^r
\\
 -\alpha K_{\alpha\beta}K^{\alpha\beta}n_\mu n_\nu
	-  \nabla_\lambda (\alpha K^\lambda_\mu n_\nu) - \nabla_\lambda (\alpha K^\lambda_\nu n_\mu)
	+  \nabla_\lambda (\alpha K_{\mu\nu}n^{\lambda}) \, ,
\end{multline}
where we dropped the bar everywhere for simplicity.
Substituting in \eqref{Ee0}-\eqref{Eer}, one can solve for $\Lambda(r)$ and $f(r)$:
\begin{multline}
f(r) 
=  \left(\frac{a'c'}{ac}-\frac{b'c'}{bc} - \frac{c''}{c}  \right)M^2_1 + \frac{1}{2}\left(\frac{a'}{a}-\frac{b'}{b}\right)(M^2_1)' - \frac{1}{2}(M^2_1)'' 
\\
- \left( \frac{3a'^2}{2a^2}-\frac{a'b'}{2ab}-\frac{a'c'}{ac} +\frac{c'^2}{c^2} - \frac{a''}{2a}\right)\alpha + \frac{a'}{2a}\alpha' \, , 
\label{fm}
\end{multline}
\begin{multline}
\Lambda(r) 
= - b^2\left(\frac{c''}{c} + \frac{a'c'}{ac}+ \frac{b'c'}{bc}+\frac{c'^2}{c^2}-\frac{1}{b^2c^2}  \right)M^2_1 - b^2\left(\frac{a'}{2a}+\frac{b'}{2b}+\frac{2c'}{c} \right) (M^2_1)' - \frac{b^2}{2}(M^2_1)''
\\ 
- b^2\left( \frac{3a'^2}{2a^2}-\frac{a'b'}{2ab}-\frac{a'c'}{ac} +\frac{c'^2}{c^2} - \frac{a''}{2a}\right)\alpha + \frac{b^2a'}{2a}\alpha'
 \, .
\label{lm}
\end{multline}
Eq. \eqref{Eet} provides a differential equation for the combination $ M^2_1 +\alpha$:
\begin{multline}
\left(\frac{a'}{a}-\frac{c'}{c} \right)(M^2_1+\alpha)' + \left( \frac{a''}{a} - \frac{c''}{c} + \frac{a'b'}{ab} + \frac{a'c'}{ac}- \frac{b'c'}{bc}-\frac{c'^2}{c^2}+\frac{1}{b^2c^2}  \right)(M^2_1+\alpha)  \\
\qquad + 2 \alpha\left( \frac{c'^2}{c^2} - \frac{a'^2}{a^2} - \frac{1}{2b^2 c^2}\right)= 0 \, .
\label{thirdtadpcondition}
\end{multline}
After introducing $\tilde M^2_1 \equiv M^2_1 + \alpha$, one can solve this equation algebraically for $\alpha$ and then plug the solution back into Eqs. \eqref{fm} and \eqref{lm}, thus obtaining  three expressions for $f, \Lambda$ and $\alpha$ in terms $\tilde M^2_1$ and its derivatives.
In other words, of the $4$ free functions of radius in the tadpole action (\ref{bSm}), only one combination 
(\textit{i.e.} $\tilde M^2_1$) is truly free, the others are fixed once $\tilde M^2_1$ and the background metric are specified.

It is worth noticing that, while the first three terms in \eqref{bSm} are generically expected in every genuine theory describing the dynamics of a scalar degree of freedom coupled to the two graviton helicities, the tadpole $\alpha(r)\bar K_{\mu\nu} K^{\mu\nu}$ is present only in higher derivative theories involving powers of the extrinsic curvature $K_{\mu\nu}\sim \nabla_\mu\nabla_\nu \Phi$. For instance, in the context of theories with second order equations of motion, this is the case of the quartic and quintic Horndeski operators \cite{Nicolis:2008in,Horndeski:1974wa,Deffayet:2011gz}. Otherwise, in theories involving at most the cubic Horndeski \cite{Nicolis:2008in,Horndeski:1974wa,Deffayet:2011gz}, i.e. with a Lagrangian of the form $P(X,\Phi)+G(X,\Phi)\square\Phi$ where $X\equiv \nabla_\mu\Phi\nabla^\mu\Phi$, the $\alpha$-tadpole is not generated. Then, the background equations are given by \eqref{fm}-\eqref{thirdtadpcondition} with $\alpha=0$. Forgetting for a moment possible couplings to additional matter fields and setting $M_1\equiv\Mpl$, it is clear that for theories belonging to the second case (i.e. with $\alpha=0$) Eq. \eqref{thirdtadpcondition} reduces to a consistency equation for the scale factors of the background metric:
\begin{equation}
\frac{a''}{a} - \frac{c''}{c} + \frac{a'b'}{ab} + \frac{a'c'}{ac}- \frac{b'c'}{bc}-\frac{c'^2}{c^2}+\frac{1}{b^2c^2} =0 \, ,
\label{consrEe}
\end{equation}
corresponding to $p_\Omega+\rho=0$ in the fluid language.

Having discussed so far the tadpole Lagrangian \eqref{bSm} and the conditions induced by the background Einstein equations, the next step is to consider operators that are quadratic in perturbations. The goal is to derive the linearized equations governing the dynamics of the $2+1$ physical degrees of freedom, which carry the information about the spectrum of the QNMs. To this end, one should first choose a parametrization for the metric perturbations $\delta g_{\mu\nu}$. Given the symmetries of the background, it turns out to be convenient to decompose them into tensor harmonics~\cite{Regge:1957td} and distinguish between ``even'' and ``odd'' (sometimes called respectively ``polar'' and ``axial'', {\it e.g.}~\cite{Chandrasekhar:1985kt}) perturbations, depending on how they transform under parity. Indeed, the spherically symmetric geometry of the background guarantees that the corresponding linearized equations of motion do not couple. However, before deriving them explicitly for a theory in the form \eqref{EFT-ss0}, we find it useful to present some general properties of QNMs.


\section{Quasi-normal modes: general considerations}
\label{QNMgeneral}

We are interested in a metric of the form
$g_{\mu\nu} = \bar g_{\mu\nu} + \delta g_{\mu\nu}$, where
$\bar g_{\mu\nu}$ is the static and spherically symmetric given by
Eq. (\ref{GMN-2}), and $\delta
g_{\mu\nu}$ represents the metric perturbations. In addition, we have a scalar degree of freedom
$\Phi = \bar\Phi (r) + \delta\Phi$, where the background scalar
profile $\bar\Phi$ is also static and spherically symmetric. 
The unitary gauge refers to the special choice of equal-$r$ surfaces
such that $\delta \Phi = 0$.

Under $(\theta, \phi)$ diffeomorphisms or rotations, 
$\delta g_{\mu\nu}$ can
be provisionally classified into scalar ($\delta g_{tt},  \delta g_{rr}, \delta
g_{tr}$) , vector ($\delta g_{ti}$, $\delta g_{ri}$) and 
and tensor ($\delta g_{ij}$) parts, where $i,j = \theta, \phi$.\footnote{
Note that we are abusing the terms scalar, vector and tensor slightly.
What really transform as scalar, vector and tensor are the full
metric components $g_{tt}, g_{rr}$, etc. The non-vanishing background $\bar
g_{\mu\nu}$ means the fluctuations $\delta g_{\mu\nu}$ would transform
nonlinearly (or more accurately, sublinearly i.e. variation of
$\delta g_{\mu\nu}$ 
under a small diffeomorphism would have terms independent of the metric
fluctuations, in addition to the expected terms linear in the metric fluctuations). 
}
Following Regge and Wheeler~\cite{Regge:1957td}, the scalars are called:
\begin{equation}
\delta g_{tt} = a^2 H_0 \quad , \quad \delta g_{tr} = H_1 \quad ,
\quad \delta g_{rr} = H_2 / b^2 \, .
\end{equation}

Each vector can be further decomposed into a scalar and
a pseudo-scalar:
\begin{equation}
\delta g_{ti} = \nabla_i  {\mathcal H}_0  +  \epsilon^j {}_i \nabla_j
{\rm h_0} \quad , \quad
\delta g_{ri} = \nabla_i  {\mathcal H}_1   +  \epsilon^j {}_i \nabla_j
{\rm h_1} \, ,
\end{equation}
where ${\cal H}_0$ and ${\cal H}_1$ are scalars, and
${\rm h_0}$ and ${\rm h_1}$ are pseudo-scalars.
A pseudo-scalar flips sign under
a parity transformation $(\theta, \phi) \rightarrow (\pi-\theta, \phi
+ \pi)$, whereas a scalar does not.\footnote{For instance, $\delta g_{t\theta}$ should flip sign under parity,
and $\nabla_\theta {\cal H}_0$ indeed flips sign as desired provided ${\cal
  H}_0$ is a scalar, whereas $\epsilon^\phi {}_\theta \nabla_\theta {\rm h}_0$
also flips sign under parity provided ${\rm h}_0$ is a pseudoscalar.
}
Here, $\nabla_i$ is the covariant
derivative defined with respect to the two-dimensional metric:
\begin{eqnarray}
ds^2_{\rm 2-sphere} = \gamma_{ij} \D x^i \D x^j = \D\theta^2 + \sin^2 \theta \, \D\phi^2 \, ,
\end{eqnarray}
and $\epsilon_{ji}, \epsilon^j {}_i , \epsilon_j {}^i, \epsilon^{ji}$ 
are the corresponding Levi-Civita
tensors:
\begin{eqnarray}
\left(\begin{array}{cc} \epsilon_{\theta\theta} & \epsilon_{\theta\phi} \\ \epsilon_{\phi\theta}
                                      &
                                        \epsilon_{\phi\phi} \end{array}\right)
                                        = {\,\rm sin} \, \theta 
\left(\begin{array}{cc} 0 & 1\\ -1
                                      &
                                        0 \end{array}\right) \quad ,
                                        \quad
\left(\begin{array}{cc} \epsilon^\theta {}_\theta & \epsilon^\theta {}_\phi \\
        \epsilon^\phi {}_\theta
                                      &
                                        \epsilon^\phi {}_\phi \end{array}\right)
                                        =
\left(\begin{array}{cc} 0 & {\rm sin\,} \theta\\ -1/{\rm sin\,}\theta
                                      &
                                        0 \end{array}\right) \nonumber
  \\
\left(\begin{array}{cc} \epsilon_\theta {}^\theta & \epsilon_\theta {}^\phi \\
        \epsilon_\phi {}^\theta
                                      &
                                        \epsilon_\phi {}^\phi \end{array}\right)
                                        =
\left(\begin{array}{cc} 0 & 1/{\rm sin\,} \theta\\ -{\rm sin\,}\theta
                                      &
                                        0 \end{array}\right) \quad ,
                                        \quad
\left(\begin{array}{cc} \epsilon^{\theta\theta} & \epsilon^{\theta\phi} \\ \epsilon^{\phi\theta}
                                      &
                                        \epsilon^{\phi\phi} \end{array}\right)
                                        = {1\over {\,\rm sin} \, \theta}
\left(\begin{array}{cc} 0 & 1\\ -1
                                      &
                                        0 \end{array}\right)  \, .
\end{eqnarray}

Just as in the case of the vectors, 
the tensor $\delta g_{ij}$ can be further decomposed: into a trace
and a traceless part, which in turn can be decomposed into a scalar and a
pseudo-scalar:
\begin{eqnarray}
\delta g_{ij} = c^2 ( {\cal K} + {1\over 2} G) \gamma_{ij} + c^2 (\nabla_i \nabla_j -
  {1\over 2} \gamma_{ij} ) G + {1\over 2} (\epsilon_i {}^k \nabla_k
  \nabla_j + \epsilon_j {}^k \nabla_k \nabla_i) {\rm h_2} \, ,
\end{eqnarray}
where $c^2({\cal K} + G/2)$ is the trace, and $c^2 G$ and ${\rm h_2}$ 
represent the analogs of
the E and B modes on the 2-sphere (our notation follows that of Regge
and Wheeler~\cite{Regge:1957td}). Here $c^2$ is part of the background metric as defined
in Eq. (\ref{GMN-2}), not to be confused with the speed of light
squared which is always set to unity.
It is also useful to note that the second covariant derivatives (in the
2-sphere sense) on a scalar function act as follows:
\begin{eqnarray} \label{2 cov dev sphere}
\begin{split}
& \qquad \qquad \quad \nabla_\theta \nabla_\theta = \partial_\theta^2 \quad , \quad \nabla_\phi \nabla_\phi
= \partial_\phi^2 + {\,\rm sin}\theta {\,\rm
  cos}\theta \partial_\theta 
\, , \\
& \nabla^2 = \partial_\theta^2 + {1\over {\,\rm
    sin}^2\theta} \partial_\phi^2 + {{\,\rm cos}\theta \over {\,\rm
    sin}\theta} \partial_\theta  \quad , \quad
\nabla_\theta \nabla_\phi = \nabla_\phi \nabla_\theta
= \partial_\theta \partial_\phi - {{\,\rm cos}\theta \over {\,\rm
    sin}\theta} \partial_\phi \, . 
    \end{split}
\end{eqnarray}

To summarize, before gauge fixing, the parity even fluctuations,
expressible in terms of 7 scalars, are\footnote{Notice that we are departing slightly from the notation of Regge and Wheeler~\cite{Regge:1957td} by denoting some of the perturbations with $\mathcal{H}_0, \mathcal{H}_1$ and $\mathcal{K}$ rather than $h_0, h_1$ and $K$ respectively. This is done in order to avoid any potential confusion with the induced metric, the perturbations in the odd sector, and the extrinsic curvature of surfaces of constant $r$.}
\begin{eqnarray}\label{even metric perturbations version 0}
\boxed{
\delta g_{\mu\nu}^{\rm even} = \left(
\begin{array}{ccc}
a^2 H_0 & H_1 & \nabla_j {\mathcal H}_0 \\
H_1 & H_2 / b^2 & \nabla_j {\mathcal H}_1 \\
\nabla_i{\mathcal H}_0 \,\,\,\, & \nabla_i {\mathcal H}_1 & \,\,\,\, c^2 {\cal K}
                                                   \gamma_{ij} + c^2
                                                   \nabla_i \nabla_j G
\end{array}\right)
}
\end{eqnarray}

The parity odd fluctuations, expressible in terms of 3 pseudo-scalars,
are:
\begin{eqnarray} \label{odd metric perturbations version 0}
\boxed{
\delta g_{\mu\nu}^{\rm odd} = \left(
\begin{array}{ccc}
0 & 0 & \epsilon^k {}_j \nabla_k {\rm h_0}\\
0 & 0 & \epsilon^k {}_j \nabla_k {\rm h_1}\\
\epsilon^k {}_i \nabla_k {\rm h_0} \,\,\,\, &  \epsilon^k {}_i \nabla_k {\rm
                                    h_1} & \,\,\,\, {1\over 2} (\epsilon_i {}^k
                                           \nabla_k \nabla_j +
                                           \epsilon_j {}^k \nabla_k
                                           \nabla_i ) {\rm h_2}
\end{array}\right)
}
\end{eqnarray}

In addition to the metric fluctuations, we have in a general gauge
the scalar fluctuation $\delta\Phi$ as well.

The background (metric and scalar) enjoys invariance under time translations and 
spatial rotations. Therefore, if we expand $\delta g_{\mu\nu}$ 
and $\delta\Phi$ in terms of the plane wave $e^{-i\omega t}$ and
the spherical harmonics $Y_{\ell m} (\theta, \phi)$, modes with
different $\omega, \ell, m$ will not mix at linear level (in the equations of motion).
This is similar to what happens around backgrounds invariant under spatial translations, where a Fourier expansion in the spatial coordinates proves to be useful because different Fourier modes do not mix at linear level.

In order to avoid introducing too many symbols, we will henceforth
abuse the notation a bit by occasionally doing the following
replacements for each scalar or pseudoscalar, {\it e.g.}:
\begin{eqnarray}
\label{H0 transform 1}
H_0 (t, r, \theta, \phi) \rightarrow H_0 (t, r, \ell, m) Y_{\ell m}
  (\theta, \phi) \, ,
\end{eqnarray}
or
\begin{eqnarray}
\label{H0 transform 2}
H_0 (t, r, \theta, \phi) \rightarrow H_0 (\omega, r, \ell, m)
  e^{-i\omega t} Y_{\ell m}
  (\theta, \phi) \, .
\end{eqnarray}
To compound the possible confusion, we will often leave out the arguments of $H_0$ altogether! 
However, in most cases, the context should be sufficient to tell apart
the different meanings---of $H_0$ in different spaces.
In cases where confusion could arise, we will make the meaning
explicit. At the level of the spherical harmonic transformed quantities,
because $Y_{\ell m} (\pi-\theta, \phi + \pi) = (-1)^\ell Y_{\ell m}
(\theta, \phi)$, we see that a scalar such as
$H_0 (t,r,\ell, m)$ picks up a factor of $(-1)^\ell$
under parity while a pseudo-scalar such as
${\rm h_0} (t,r,\ell, m)$ picks up a factor of $(-1)^{\ell+1}$ under parity.
As discussed around Eq. (\ref{EFT-ss0}), the quadratic action with at most two derivatives respects parity.
Thus, the parity even and odd modes do not mix.

Without loss of generality, it is customary to set $m=0$.
In a spherically symmetric background,
the radial and time dependence of the perturbations
is sensitive to $\ell$ but not $m$. This is because there is no
preferred z-axis around which azimuthal rotations are defined,
and $m \ne 0$ modes can be obtained from an $m=0$ 
mode by simply rotating the z-axis.

For $m=0$ (i.e. $\partial_\phi = 0$), 
and a gravitational wave that propagates in the radial
direction, one can see that
\begin{eqnarray}
\delta g_{ij}^{\rm GW} = 
{1\over 2} \left( \begin{array}{cc}
 (\partial_\theta^2 - {{\,\rm cos}\theta \over {\,\rm
    sin}\theta} \partial_\theta) c^2 G & \,\,\,\, ( {\,\rm
  cos}\theta \partial_\theta - {\,\rm sin}\theta \partial_\theta^2 ) {\,\rm h_2}
\\
 ( {\,\rm
  cos}\theta \partial_\theta - {\,\rm sin}\theta \partial_\theta^2 ) {\,\rm h_2}&
\, \, \,\, -  {\,\rm sin}^2\theta (\partial_\theta^2 - {{\,\rm cos}\theta \over {\,\rm
    sin}\theta} \partial_\theta) c^2 G 
\end{array} \right) \, .
\end{eqnarray}
Thus, $G$ and ${\rm h_2}$ play the roles of the two planar
polarizations of the graviton (in the large $r$ limit such that a spherical
wave is locally well approximated by a plane wave).
Note that in the even sector, the angular components
of the metric take the form $c^2 {\cal K} \gamma_{ij} + \nabla_i \nabla_j
c^2 G$ which is better rewritten as $c^2 ({\cal K} + \nabla^2 G /
2)\gamma_{ij} + (\nabla_i \nabla_j - [\gamma_{ij}/2] \nabla^2)
c^2 G$, and we have ignored the trace part when focusing on
gravitational waves. 

Summarizing the discussion above, we have the metric fluctuations $\delta g_{\mu\nu}$
labeled according to Eqs. (\ref{even metric perturbations version 0})
and (\ref{odd metric perturbations version 0}), and the scalar fluctuation $\delta\Phi$,
in a general gauge. The next step is to choose a gauge. As will be detailed below, the
gauge we adopt is the Regge-Wheeler-unitary gauge, meaning:
\begin{eqnarray}
\delta\Phi = 
{\rm h_2} = 
{\mathcal H}_0 = G = 0 \, .
\end{eqnarray}
Regge and Wheeler \cite{Regge:1957td} 
also set ${\mathcal H}_1$ to zero, but our unitary gauge choice of $\delta\Phi = 0$
makes that impossible in general.

Once this is done, a Schr\"odinger-like equation can typically be obtained (after quite a bit of algebra!)
separately for the odd and the even sectors:
\begin{equation}
\partial_{\tilde r}^2 Q + W Q = 0 \quad ,
\label{SclikEq}
\end{equation}
where $\tilde r$ is some redefined radial coordinate chosen in such a way that the speed of propagation equals one (see App. \ref{schf} for more details); the variable $Q$ may have one or more components, obtained by combining perturbations and their derivatives; $W$ is a function of $\tilde r$ as well as $\omega$ and $\ell$ (or a matrix of functions, if $Q$ has more than one component). An example is the Regge-Wheeler equation given in Eq. (\ref{RWequation}) for odd perturbations in GR.

The spectrum of QNMs is usually calculated by solving equation \eqref{SclikEq}
numerically, imposing the appropriate boundary conditions at the horizon and at spatial infinity. However, very useful insights can be obtained by combining the EFT we introduced in Sec. \ref{Sec:EFT} with approximate analytic methods.

\section{WKB approximation and light ring expansion}\label{sec:WKB}

There are in principle three possible ways in which Eq. \eqref{SclikEq} could differ from the standard GR result: (1) the coordinate $\tilde{r}$ could differ from the tortoise coordinate defined in GR---see Eq. (\ref{rstar}); (2) the variable $Q$ could be modified (for instance, in the even sector of scalar tensor theories $Q$ would have two components, in which case $W$ would be a matrix); (3) the QNM potential $-W$ could have a different shape. Ultimately, the spectrum of QNM is completely determined by the shape of~$W(\tilde r)$.  In the following, we will focus on the simplest case where $W$ is just a single function rather than a matrix of functions.

Schutz and Will~\cite{Schutz:1985zz} showed that the quasi-normal spectrum associated with the equation \eqref{SclikEq} can be approximated analytically using the WKB method. Their main result is the relation
\begin{equation}
\label{WKB}
{W \over (2 \partial_{\tilde{r}}^2 W)^{1/2}}  \Big|_{\tilde{r} = \tilde{r}_*} 
=  - i \left( n + {1\over 2} \right) \, ,
\end{equation}
where $n = 0, 1, 2, ...$ and $\tilde{r}_*$ is the position of the maximum of $-W$.\footnote{With a certain abuse of terminology, we shall use sometimes the notion of ``light ring'' to refer to $\tilde{r}_*$. Even if, strictly speaking, the two positions tend to coincide only in the eikonal limit $\ell\rightarrow\infty$, for a generic $\ell\sim$ few they differ by an amount that is within the WKB accuracy. This makes the abuse consistent for practical purposes and justifies the definition ``light ring expansion'' to denote the procedure outlined below.}
Since $W$ depends on the frequency $\omega$, this expression defines implicitly the complex quasi-normal frequencies.
This is the lowest order WKB result (accurate at the few percent
level), and can be improved upon
if desired~\cite{Iyer:1986np} (see discussion below). A side remark on conventions: the sign on the RHS of eq. \eqref{WKB} is consistent with the fact that the time dependence of our solution is of the form $e^{- i \omega t}$---see {\it e.g.} Eq. \eqref{H0 transform 2}. Thus, we are following the convention used in~\cite{Iyer:1986np}, and our equation differs by a sign compared to the one in~\cite{Schutz:1985zz}, where the  solutions were proportional to $e^{ i \omega t}$.

The WKB result \eqref{WKB} shows that the quasi-normal frequencies are  actually sensitive to a small region of the potential around the light ring.\footnote{This is true under the assumption that the asymptotic behavior of $W$ allows us to impose standard boundary conditions.  Notice that such an approach is supported by explicit examples \cite{Blazquez-Salcedo:2016enn,Witek:2018dmd} where the QNM spectrum is studied perturbatively in the coupling constant between the scalar field and the Gauss-Bonnet term. Conversely, we refer \textit{e.g.} to \cite{Cardoso:2016rao,Cardoso:2017njb} for a discussion about non-perturbative effects like echoes and resonances. } Thus, in principle one only needs to know the values of the EFT coefficients and their derivatives at the light ring. The catch however is that $\tilde{r}_*$ is the light ring of $W$, whose calculation would in principle require full knowledge of the EFT coefficients. This problem can be bypassed by assuming that our background is ``quasi-Schwarzschild''---an assumption that is certainly well supported by present observations. 

To make this statement more precise, it is convenient to work with a radial coordinate such that $c(r) = r$, {\it i.e.} such that $4 \pi r^2$ is the surface area of a sphere with radius $r$. Notice that the coordinate $r$ usually differs from the coordinate $\tilde r$ introduced above. The position of the light ring is unique, but will  be denoted with $r_*$ or $\tilde r_*$ depending on which coordinate system we are using. Then, our quasi-Schwarzschild approximation amounts to assuming that the location of the light ring $r_*$ doesn't differ much from the GR value, {\it i.e.} $r_* = r_{*,{\rm GR}} + \delta r_*$ with $\delta r_*/r_{*,{\rm GR}} \ll 1$. Under this assumption, we can approximate the RHS of Eq. \eqref{WKB} by turning the derivatives with respect to $\tilde r$ into derivatives with respect to $r$ and expanding up to first order in $\delta r_*$ to get
\begin{equation} \label{light ring exp 1}
	\left(1 + \delta r_* \partial_r \right) \left\{ W \left[ \frac{\partial r}{\partial \tilde r} \frac{\partial}{\partial r}\left( \frac{\partial r}{\partial \tilde r} \frac{\partial W}{\partial r} \right)\right]^{-1/2} \right\}_{r = r_{*,{\rm GR}}} = - i \left( n + {1\over 2} \right) .
\end{equation}
The shift $\delta r_*$ can also be calculated at the GR light ring by expanding its defining property $\left.\partial_r W \right|_{r = r_{*,{\rm GR}} + \delta r_*} = 0$ up to first order in $\delta r_*$ to obtain
\begin{equation}
	\delta r_* = - \frac{\partial_{r} W}{\partial_{r}^2 W}\Big|_{r = r_{*,{\rm GR}}} \, .
\end{equation}

\begin{figure}[t!]
\begin{center}
       \includegraphics[width=0.7\textwidth]{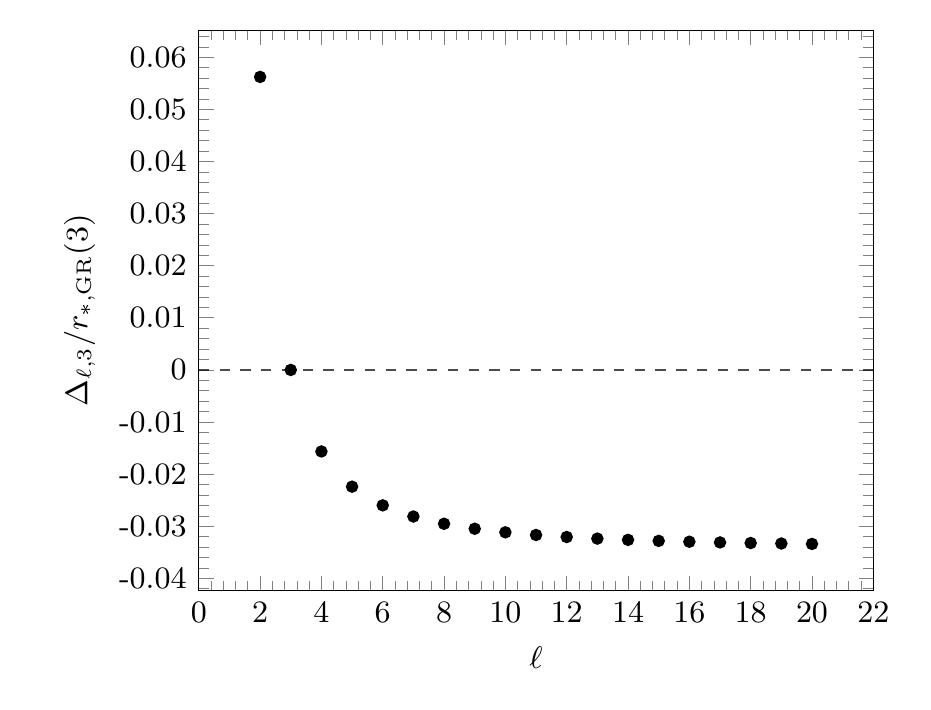}
\end{center}
\caption{Plot of the fractional change in position of the GR light ring $r_{*, {\rm GR}}(\ell)$ corresponding to a change in angular momentum $\ell$. We are considering the change with respect to the fiducial value $r_{*, {\rm GR}}(3)$, {\it i.e.} we have defined $\Delta_{\ell, 3} = r_{*, {\rm GR}}(\ell) - r_{*, {\rm GR}}(3)$. This shows that the location $r_{*, {\rm GR}}$ of the light ring in GR  depends weakly on the angular momentum number $\ell$.}
\label{fig2}
\end{figure}

There is one further subtlety that we need to address, and that is the fact that the position of the light ring depends on the angular momentum number $\ell$. Thus, at this stage we still need to know the values of the EFT coefficients, the background metric components, and their derivatives at different points for different values of $\ell$. Fortunately, the position of the light ring in GR depends only mildly on $\ell$. We can therefore choose a fiducial value of $\ell$---for instance, $\ell =3$---and expand Eq. \eqref{light ring exp 1} up to first order in $\Delta_{\ell, 3} \equiv r_{*,{\rm GR}}(\ell) - r_{*,{\rm GR}}(3)$ and $\delta r_*$ to find
\begin{equation} 
	\left[1 + (\delta r_* +\Delta_{\ell,3})\partial_r \right] \left\{ W \left[ \frac{\partial r}{\partial \tilde r} \frac{\partial}{\partial r}\left( \frac{\partial r}{\partial \tilde r} \frac{\partial W}{\partial r} \right)\right]^{-1/2} \right\}_{r = r_{*,{\rm GR}}(\ell=3)} = - i \left( n + {1\over 2} \right) .  \label{light ring exp 2}
\end{equation}
Figure \ref{fig2} shows that the accuracy of this approximation is  comparable if not better than the accuracy of the lowest order WKB expansion, to be discussed in a moment.

The LHS of Eq. \eqref{light ring exp 2} now depends  on EFT coefficients, the metric components, and their derivatives evaluated  {\it at the same point} for all values of $\ell$. This allows us to express the QNM spectrum in terms of a finite number of free parameters (whose precise number depends on the number of operators included in the effective action). In going from the lowest order WKB result \eqref{WKB} to the Eq. \eqref{light ring exp 2} we have assumed that background is quasi-Schwarzschild (meaning that the position of the light ring is close to its Schwarzschild value) and we have performed an expansion analogous to the slow-roll expansion in inflation. We will call this expansion the {\it light ring expansion}.

Our discussion so far was based on the lowest order WKB approximation, but can be easily extended to higher order. The next-to-leading order WKB corrections were calculated by Iyer and Will \cite{Iyer:1986np}, and contribute the following term to the righthand side of Eq. \eqref{WKB}: 
\begin{eqnarray} \label{WKB nol}
\frac{i}{(2 \partial_{\tilde{r}}^2 W)^{1/2}} \left\{ \frac{1+4 \alpha^2}{32}\,  \frac{\partial_{\tilde{r}}^4 W}{\partial_{\tilde{r}}^2 W}-\frac{7+60 \alpha^2}{288} \left[\frac{\partial_{\tilde{r}}^3 W}{\partial_{\tilde{r}}^2 W}\right]^2\right\}_{\tilde{r}=\tilde{r}_*} ,
\end{eqnarray}
with $\alpha \equiv n + 1/2$. We can estimate the size of these corrections using the fact that the lowest order WKB result implies $W \tilde{r}_*^2 \sim \mathcal{O}(1)$. For the lowest overtone number, $n=0$, the terms in \eqref{WKB nol} provide a correction $\sim 5 \%$. Hence, the lowest order result in Eq. \eqref{light ring exp 2} is applicable when the corrections to the QNM spectrum due to modifications of GR are small, but large enough that
the higher order WKB corrections need not be taken into account.

The impact of next-to-leading and next-to-next-to-leading order WKB corrections in GR was calculated explicitly in~\cite{Iyer:1986nq}. There, it was shown that the accuracy of the WKB approximation decreases for larger $n$, but actually increases for larger $\ell$. Assuming that the same holds true in scalar tensor theories, this is quite encouraging because future experiments will have access to higher values of $\ell$ but not of $n$~\cite{Baibhav:2018rfk}.

\section{The odd sector}\label{Sec:odd}

After the general discussion about QNMs and the light ring approximation in the last two sections, we will now turn our attention to studying the odd sector. As we mentioned earlier, the odd modes are easier to study than the even ones since, based on our assumptions, they amount to a single propagating degree of freedom. The most general odd-parity metric perturbation, shown in Eq. \eqref{odd metric perturbations version 0}, can be rewritten more explicitly as follows~\cite{Regge:1957td}:
\begin{equation}
\delta g_{\mu\nu} ^{\rm odd}= \begin{pmatrix}
0 &0  &- {\rm h}_0 \csc \theta  \partial_\phi & {\rm h}_0 \sin \theta  \partial_\theta  \\
0 & 0  & - {\rm h}_1 \csc \theta  \partial_\phi& {\rm h}_1 \sin \theta  \partial_\theta \\
- {\rm h}_0 \csc \theta  \partial_\phi & - {\rm h}_1 \csc \theta  \partial_\phi& \frac{1}{2} {\rm h}_2 \csc \theta \mathcal{X} &  -\frac{1}{2} {\rm h}_2 \sin \theta \mathcal{W}\\
 {\rm h}_0 \sin \theta  \partial_\theta&  {\rm h}_1 \sin \theta  \partial_\theta&  -\frac{1}{2} {\rm h}_2 \sin \theta \mathcal{W} &  -\frac{1}{2} {\rm h}_2 \sin \theta \mathcal{X} 
\end{pmatrix} Y_{\ell m} e^{- i \omega t}\, , \label{odd metric perturbations}
\end{equation}
where ${\rm h}_0$, ${\rm h}_1$, ${\rm h}_2$ are functions of $r$ alone, and we have defined the differential operators
\begin{equation}
\begin{aligned}
\mathcal{X} &= 2(\partial _ \theta \partial _ \phi - \cot \theta \partial _ \phi  )\,,
\\
\mathcal{W} &= (\partial _ \theta \partial _ \theta - \cot \theta \partial _ \theta  - \csc ^2 \theta  \partial _ \phi\partial _ \phi  )\,.
\end{aligned}
\end{equation}
Under a gauge transformation of the form
$\tilde x_\mu = x_\mu + \xi_\mu$, with
\begin{equation}
\xi_\mu = (0, 0, \delta {1\over {\,\rm sin}\theta} {\partial_\phi}, - \delta \, {\rm sin}\theta \, \partial_\theta )
Y_{\ell m} e^{-i\omega t}\, ,
\end{equation}
the metric perturbations transform as
\begin{eqnarray}
{\rm \tilde h_0} = {\rm h_0} - i\omega \delta \quad , \quad
{\rm \tilde h_1} = {\rm h_1} + \delta' - {2c' \over c} \delta
\quad , \quad
{\rm \tilde h_2} = {\rm h_2} - 2\delta \, ,
\end{eqnarray}
where $(\phantom{x})' \equiv \partial_r(\phantom{x})$. 

In the rest of this section we will adopt the Regge-Wheeler gauge, where ${\rm h}_2(r) =0$. Notice that this choice is perfectly compatible with the unitary gauge in the even sector, and that, as expected, the physical conclusions we derive in this section are ultimately independent of the choice of gauge. 

Our starting point is the effective action for perturbations in unitary gauge in Eq. \eqref{EFT-ss0}. Owing to the simplicity of the odd sector, only a subset of the operators shown there actually affect the dynamics of odd perturbations. In fact, recall that the background metric \eqref{GMN-2} is even by construction, and so is every tensor evaluated on the background. The operators $\delta g^{rr}$ and $\delta K$ are also even, and thus one can safely disregard any term that contains them. The resulting action reads:
\begin{multline}\label{EFT-ss0-odd}
 S_{\rm odd} =  \int\D^4x \, \sqrt{-g} \bigg[
\frac{1}{2}M^2_1(r) R -\Lambda(r) - f(r)g^{rr} - \alpha(r)\bar K_{\mu\nu} K^{\mu\nu}
\\
  + M_{10}^2(r)\delta K_{\mu\nu}\delta K^{\mu\nu} 
+ M_{12}(r) \bar  K_{\mu\nu}\delta K^{\mu\rho} \delta {K^{\nu}}_\rho 
+ \ldots
\bigg] \, .
\end{multline}
As one can see, at quadratic level and up to second order in derivatives, the effective action for the odd modes contains six functions $\{M^2_1(r),\Lambda(r),f(r),\alpha(r), M_{10}^2(r),M_{12}(r)\}$, three of which can be expressed in terms of a fourth one and the background metric components by using the background equations of motion---see discussion in Sec.~\ref{sec:tadpoles}.

\subsection{Leading order in derivatives}

We will at first restrict our attention to the lowest-order terms in the derivative expansion, which appear in the first line in Eq. \eqref{EFT-ss0-odd}. 
Here, by lowest-order in derivatives, we refer to expanding the non-Einstein-Hilbert terms in number of derivatives and keeping the leading order. In this case in particular, the surviving terms are
$\sqrt{-g} \Lambda (r)$ and $\sqrt{-g} f(r) g^{rr}$ which carry no derivatives
on metric fluctuations. These are the terms that break the radial-diffeomorphism invariance. 
We will also perform a conformal transformation of the metric to set $M_1^2(r) = \Mpl^2$. As we already pointed out, this transformation would affect the coupling with other matter fields (for instance, those that make up the detector). At this stage, however, we will not concern ourselves with that, and  therefore we will work with the effective action
\begin{equation}
S =  \int\D^4x \, \sqrt{-g} \left[
	\frac{\Mpl^2}{2}R
	-\Lambda(r)	- f(r)g^{rr}
\right] \, .
\label{appEFT-ex1}
\end{equation}
The functions $\Lambda(r)$ and $f(r)$ are given in terms of the background metric coefficients by Eqs. \eqref{lm} and \eqref{fm} with $\alpha =0$ and $M_1 = \Mpl$.
By varying our action with respect to the inverse metric, we find that the $(\theta, \phi)$ component yields 
\begin{equation}
{\rm h}_0=	\frac{i a^2 b^2 }{\omega }  \left[ {\rm h}_1 \left(\frac{a'}{a}+\frac{b'}{b} \right)+ {\rm h}_1'
\right]\,.
\end{equation}
This constraint can be combined with the $(r, \phi)$ component to derive a second order equation of motion for ${\rm h}_1$. By following the procedure outlined in App. \ref{schf}, this equation can be cast in a Schr\"odinger-like form, i.e.
\begin{equation}\label{sch-odd-e1}
	\frac{\D^2}{{\D \tilde r}^2}\Psi(\tilde r)+ W(\tilde  r )\Psi(\tilde r)=0\, 
\end{equation}
where the potential is given by
\begin{equation}\label{rwpot}
W(\tilde r(r))=	 \omega^2 + a^2 b^2 \left[ \frac{c''}{c} - 2 \frac{c^{\prime 2}}{c^2}+ 
	\frac{ a' c'}{ac}
	+\frac{b' c'}{bc}
	-\frac{(\ell+2)(\ell-1)}{b^2c^2}\right ]\, ,
\end{equation}
the radial coordinate $\tilde r$ is defined as
\begin{equation}\label{tortc}
\tilde r(r)= \int_{r_c} ^ r \frac{\D l}{a(l) b(l)} 
\end{equation}
for some fiducial $r_c$, and the variable $\Psi$ is related to ${\rm h}_1$ by an overall rescaling:
\begin{equation}\label{odd-field-red}
	{\rm h}_1(r) \rightarrow \Psi(\tilde r (r))= \exp\left [  \int_{r_c} ^ r \left ( \frac{a'(l)}{a(l)}+\frac{b'(l)}{b(l)}-\frac{c'(l)}{c(l)} \right )\D l  \right ]{\rm h}_1(r)\,.
\end{equation}
These results are consistent with what was found in \cite{Kobayashi:2012kh}. Furthermore, in the limit where the background solution is exactly Schwarzschild, i.e. $a(r)=b(r) =(1-2 G \mathcal{M}/r)^{1/2}$ and $c(r)=r$, the coordinate $\tilde r$ reduces to the usual tortoise coordinate in GR (see Eq. \eqref{rstar}), \eqref{odd-field-red} matches the field redefinition in \cite{Regge:1957td}, and $V(\tilde r )$ reduces to the Regge-Wheeler potential.

A few additional comments are in order. First, we should stress that without loss of generality one can always choose a radial coordinate such that $c(r)=r$, in which case the potential \eqref{rwpot} is completely determined by the two functions $a(r)$ and $b(r)$.
Moreover, in the Einstein frame and at the order in derivatives we are considering, these two functions are in turn constrained by the differential equation \eqref{consrEe}. 
Finally, remembering that $\ell(\ell+1)$ is the eigenvalue of the angular part of the Laplacian, from \eqref{sch-odd-e1} and \eqref{rwpot}, it is easy to see that the squared propagation speeds in the radial and angular directions are $c_r ^2= c_\Omega ^2 =1$. 

\subsection{Worked examples}

Before discussing the higher derivative corrections appearing in the second line of \eqref{EFT-ss0-odd}, we will pause for a moment to illustrate the usefulness of our EFT approach in a couple of different scenarios. First, we will consider a particular scalar tensor theory which admits an analytic black hole solution with scalar hair. In this case, we will show how, by matching the action of this particular model onto our effective action \eqref{appEFT-ex1}, one can bypass the entire derivation of the Schr\"odinger equation for QNMs and obtain the effective potential directly from Eq. \eqref{rwpot}. Although in this section we are focusing on the odd modes, this strategy becomes particularly convenient in the case of even modes.  The second idealized scenario we will consider is one in which the spectrum of QNMs is known from observations. We will then show how this information can be used to constrain the arbitrary functions that appear in our effective action. At the lowest order in the derivative expansion (and only at this order), this procedure is equivalent to constraining the background metric coefficients $a(r)$ and $b(r)$.

\subsubsection{From the background solution to the QNM spectrum}

 In this section, we will consider the black hole solutions with scalar hair found in \cite{Dennhardt:1996cz} for a scalar-tensor theory theory which can be described using the leading order action \eqref{appEFT-ex1}. In this section only, we decide to set $G=(8 \pi)^{-1} $ 
  for simplicity  ($\Mpl=1$).  The action for such theory is of the form
\begin{equation} \label{action toy model odd}
S =  \int\D^4x \, \sqrt{-g} \left[ \frac{ 
R}{2}  + \frac{1}{2} g^{\mu \nu} \partial_\mu  \Phi  \partial_\nu  \Phi  - \mathcal{V}(\Phi)\right ],
\end{equation} 
with a potential given by
\begin{equation}\label{Vcoul}
\mathcal{V}(\Phi)\ =\ \frac{3(q+2\mathcal{M})}{|q|^3}  \Bigl[
\left(3+\Phi^2\right)\cdot\sinh|\Phi|-3|\Phi|\cdot\cosh\Phi\Bigr] .
\end{equation}
We should emphasize that such an action is not very well-motivated from an effective field theory viewpoint, due to the {\it ad hoc} form of the potential, and especially due to the ghost-like kinetic term. However, this theory will serve our purposes as an interesting toy-model, since it admits an analytic black hole solution with scalar hair. Such a solution is parametrized by two numbers, the asymptotic mass $8 \pi \mathcal M$ of the black hole and its scalar charge $q$, and it is such that~\cite{Dennhardt:1996cz}
\begin{subequations} \label{toy hairy solution}
\begin{align} 
	\bar \Phi (r) &= \frac{q}{r}\,
	\\
	a^2 (r) &= b^2(r) = \frac{r^2 (6 \mathcal{M}+3 q) e^{-\frac{q}{r}}}{4 q^3}-e^{\frac{q}{r}} \left[\frac{r^2 (6 \mathcal{M}+3 q)}{4 q^3}-\frac{r (6 \mathcal{M}+3 q)}{2 q^2}+\frac{6 \mathcal{M}+q}{2 q}\right],	\\
	c^2(r)&= r^2 e^{-\frac{q}{r}} .
\end{align}
\end{subequations}
Notice that the existence of a horizon requires $q>-2\mathcal{M}$, and that in the limit $q \rightarrow 0$ this solution reduces to the usual Schwarzschild solution. However, for non-vanishing $q$ can in principle deviate significantly from the Schwarzschild solution, and therefore so can the corresponding spectrum of QNMs. 

By working in unitary gauge, it is easy to see that the action \eqref{action toy model odd} is precisely of the form \eqref{appEFT-ex1} with $\Lambda(r) = \mathcal{V}(\bar \Phi(r)) $ and $f(r) = \bar \Phi'(r)^2/2$. Thus, we can immediately plug the expressions for $a(r)$, $b(r)$ and $c(r)$ given in Eqs. \eqref{toy hairy solution} into the QNM effective potential \eqref{rwpot} to obtain
\begin{equation}\label{dev0}
	\begin{aligned}
	W(\tilde r(r))= \omega^2 + \frac{e^{-\frac{2 q}{r}}}{64 q^4 r^4} \left\{e^{\frac{2 q}{r}} \left[2 q^2 (6 \mathcal{M}+q)+3 r^2 (2 \mathcal{M}+q)-6 q r (2 \mathcal{M}+q)\right]-3 r^2 (2 \mathcal{M}+q)\right\} \qquad
	\\
	\left\{e^{\frac{2 q}{r}}
	\left[-r^2 ((4 \ell+1) (4
   \ell+3) q+6 \mathcal{M})+6 q^2 (6 \mathcal{M}+q)+6 q r (18 \mathcal{M}+q)\right]+3 r^2 (2 \mathcal{M}+q)\right\}. \qquad 
	\end{aligned}
\end{equation}
As expected, this expression reduces to the Regge-Wheeler effective potential~\cite{Regge:1957td} in the limit $q\to 0$. It is also important to point out that, for any value of $q$, the potential $V(r)\equiv \omega^2 - W(r)$ vanishes exactly at the  horizon, and $V(r)>0$ for $r>r_{\rm hor}$. This condition by itself is sufficient to ensure the stability of the odd sector.

Various plots of $V(\tilde r(r))$ for $\ell=2,3,4$ and different values of $q$ are shown in the right panels of Fig. \ref{img2}. The corresponding left panels show the values of real and imaginary parts of the QNM frequencies with $n=0$. For simplicity, these values were derived using the WKB result \eqref{WKB}, but of course they could also be calculated numerically using the exact potential \eqref{dev0}.
\afterpage{
\begin{figure}[t!]
  \begin{tabular}{ccc} 
   \begin{minipage}{0.5\hsize}
     \begin{center}
       \includegraphics[width=0.95\textwidth]{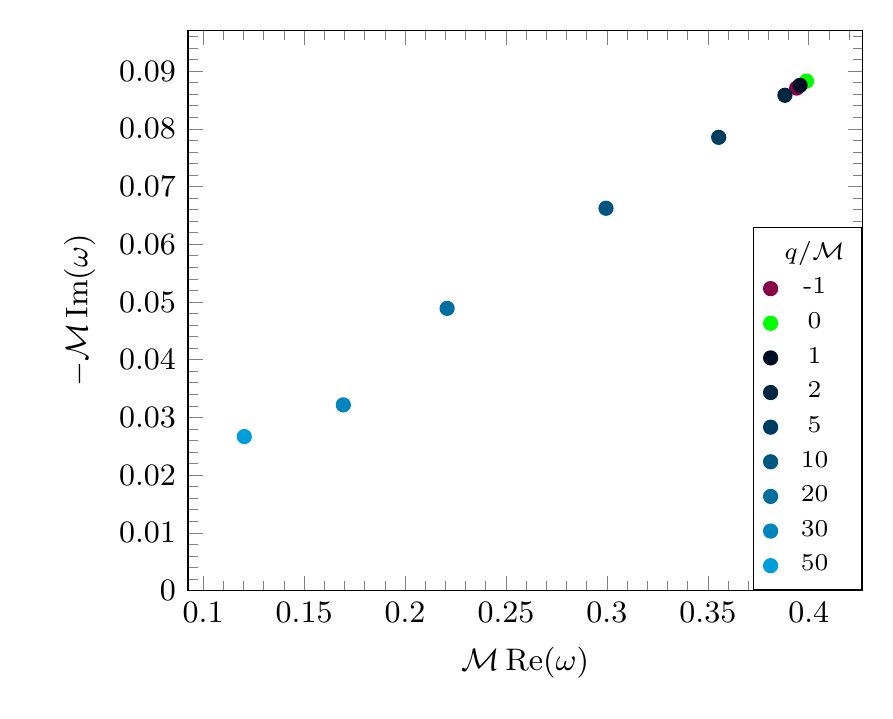}
  \end{center}
   \end{minipage}
   \begin{minipage}{0.5\hsize}
     \begin{center}
       \includegraphics[width=0.95\textwidth]{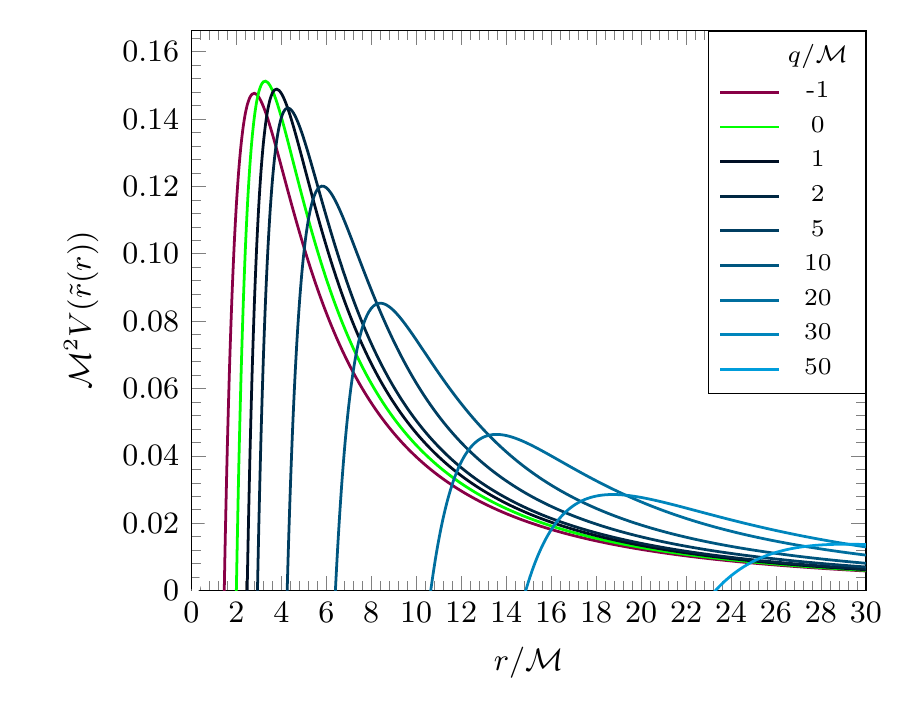}
  \end{center}
   \end{minipage}
  \end{tabular}
  \begin{tabular}{ccc} 
   \begin{minipage}{0.5\hsize}
     \begin{center}
       \includegraphics[width=0.95\textwidth]{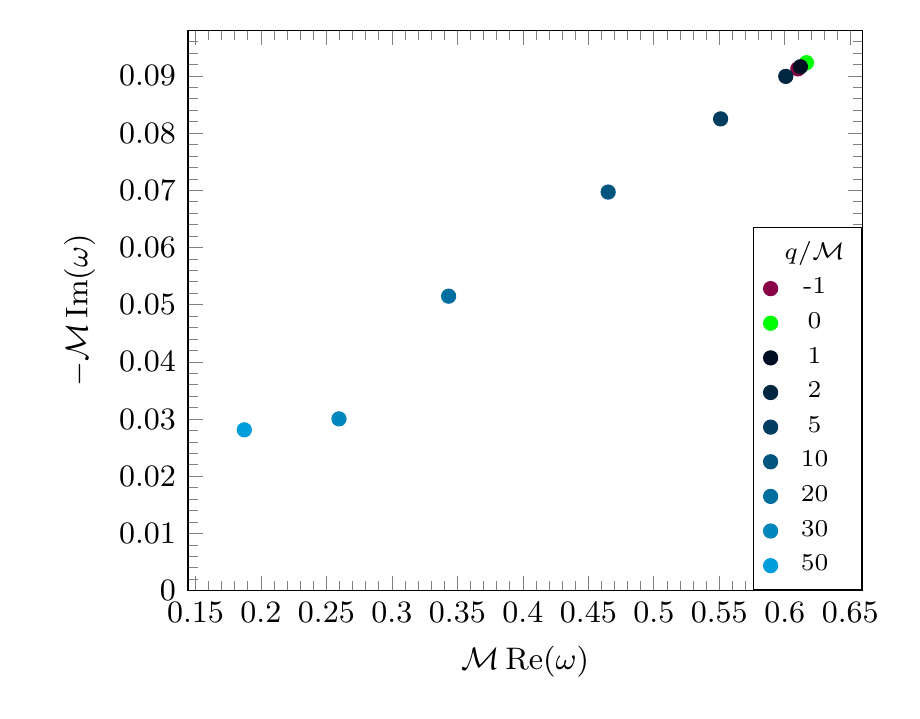}
     \end{center}
   \end{minipage}
   \begin{minipage}{0.5\hsize}
     \begin{center}
       \includegraphics[width=0.95\textwidth]{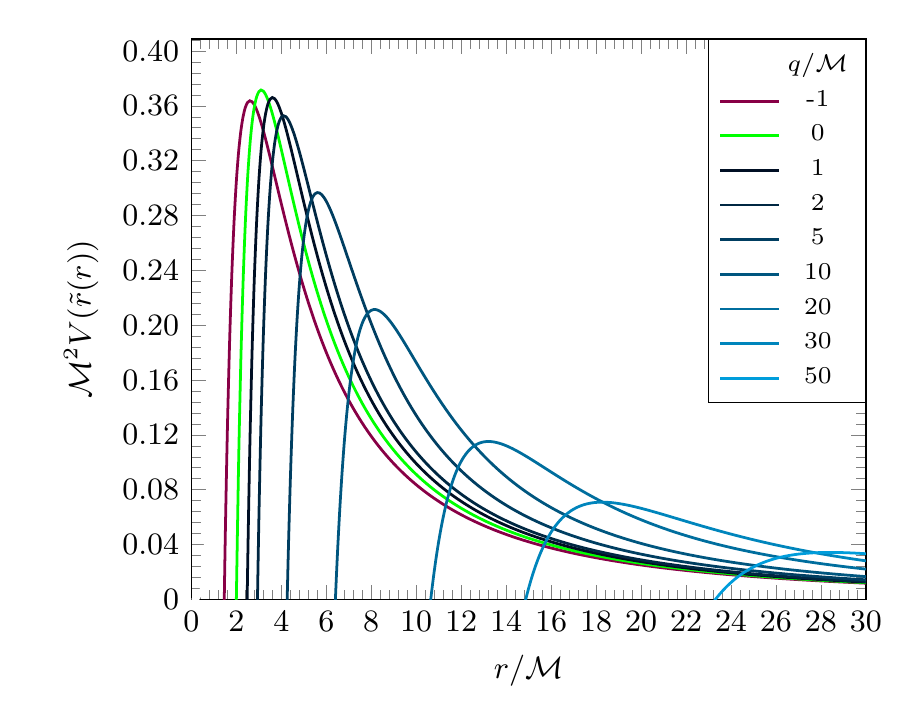}
  \end{center}
   \end{minipage}
     \end{tabular}
     \begin{tabular}{ccc} 
   \begin{minipage}{0.5\hsize}
     \begin{center}
       \includegraphics[width=0.95\textwidth]{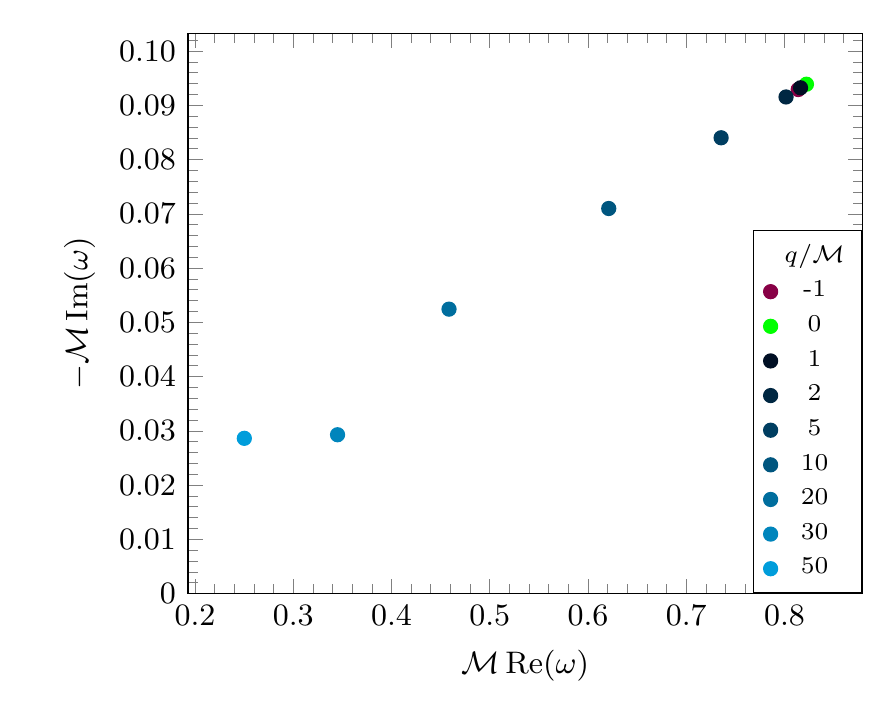}
     \end{center}
   \end{minipage}
   \begin{minipage}{0.5\hsize}
     \begin{center}
       \includegraphics[width=0.95\textwidth]{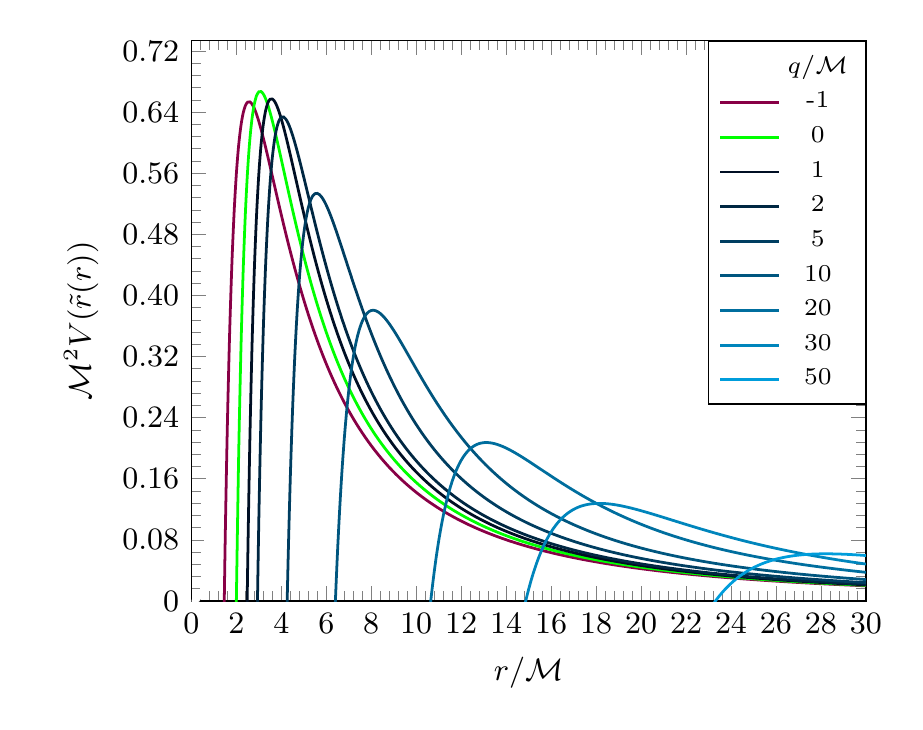}
  \end{center}
   \end{minipage}
  \end{tabular}
  \caption{{\it Left panel:} Odd spectrum of QNM for $n=0$ and some representative values of $q$. {\it Right panel:} The plot of the corresponding potential $V(\tilde r(r))$ . {\it From top to bottom:} $\ell =2$, $\ell =3$, $\ell =4$. 
\label{img2}}
\end{figure}
\clearpage}
\subsubsection{From the QNM spectrum to the effective potential: the inverse problem} 
\label{sec:the inverse problem}

Let us now consider scenario where the QNM spectrum is known empirically, and discuss to what extent this information can be used to constrain the coefficients appearing in our effective action. This procedure goes under the name of
``inverse problem"---see for example \cite{Volkel:2018czg,Konoplya:2018ala}. In our analysis, we will resort to a WKB approximation, and restrict ourselves to the case where the background metric is a ``small'' deviation from Schwarzschild. This, in turn, can be reasonably expected to lead to a ``small'' displacement of the light ring. As shown in Sec. \ref{sec:WKB}, in this limit one can perform a light-ring expansion to the express the QNM spectrum in terms of a finite number of parameters, which are the values of the EFT coefficients, the background metric components and a few of their derivatives at the GR light ring. 

The advantage of this approach is that one could in principle use observation to place model-independent constraints on these parameters. This should be reminiscent of what happens in slow-roll inflation. There, all the complexities of the inflaton action are reduced to a handful of slow-roll parameters, which in turn determine observable quantities such as spectral indices and scalar-tensor ratio. There are however also a couple of downsides to our approach. First, small deviations from the Regge-Wheeler potential likely correspond to small deviations from the QNM spectrum predicted by GR, making their detection in principle more difficult. Second, because we are essentially placing constraints on the effective potential and its first few derivatives at a single point, there will be in general a degenerate set of background solutions compatible with any given QNM spectrum.

Our result \eqref{light ring exp 2}, valid at first order in the light ring expansion and for a radial coordinate such that $c(r) = r$, implies that the tower of complex frequencies $\omega_{n, \ell}$ is completely determined by the values of $W, \partial_r W, \partial_r^2 W, \partial_r^3 W, \partial r/ \partial \tilde{r},\partial_r (\partial r/ \partial \tilde{r})$ and $\partial_r^2 (\partial r/ \partial \tilde{r})$ at the GR light ring. According to the results \eqref{rwpot} and \eqref{tortc} (which we derived from the effective action \eqref{appEFT-ex1}, valid at lowest order in the derivative expansion) these quantities depend only on $a(r), b(r)$ and their derivatives at the GR light ring. To be more precise, second and higher derivatives of $a(r)$ can always be removed by using the constraint \eqref{consrEe} recursively. Thus, using Eq. \eqref{light ring exp 2} we can calculate the QNM frequencies in terms of the following 7 parameters:

\begin{equation} \label{7 parameters}
	\{a (r),a'(r),b (r),b'(r),b''(r),b ^{(3)}(r),b ^{(4)}(r) \} \big| _{r=r_{*, {\rm GR}}(\ell =3)}.
\end{equation}

Since the QNM frequencies $\omega_{n \ell}$ are complex, in principle one has {\it two} real measurements for each mode that is observed.  In practice, though, it is challenging to observe overtones with $n>0$ for a single event, though a stacking of multiple events might prove helpful. One is also limited to the frequencies $\omega_{0, \ell}$ for a finite range of values of $\ell$. However, a space-based experiment like LISA could be sensitive to modes with angular momentum as large as $\ell = 7$ (see \cite{Baibhav:2018rfk} for a discussion), and these modes would be sufficient to estimate the 7 parameters in \eqref{7 parameters}.

Finally, notice that by using Eqs. \eqref{fm}, \eqref{lm} and \eqref{consrEe} we could translate these results into direct constraints on the values of the EFT coefficients $f(r)$, $\Lambda(r)$, and their first few derivatives at $r_{*,{\rm GR}}(3)$.

\subsection{Next-to-leading order}
\label{sec:nlorder}

In the previous section we worked at leading order in derivatives and derived the equation of motion governing the single physical degree of freedom in the odd sector. At this order, we showed that the QNM spectrum is completely determined by the background metric coefficients. We will now turn our attention to the higher derivative corrections appearing in the effective action \eqref{EFT-ss0-odd}. These introduce three additional parameters---$\alpha(r), M_{10}^2(r)$, and $M_{12}(r)$---the last two of which cannot be expressed in terms of the background metric coefficients. This kind of behavior is of course commonplace in effective theories: the next order in the derivative expansion always introduces new free parameters (functions, in our case). 

Once again, we choose to work in the Regge-Wheeler gauge. At higher order, it becomes more convenient to work directly at the level of the action. To this end, we expand \eqref{EFT-ss0-odd} up to quadratic order in the odd-type metric perturbations. We focus on the modes with $m=0$, since all the others with $m \neq 0$ satisfy the same equations of motion due to the spherical symmetry of the background. After series of integration by parts, find the following result: 
\begin{equation}\label{japa1}
	S_{{\rm odd}, m=0} ^{(2)}= \sum _{\ell =0}^{\infty} \int \D t \D r 
	 \left[u_1
	 {\rm h}_0^2
	 +u_2
	  {\rm h}_1^2+
	 u_3
	 \left(\dot {{\rm h}}_1^2
	 -2\dot {{\rm h}}_1{\rm h}_0^\prime 
	 +{\rm h}_0^{\prime 2}
	 +2v
	 \dot {{\rm h}}_1{\rm h}_0
	 \right) \right]\,,
\end{equation}
where $u_i (r)$ and $v (r)$ are defined as follows: 
\begin{subequations}\label{u and v}
\begin{align}
&\begin{aligned}
u_1 (r)= 
\frac{\ell (\ell+1)}{4 a^4 b c^3}\left\{
 a^2 c  \left[ 2a b c b' c'+2 ab^2 \partial_r \left(c c'\right)+a(\ell-1) (\ell+2)-2 b^2 c a' c'\right] \Mpl^2\right .
\\
\left.
   -2 a^2 b^2 c^2 \partial_r (a c) (M^{2}_{10})'
   -2 a b^3 c  \left(\partial_r (a c)\right)^2 M^{\prime}_{12}\right.
   \\
\left.  +2 a b c   \left[a b c a' c'+4 b c^2 a'^2+a^2 c b' c'+a^2 b c c''+3 a^2 b c'^2\right] \right.
   \alpha  
   +2 a^3 b^2 c^2 c' \alpha'
   \\
 \left.  -2 a b c \left[a c b'  \partial_r (a c)+ a bc \left(a c'' -a' c'\right)+bc^2 \left(a a''-2 a'^2\right)-a^2b c'^2\right] M_{10}^2  \right .
   \\
   \left.+2b^2c \, \partial_r(ac) \left[ -2ab' \partial_r(ac) + bc(a'^2-2aa'')-ab(3a'c'+2ac'') \right] M_{12}
      \right\}\, ,
   \end{aligned}
\\
&u_2 (r)=\frac{\ell (\ell+1) (\ell-1) (\ell+2) a b }{4 c^3}\left[2 b  c' M_{12}-c \left(\Mpl^2-2 M_{10}^2\right)\right]\, ,
\\
&u_3 (r)=-\frac{\ell (\ell+1) b }{4
   a^2 c}\left[ \partial_r (a c) b M_{12} - a c \left(\Mpl^2-2 M_{10}^2\right)\right]\, ,
\\
&\begin{aligned}
v (r)=\frac{\ell (\ell+1)b}{ 2a^3 c^2 u_3} \left[a^2cc' ( \Mpl^2 +\alpha) - ac \, \partial_r(ac) M_{10}^2 - b \left( \partial_r(ac)\right)^2 M_{12}
  \right]\, .
   \end{aligned}
\end{align}
\end{subequations}
Fixing the gauge at the level of the action might rightfully be a source of apprehension for some readers. In this case, though, we have checked explicitly that the equation of motion one ``misses'' by doing so becomes redundant with this gauge choice. Hence, this particular gauge can be fixed at the level of the action.
  
The modes with $\ell =0, 1$ must be treated separately. For simplicity, we will therefore focus our attention on the modes with $\ell \geqslant 2$.
Following the same procedure used in \cite{DeFelice:2011ka}, we introduce an auxiliary field ${\rm q}(t,r)$ and rewrite the action above as follows: 
\begin{multline}
	S_{{\rm odd},m=0,\ell \geqslant 2} ^{(2)} = \int  \D t \D r
	 \Big \{\left(u_1
	 -\partial_r (u_3 v )-u_3v^2 \right)
	 {\rm h}_0^2
	 +u_2
	  {\rm h}_1^2 \\
	   +
	\left . u_3
	 \left[2{\rm q}\left(
	 \dot {{\rm h}}_1
	 -{\rm h}_0^\prime 
	 +v
	 {\rm h}_0
	 \right) -{\rm q}^2 \right] \right\}\,.  
\end{multline}
It is easy to show that by solving for ${\rm q}$ one recovers the action in \eqref{japa1}. Varying instead the action with respect to ${\rm h}_0$ and ${\rm h}_1$ yields the algebraic constraints
\begin{equation}\begin{aligned}
{\rm h}_0=\frac{\partial _ r\left(u_3 {\rm q}\right) +u_3 v  {\rm q}}{\partial _ r\left(u_3 v \right)+v^2u_3-u_1 } \, , \qquad \qquad \qquad {\rm h}_1=\frac{u_3}{u_2} \dot {\rm q}\, ,
\end{aligned}
\end{equation}
which can be used to find the effective action for the master variable ${\rm q}$: 
\begin{equation}\begin{aligned}
	S_{{\rm odd},m=0,\ell \geqslant 2}^{(2)}= \int  \D t \D r
	 \left[ \mathcal{G}^{00} \dot {\rm q}^2+ \mathcal{G}^{rr}  {\rm q}^{\prime 2}
	 +\mathcal{G}^{{\rm q}{\rm q}}  {\rm q}^2\right]\,.
\end{aligned}
\end{equation}
with
\begin{subequations}
\begin{align}
\mathcal{G}^{00}(r) =& -\frac{u_3^2}{u_2}\, ,\\
 \mathcal{G}^{rr}(r) =& \frac{u_3^2}{-u_1+v u_3'+u_3 \left(v'+v^2\right)}\, , \\
\mathcal{G}^{{\rm q}{\rm q}}(r) =& \frac{(\mathcal{G}^{rr} )^2}{u_3^3} \left[u_3' \left(2 u_3' v'-u_1'\right)+u_3 \left(-v u_1'-u_3'' v'+u_3' v''+v u_3'
   v'\right) + u_1 u_3''\right .
   \\
  & \qquad \qquad \qquad \qquad \left . +2 v u_1 u_3'+u_1 u_3 \left(3 v'+v^2\right)-u_1^2+u_3^2 \left(v v''-2
   v'^2\right)\right]\, . \nonumber 
\end{align}
\end{subequations}

By varying the action above one finds a second order equation of motion for $q$, which can be again cast in a  Schr\"odinger-like form by following the procedure discussed in Appendix \ref{schf}. Since the final result is not particularly illuminating, we will not report it here.


\section{The even sector}

We now turn our attention to the even sector. Unlike what happens for the odd sector, now all the operators in the EFT \eqref{EFT-ss0} in principle contribute to the quadratic action for the even perturbations. Following the most general parity-even perturbation of the metric in Eq. \eqref{even metric perturbations version 0} can be written more explicitly as follows:
\begin{equation}
\delta g_{\mu\nu} = \begin{pmatrix}
a^2H_0 & H_1  & \mathcal{H}_0 \partial_\theta & \mathcal{H}_0 \partial_\phi \\
H_1 & H_2/b^2  & \mathcal{H}_1 \partial_\theta & \mathcal{H}_1 \partial_\phi \\
\mathcal{H}_0 \partial_\theta & \mathcal{H}_1 \partial_\theta & c^2[ \mathcal{K} + G \nabla_\theta \nabla_\theta ]& c^2 G \nabla_\theta \nabla_\phi\\
\mathcal{H}_0 \partial_\phi & \mathcal{H}_1 \partial_\phi & c^2 G \nabla_\phi \nabla_\theta & c^2[\sin^2\theta \mathcal{K}+ G \nabla_\phi \nabla_\phi ]  
\end{pmatrix} Y_{\ell m} \, , \label{even metric perturbations}
\end{equation}
where $H_0, H_1, H_2, \mathcal{H}_0, \mathcal{H}_1, \mathcal{K}$ and $G$ are functions of $(t,r)$, and $\nabla_{\theta, \phi}$ are covariant derivatives on the 2-sphere of radius one. Explicit expressions for the second derivatives are given in Eq.~\eqref{2 cov dev sphere}. 

In Appendix \ref{sec:choosing gauge} we show explicitly how one of the odd perturbations can always be set to zero by an appropriate choice of coordinates. Something similar holds true for even metric perturbations. In fact, any infinitesimal ``even diffeomorphism'' $x^\mu \to x^\mu + \xi^\mu$ can be parametrized~as
\begin{equation}\label{even diff}
	\xi^\mu = \left(\alpha(t,r) , \beta(t,r), \gamma (t,r) \partial_\theta, \frac{\gamma (t,r) \partial_\phi}{\sin^2 \theta} \right) Y_{\ell m} (\theta, \phi) .
\end{equation}
Under this coordinate change, the metric perturbations in Eq. \eqref{even metric perturbations} transform as follows:
\begin{subequations}\label{even gauge change}
\begin{align} 
	\tilde H_0 &= H_0 +  2 \dot \alpha + 2 \tfrac{a'}{a} \beta \\
	\tilde H_1 &= H_1 + a^2 \alpha'  - \dot \beta /b^2 \\
	\tilde H_2 &= H_2 + 2 \tfrac{b'}{b} \beta - 2 \beta' \\
	\tilde{\mathcal{H}}_0 &= \mathcal{H}_0 + a^2 \alpha -  c^2 \dot \gamma \\
	\tilde{\mathcal{H}}_1 &= \mathcal{H}_1 - \beta/b^2 - c^2 \gamma' \\
	\tilde G &= G - 2 \gamma\\
	\tilde{\mathcal{K}} &= \mathcal{K} - 2 \tfrac{c'}{c} \beta.
\end{align}
\end{subequations}

By choosing to work in unitary gauge, we are choosing the function $\beta(r)$ in such a way as to ensure that the scalar perturbation vanishes. We can then use the remaining coordinate freedom and fix the gauge completely by demanding that 
\begin{equation} \label{unitary gauge metric pert}
	\mathcal{H}_0 = G = 0.
\end{equation}
We refer to this gauge choice of $\delta\Phi = \mathcal{H}_0 = G = 0$ as Regge-Wheeler-unitary gauge. Once again, in what follows we will fix this gauge at the level of the action. We have checked explicitly that this is allowed, in that the equations of motion for $\delta \Phi$, $\mathcal{H}_0$ and $G$ become redundant once these variables are set to zero.  

Two other popular gauge choices for the even sector are $\mathcal{H}_0 = \mathcal{H}_1 = G =0$ (this is the original Regge-Wheeler gauge; see {\it e.g.}~\cite{Regge:1957td,Zerilli:1970se,Tattersall:2017erk}) and $\mathcal{H}_0 = \mathcal{K} = G =0$ ({\it e.g.} \cite{Kobayashi:2014wsa}). The perturbations in these gauges can be easily obtained from ours by using Eqs. \eqref{even gauge change} with $\alpha=\gamma =0$ and $\beta = b^2 \mathcal{H}_1$ in the first case, $\beta = \tfrac{c}{2c'} \mathcal{K}$ in the latter. The scalar perturbation then is given by $ \tilde{\delta \Phi} = - \beta \bar \Phi'$, where $\bar \Phi (r)$ is the background configuration of the scalar field.

\subsection{Leading order in derivatives}

At lowest order in the derivative expansion, the effective action \eqref{EFT-ss0} reduces to
\begin{equation} \label{action even lowest order}
S =  \int\D^4x \, \sqrt{-g} \bigg[
\frac{1}{2} M^2_1(r) R -\Lambda(r) - f(r)g^{rr} + M_2^4(r)(\delta g^{rr})^2 \bigg] .
\end{equation}
Unlike in the odd sector, here there are three operators that can in principle modify the linear behavior of even quasi-normal modes. For simplicity, in this section we will work with a radial coordinate such that $c^2(r) = r^2$, and we will perform once again a conformal transformation of the metric to set $M_1^2(r) = \Mpl^2$. 
Also, due to the spherical symmetry of the background, modes with different values of $m$ satisfy the same equations of motion. For this reason, we can choose to focus on the modes with $m=0$, in which case the functions $H_0, H_1, ...$ are all real.  

Upon judicious and repeated use of the background equations, we find the following quadratic action for the modes with $m=0$:
\begin{align} \label{quadratic Lagrangian even modes}
	S^{(2)}_{{\rm even}, m=0} =& \sum_{\ell = 0}^{\infty} \int \D t \D r \tfrac{a r^2}{2b} \Mpl^2 \bigg\{ H_0 \Big[  - \tfrac{J b^2}{r^3} \left( 1+ r \tfrac{b'}{b} \right) \mathcal{H}_1 - \tfrac{1}{2 r^2} \left( J + 2 b^2 \left( 1 + r \tfrac{a'}{a} + r \tfrac{b'}{b}\right) \right) H_2 \nonumber  \\
	& \quad +\tfrac{(2-J)}{2 r^2} \mathcal{K} - \tfrac{b^2}{r} H_2' + \tfrac{Jb^2}{r^2} \mathcal{H}_1' + \tfrac{b^2}{r} \left( 3 + r \tfrac{b'}{b} \right) \mathcal{K}' + b^2 \mathcal{K}'' \Big] + \tfrac{J b^2}{r^2 a^2} \left( \tfrac{1}{2} H_1^2 + \tfrac{a^2}{r^2} \mathcal{H}_1^2\right)\nonumber  \\
	& \quad + \tfrac{b^2}{2r^2} \left( 1 + r \tfrac{a'}{a} + r \tfrac{b'}{b} + \tfrac{4 b^2 r^2M_2^4}{\Mpl^2} \right) H_2^2 + \tfrac{(J-4)a-2r^2b b'a'}{2 r^2 a}H_2 \mathcal{K}  \\
	& \quad + \tfrac{b^2}{a^2} H_1 \left[ \tfrac{2}{r} \dot H_2 - \tfrac{J}{r^2} \dot{\mathcal{H}}_1 -2 r \dot{\mathcal{K}}' -\tfrac{2}{r} \left( 1 - r \tfrac{a'}{a} \right) \dot{\mathcal{K}} \right] + \tfrac{J b^2}{r^2} \mathcal{H}_1 (\dot H_2 + \dot{\mathcal{K}}) \nonumber \\
	& \quad + \tfrac{b^2}{r} H_2 \Big[ -\tfrac{J}{r^2}\left( 1-r \tfrac{b'}{b}\right) \mathcal{H}_1 + \tfrac{1}{r b^2} \left( 1 + \tfrac{r^2 b b' a'}{a}\right) \mathcal{K} - \left( 1 + r \tfrac{a'}{a} \right) \mathcal{K}' \nonumber \\
	& \quad + \tfrac{J}{r} \mathcal{H}_1' +\tfrac{r}{a^2}{b^2} \ddot{\mathcal{K}}\Big]  - \tfrac{1}{2a^2} \dot{\mathcal{K}}^2 + \tfrac{b^2}{2} \mathcal{K}^{\prime 2} + \tfrac{Jb^2}{2r^2 a^2} \dot{\mathcal{H}}_1^2 \bigg\}, \nonumber 
\end{align} 
where we have defined $J \equiv \ell (\ell +1)$ for notational convenience. Despite its complicated appearance, this Lagrangian propagates only two physical degrees of freedom, which we can loosely think of the scalar mode plus one of the graviton polarizations. We will now show explicitly how to obtain a Lagrangian that contains only these two degrees of freedom. The modes with $\ell =0, 1$ require once again special treatment---see {\it e.g.}~\cite{Kobayashi:2014wsa}. Hence, for simplicity we will restrict our attention to the modes with $\ell \geqslant 2$ in what follows.

The form of the Lagrangian \eqref{quadratic Lagrangian even modes} makes it clear that $H_0$ is just a Lagrange multiplier enforcing a ``Hamiltonian'' constraint among the remaining variables. We can render such constraint algebraic in $\mathcal{H}_1$ by trading $H_2$ for a new variable $\psi$ defined as follows,
\begin{equation} \label{psi definition}
	\psi = - \frac{r a b}{2} \left( H_2   -r \mathcal{K}'  -\frac{J \mathcal{H}_1}{r}\right),
\end{equation}
 Notice that this field redefinition is such that the term $\mathcal{K}''$ also disappears from the constraint equation. A similar change of variable was performed in a different gauge in~\cite{Kobayashi:2014wsa}. Then, this constraint can be easily solved for $\mathcal{H}_1$ to express it in terms of $\psi, \mathcal{K}$ and their radial derivatives. Using the background equations of motion, the solution can be expressed as
\begin{equation} \label{sol h_1}
	\mathcal{H}_1 = \frac{4 b^2 r \psi'+2 J \psi +(2-J) r a b \mathcal{K}}{J b \left[a(J-2 b^2) +2 r b^2 a'\right]} - \frac{r^2}{J} \mathcal{K}'.
\end{equation}

Moreover, the only term quadratic in $H_1$ that appears in the Lagrangian \eqref{quadratic Lagrangian even modes} does not contain any derivatives. This means that the equation of motion for $H_1$ can be solved immediately, and using the field redefinition \eqref{psi definition} we find
\begin{equation}
	H_1 = - \dot{\mathcal{H}}_1 + \frac{4\dot \psi }{Jab}+\frac{2r}{J} \left(1 -r \frac{a'}{a}\right) \dot{\mathcal{K}}.
\end{equation}

By plugging the solution \eqref{sol h_1} into this result, one can express $H_1$ just in terms of $\psi, \mathcal{K}$, and their derivatives. Notice that such an expression contains mixed second derivatives of the form $\dot \psi'$ and $\dot{\mathcal{K}}'$, and therefore one might fear that the quadratic Lagrangian \eqref{quadratic Lagrangian even modes} contains terms with four derivatives when expressed in terms of $\psi$ and $\mathcal{K}$ alone. However, because the coefficients in front of the $H_1^2$ and $\dot{\mathcal{H}}_1^2$ terms are identical, these higher derivative terms cancel out. After some integrations by parts, terms with three derivatives cancel out as well,\footnote{This is not surprising given that the action \eqref{appEFT-ex1} simply corresponds to a $P(X,\Phi)$ theory in the covariant formulation, which clearly has second order equations of motion.} and we are left with a Lagrangian of the form
\begin{equation} \label{even quadratic action}
	S^{(2)}_{{\rm even}, m=0, \ell \geqslant 2} =  \int \D t \D r \tfrac{1}{2} \Mpl^2\left( \mathcal{A}^{ij} \dot \chi_i \dot \chi_j - \mathcal{B}^{ij} \chi_i' \chi_j' - \mathcal{C}^{ij} \chi_i \chi_j' - \mathcal{D}^{ij} \chi_i \chi_j \right),
\end{equation}
where $\chi_i \equiv (\psi, \mathcal{K})$ and
\begin{subequations}
\begin{align}
	\mathcal{A}^{ij} =& \begin{pmatrix}
 \frac{8 \left[J r (b a'- a 
   b')
   + a b \left(J-2\right)\right]}{J a^2 \left[2 r b^2
   a'+a \left(J-2 b^2\right)\right]^2} &
   -\frac{2 \left(J-2\right) r}{J a \left[2 r
   b^2 a'+a \left(J-2 b^2\right)\right]} \label{Aij}\\
  -\frac{2 \left(J-2\right) r}{J a \left[2 r
   b^2 a'+a \left(J-2 b^2\right)\right]}  &
   \frac{\left(J-2\right) r^2}{2 J a b} \\
\end{pmatrix} , \\
\label{Bij}	\mathcal{B}^{ij} =& a^2 b^2 \mathcal{A}^{ij} + \begin{pmatrix}
 \frac{-32  r^2 a b^5 M_2^4/\Mpl^2}{ \left[2 r
   b^2 a'+a \left(J-2 b^2\right)\right]^2} & 0 \\
0 & 0 \\
\end{pmatrix} ,\\
	\mathcal{C}^{ij} =& \begin{pmatrix} 0 & 0 \\ -\frac{4 \left(J-2\right) a b \left[J r (b a'-a b') +ab\left(J-2\right)-4 J r^2 a b^3 M^4_2/\Mpl^2\right]}{J\left[2 r b^2
   a'+a \left(J-2 b^2\right)\right]^2} & 0 \end{pmatrix} , \\
   \mathcal{D}_{11} =& -\tfrac{8}{r^2 a^2 \left[2 r b^2 a'+a \left(J-2 b^2\right)\right]^3} \bigg\{ {\scriptstyle 4 r^2 a b^3  \left[-r^2 a b^2 a^{\prime 2} \left(-4 r b b'+2 b^2+J\right)  +2 r a^2 b a' \left(2 J r
   b'-4 r b^2 b'-b^3+b\right) +2 r^3 b^4 a^{\prime 3} \right. } \nonumber \\
   & {\scriptstyle \left.+a^3 \left(-4 J r b b'+4 r b^3 b'+\left(J-2\right)
   b^2+2 b^4-J\right)\right]M_2^4 /\Mpl^2  -4 r^3 a^2 b^5 \left(a-r a'\right) \left[2 r b^2 a'+a \left(J-2
   b^2\right)\right](M_2^4)'/\Mpl^2 } \nonumber \\
   & {\scriptstyle  + r^2 a^2 b^2 \left(r a'-a\right) b'' \left[2 r
   b^2 a'+a \left(J-2 b^2\right)\right]+r^2a^2 b \left(a-r a'\right) b^{\prime 2}
   \left[2 r b^2
   a' - a \left(J+2 b^2\right)\right] } \\
   & {\scriptstyle + r a b' \left[-r^2 a b^2 a^{\prime 2}
   \left(4 b^2+3 J\right)+r a^2 b^2 a' \left(2
   b^2+J+4\right)+2 r^3 b^4 a^{\prime 3}+(J-2) a^3 \left(2
   b^2+J\right)\right] }\nonumber \\
   & {\scriptstyle + b \left[r^3 a b^2 a^{\prime 3}
   \left(3 J-4 b^2\right)-(J+2) r^2 a^2 b^2 a^{\prime 2}-r a^3
   a' \left(3 J b^2-4 b^4+(J-3) J\right)+2 r^4 b^4
   a^{\prime 4}+a^4 \left((J+2) b^2-2 b^4-J^2+3 J-4\right)\right], } \nonumber \\
	\mathcal{D}_{12} =& \mathcal{D}_{21} = -\tfrac{2 \left(J-2\right) \left[a^2 (J-2) +2rb(a'r -a) (ab' -ba' + 2ra b^3 M_2^4 /\Mpl^2 )\right]}{ r \left[2 r b^2 a'+a \left(J-2 b^2\right)\right]^2}, \\
	\mathcal{D}_{22} =& \tfrac{\left(J-2\right)^2 a \left[2 J  r a b (b a'-a b')+ 4 b^2 (a^2 r^2 a^{\prime 2})-2(J+2)a^2b^2+a^2J^2-4J r^2 a^2 b^4 M_2^4 /\Mpl^2 \right]}{2 J  b \left[2 r b^2 a'+a \left(J-2 b^2\right)\right]^2} .
\end{align} 
\end{subequations}

Starting from a Lagrangian of the form \eqref{even quadratic action}, we can obtain the radial speeds of propagation for the two even modes by diagonalizing the matrix
\begin{equation}
c_r^2 \equiv \frac{1}{a^2b^2}\mathcal{A}^{-1}\mathcal{B} \, .
\label{csmatrix}
\end{equation}
Notice the extra factor of $1/(a^2 b^2)$, which is needed because the $(t,t)$ and $(r,r)$ components of the background metric are non-trivial. 
Using the particular form of the  kinetic coefficients in Eqs. \eqref{Aij} and \eqref{Bij}, we find that the two eigenvalues of the matrix \eqref{csmatrix} are 
\begin{equation}
c_{r,1}^2 = 1 \, ,
\qquad \qquad \quad 
c_{r,2}^2 = \frac{f - 4M_2^4}{f} \, ,
\label{eigencs}
\end{equation}
where $f(r)$ is the tadpole coefficient \eqref{fm} with $M_1\equiv\Mpl$, $\alpha\equiv0$ and $c(r)=r$. Thus, we see that the last operator in \eqref{action even lowest order} breaks the degeneracy between the two sound speeds. Absence of superluminality as well as  gradient instabilities in the radial direction then implies $0< 4M_2^4 /f < 1$. At this order in the derivative expansion, the sound speeds in \eqref{eigencs} can also be recovered by working in the decoupling limit, as we will show in the next section. This is no longer true when higher derivative operators such as $\delta g^{rr}\delta K$ are included in the  action.

By varying the action \eqref{even quadratic action} with respect to $\psi$ and $\mathcal{K}$ one finds a system of two coupled, linear differential equations, which can be further simplified with an appropriate rescaling of the coordinate $r$ and a redefinition of the field $\psi$ and $\mathcal{K}$. A discussion along this line will be presented elsewhere, while in the following we will focus only on a very specific limit such that the dynamics of the scalar degree of freedom decouples from the gravity sector.


\subsection{Goldstone mode and decoupling limit}
\label{sec:declim}

In constructing the unitary gauge action \eqref{EFT-ss0} we have chosen a specific foliation  of the spacetime by fixing radial diffeomorphisms in such a way to set to zero the perturbations of the scalar field. In turn, they have shown up in the metric tensor \eqref{even metric perturbations}.
An alternative but equivalent choice, which turns out to be particularly convenient to decouple scalar and metric perturbations, can be made by restoring the full diffeomorphism invariance using the so-called St\"uckelberg trick. This amounts to performing a broken radial diffeomorphism of the form $r\rightarrow r+\pi(r,x^a)$ in the action \eqref{EFT-ss0}, and promoting the gauge parameter $\pi$ to a full-fledged field. The field $\pi$ then admits a natural interpretation as the Goldstone boson that realizes non-linearly the spontaneously broken $r$-translations. After restoring full diff-invariance, one can then fix the gauge by imposing conditions on the metric perturbations alone, as discussed below Eq. \eqref{unitary gauge metric pert}.

The explicit transformation laws of the various geometric ingredients appearing in \eqref{EFT-ss0} under a broken radial diffeomorphism are summarized in Appendix \ref{stueck}. For the purposes of the present discussion, it is sufficient to remind the reader of the following result:
\begin{equation}
g^{rr} \rightarrow  g^{rr}(1+ 2\pi' + {\pi'}^2)
		+ 2g^{a r}\partial_a\pi + 2g^{a r}\pi'\partial_a\pi
		+ (\partial_a\pi)(\partial_b\pi)g^{ab} \, ,
\label{StuckForm-g000}
\end{equation}
Without loss of generality, in the remaining of this section we will work with a radial coordinate such that $b\equiv1$ in the background metric \eqref{GMN-2}.

For simplicity we will restrict our attention to the leading order action \eqref{action even lowest order}. The inclusion of higher derivative operators is discussed in Appendix \ref{app:declimcubic}. The tadpole coefficients \eqref{fm} and \eqref{lm} then reduce to
\begin{equation}
\Lambda(r) 
= -\left(\frac{c''}{c} + \frac{a'c'}{ac}+\frac{c'^2}{c^2}-\frac{1}{c^2}  \right)\Mpl^2 \, \qquad \qquad 
 f(r) 
=  \left(\frac{a'c'}{ac} - \frac{c''}{c}  \right)\Mpl^2  \, .
\label{flm2}
\end{equation}
After performing the St\"uckelberg transformation \eqref{StuckForm-g000}, the action \eqref{action even lowest order} takes on the form
\begin{equation}
\begin{split}
S & =  \int\D^4x \, \sqrt{-g} \bigg[
	\frac{\Mpl^2}{2}R -\Lambda(r+\pi)
\\
&
	- f(r+\pi)\left(g^{rr}(1+ 2\pi' + {\pi'}^2)
		+ 2g^{a r}\partial_a\pi + 2g^{a r}\pi'\partial_a\pi
		+ (\partial_a\pi)(\partial_b\pi)g^{ab} \right)
\\
&
	+ M^4_2(r+\pi)\left(\delta g^{rr}+g^{rr}( 2\pi' + {\pi'}^2)
		+ 2g^{a r}\partial_a\pi + 2g^{a r}\pi'\partial_a\pi
		+ (\partial_a\pi)(\partial_b\pi)g^{ab}  \right)^2 
\bigg]  .
\end{split}
\label{appEFT-ex2}
\end{equation}

Despite this action's complicated appearance, one can usually find a regime---known as decoupling limit---where the kinetic mixing between the scalar mode and the graviton helicities becomes negligible compared to their kinetic terms. For instance, let's consider the mixing term $2f(r)\delta g^{rr}\pi'$ in Eq. \eqref{appEFT-ex2}. After introducing the canonically normalized fields  $\pi_c\equiv \pi \sqrt{2f}$ and $\delta g^{rr}_c\equiv \delta g^{rr}\Mpl$, this terms reads $\frac{\sqrt{2f}}{\Mpl}\delta g_c^{rr}\pi'_c$. Thus, for energies above $E_\text{mix}\equiv \frac{\sqrt{2f}}{\Mpl}$ it can be safely neglected compared to the kinetic terms for $\delta g_c^{\mu\nu}$ and $\pi_c$.\footnote{For simplicity we do not distinguish between energy and momentum scales. Since, strictly speaking, this is truly legitimate only for luminal propagation, we will tacitly assume here that the speed of sound is not too small.} As a result, in this regime one study the perturbations $\delta g^{\mu\nu}$ and $\pi$ separately. Focusing on the latter, we set the metric to its background value \eqref{GMN-2} to obtain the following quadratic action for $\pi$:
\begin{equation}
S_\pi^{(2)}  =  \int \D t \D r \D\Omega \, ac^2\left\lbrace   \left[\frac{\partial_r\left(ac^2 f' \right)}{ac^2}
	 -\frac{\left( 
	 \Lambda'' + f'' \right)}{2} \right]\pi^2
	 	-\left( f - 4M_2^4 \right){\pi'}^2 - f (\partial_a\pi)(\partial^a\pi)
\right\rbrace   .
\label{appEFT-ex2-2}
\end{equation}
The sound speeds of $\pi$ in the angular and radial directions can be immediately read off from \eqref{appEFT-ex2-2}:
\begin{equation}
c_{\pi,r}^2 = \frac{f - 4M_2^4}{f} \, ,
\qquad \qquad 
c_{\pi,\Omega}^2 = 1 \, .
\label{csdclod}
\end{equation}
As anticipated, $c_{\pi,r}^2$ coincides with one of the two eigenvalues found in \eqref{eigencs}.

In the usual covariant language, the theory described by the action \eqref{action even lowest order} corresponds to a $P(X,\Phi)$-theory. It is a straightforward exercise to show that the action \eqref{appEFT-ex2-2} can also be obtained from the Lagrangian $\mathcal{L}=P(X,\Phi)$ by expanding $\Phi=\bar\Phi+\delta\Phi$. The relation between the St\"uckelberg field $\pi$ and the scalar fluctuation $\delta\Phi$ is given by the equation $\bar\Phi(r+\pi)=\bar\Phi(r)+\delta\Phi(r,x^a)$. In particular, the tadpole conditions \eqref{flm2} reduce to
\begin{equation} \label{P(X phi) tadpoles}
\Lambda \equiv -P+XP_X \, ,
\qquad \qquad 
f \equiv -XP_X \, ,
\qquad \qquad 
M_2^4 \equiv \frac{1}{2}X^2P_{XX} \, .
\end{equation}

After expanding the action \eqref{appEFT-ex2} in powers of $\pi$, the linear term vanishes because it is proportional to the background equation of motion for $\Phi$, which in light of the results \eqref{P(X phi) tadpoles} reads:
\begin{equation}
\frac{2}{ac^2}\partial_r\left( ac^2 f \right) -  \frac{2 f\bar\Phi''}{\bar\Phi'} + P_{,\Phi}\bar\Phi' =0 \, .
\label{shift01}
\end{equation}

Quite remarkably, if the theory is shift symmetric (i.e. $P_{,\Phi}=0$), the equation \eqref{shift01} can be solved analytically irrespective of the functional form of $P(X)$. Indeed, integrating twice Eq. \eqref{shift01} with respect to $r$, one finds
\begin{equation}
\bar\Phi(r)  = \Phi_0 + \Phi_1 \int^r\D \tilde r \, a(\tilde r)c^2(\tilde r)f(\tilde r) \, ,
\end{equation}
where $\Phi_0$ and $\Phi_1$ are arbitrary integration constants, and $f$ is given by Eq. \eqref{flm2}.\footnote{We remind that the analogue for FLRW backgrounds is (see {\it e.g.} \cite{Finelli:2018upr})
\begin{equation}
\bar\Phi(t) = \Phi_0 + \Phi_1 \int^t\D \tilde t \, a^3(\tilde t )\dot{H}(\tilde t ) \, .
\nonumber
\end{equation}
}
This occurs because, for shift symmetric theories, the background values of $P$ and all its derivatives can be unambiguously fixed in terms of the background metric only \cite{Finelli:2018upr}.

\section{Outlook}
\label{sec:outlook}

The breakthrough detection of gravitational waves from binary black hole or neutron star mergers presents us with the opportunity to test gravity in a new regime. Although GR with a cosmological constant (CC) seems so far to provide the correct description of gravitational interactions at long distances, the lack of a plausible field theory mechanism for the smallness of the CC could be an indication of additional degrees of freedom in the gravitational sector. It has even been recently conjectured that, if string theory is the correct UV completion of gravity, the present accelerated expansion of the universe cannot be the result of a positive cosmological constant~\cite{Obied:2018sgi}, thus suggesting that additional dark energy fields must be present. 

The new observational window opened up by gravitational waves could potentially uncover,  or at the very least constrain, the existence of this additional sector. A promising observable is the spectrum of the QNMs emitted by the BH remnant during the ringdown phase of a coalescence. The presence of additional degrees of freedom can in fact modify the frequencies compared to the predictions of GR, depending on the additional interactions. In the absence of a ‘best motivated’ proposal for the dynamics of the new sector, however, the best way to characterize how different models affect the QNMs is to follow an EFT approach. 

In this paper we have made a first step towards constraining in a model-independent way modifications of GR. We focused our attention on alternative theories of gravity that satisfy two main assumptions: \textit{i)} include one additional light scalar degree of freedom besides the graviton, and \textit{ii)} admit BH solutions with a scalar hair.   The second assumption is crucial to generate a significant (and hopefully testable) departure from the QNM spectrum predicted by GR. In fact, without a scalar hair it can be shown that, at second order in derivatives, the GR frequencies are always a subset of the QNM spectrum of black hole solutions in scalar-tensor theories~\cite{Tattersall:2017erk}.

Based on these assumptions, we derived the most general EFT up to quadratic order in perturbations around static and spherically symmetric backgrounds with a scalar hair.  
To constrain the coefficients in the effective Lagrangian it is necessary to measure the frequencies of at least two QNMs. Such a result is presently not achievable at LIGO/Virgo, but should be within reach when the upgraded detectors will reach  design sensitivity, and certainly with third-generation or space-based detectors (for a review on BH spectroscopy prospects see \cite{Berti:2018vdi} and references therein). Such experiments will be able to probe quasinormal modes with higher $\ell$, for which the WKB approximation  employed in this paper is expected to perform even better. 
In principle, there exist other ways, which do not rely on the ringdown phase, to obtain independent bounds on the EFT parameters, \textit{e.g.} the one given by the study of dissipative effects in the form of dipole radiation during the inspiral stage \cite{Barausse:2016eii}. However, this would deserve a separate discussion, beyond the purposes of the present work,  and is left for the future.

To make contact with actual observations there are still a few steps missing. An important issue that should be addressed in full generality is the coupling to matter fields.
Our calculations for both even and odd sectors are done in the frame where
the $M_1(r) = \Mpl$ (the so called Einstein frame). The question is whether/how matter fields, for instance the LIGO mirrors, are coupled to the scalar field $\Phi$ in addition to the expected coupling to the Einstein frame metric.
The observational implications fall into two classes depending on this coupling:
\begin{itemize}

\item If the scalar-matter coupling is absent or very weak, only the tensor quasi-normal modes would be observed. The scalar could still indirectly affected the observational signals through the deviation in the tensor spectrum from GR, and through the breaking of isospectrality between even and odd modes.

\item If the scalar-matter coupling is at gravitational strength or larger, the most prominent observational signal would be the scalar mode itself---this is the extra mode in the even sector. It is distinguished from the tensor modes not only in terms of its spectrum, but also in terms of how it affects the detectors. Interferometric detectors with different orientations can be combined to tell apart scalar from tensor modes. Of course, the scalar could also make its presence known through a deviation of the tensor spectrum from GR, and through the breaking of isospectrality.

\end{itemize}

Another important issue to address is that the EFT for perturbations must be generalized from static backgrounds to spinning ones. The reduced number of isometries of the background will translate into a larger number of operators in the action and, therefore, of free parameters to be constrained by observations. 

The number of independent operators can be reduced, even in the case of static backgrounds, if one is willing to make additional assumptions about the underlying scalar-tensor theory. In the present paper, we have included in the effective action all  operators compatible with the symmetries that contribute at quadratic order and contain at most two derivatives. Some of these operators are actually generated by terms in the covariant scalar-tensor theory that depends on second derivative of the scalar field. This can lead to a much richer phenomenology, but at the same time it is potentially dangerous because it  can propagate an additional degree of freedom giving rise to instabilities. Additional relations on the coefficients can be enforced to prevent its appearance, extending what is already known in the case of time dependent backgrounds~\cite{Langlois:2017mxy}. Further conditions can be derived by imposing additional symmetries ({\it e.g.} shift symmetry on the scalar~\cite{Finelli:2018upr}), or by considering positivity constraints that follow from unitarity and analyticity of scattering amplitudes (as was done for instance in the case of inflation in~\cite{Baumann:2015nta}). All these topics will be discussed elsewhere. 

To conclude, we should mention that the formalism developed in this paper lends itself also to a few  more formal applications. First, because our effective Lagrangian is solely determined by the background isometries, it can also be used to describe perturbations around metrics other than black holes, provided they are static and spherically symmetric. For instance, our approach can be used to investigate the stability of wormholes in theories with a scalar field \cite{Franciolini:2018aad}. It can also offer a different perspective on the (long) list of no-hair theorems, allowing to characterize the conditions for hairy BHs in a more model independent way. The formal applications of this approach will be discussed in a companion paper~\cite{us-in-preparation}.

\acknowledgments

We thank Imre Bartos, Emanuele Berti, Vitor Cardoso, Andy Cohen, Pedro Ferreira, Austin Joyce, Xinyu Li, Michele Maggiore, Szabi Marka, Alberto Nicolis, Paolo Pani, Federico Piazza and Sam Wong for discussions.
The work of R.P. was mostly supported by the US Department of Energy (HEP) Award DE-SC0013528. L.S. is supported by the Netherlands organization for scientific research (NWO). E.T. is partially supported by the MIUR PRIN project 2015P5SBHT.
L.H. is supported in part by NASA grant NXX16AB27G and DOE grant DE-SC011941.

\appendix

\section{Construction of the unitary gauge action}
\label{Sec:constructionEFT}

In this Appendix we present the construction of the effective theory for perturbations in the unitary gauge, defined by $\delta\Phi\equiv0$.  The breaking of radial diffeomorphism invariance induced by $\delta\Phi\equiv0$ makes a radial foliation more natural with respect to the standard time-foliation of the ADM decomposition: in the following, we will introduce the corresponding geometric ingredients and derive analogue equations to the standard ADM case.

\subsection{Notation and radial foliation}

We start introducing a foliation of the spacetime manifold defined by the the family of hypersurfaces that satisfy  $\Phi=\text{constant}$. The orthogonal unit vector is defined by
\begin{equation}
n^\mu \equiv \frac{\nabla^\mu \Phi(x)}{\sqrt{\nabla_\mu \Phi\nabla^\mu \Phi}} \, ,
\label{ouv}
\end{equation}
satisfying $n_\mu n^\mu=1$.
In analogy with the ADM decomposition, the metric can be written as
\begin{equation}
\D s^2  = N^2\D r^2 + h_{ab}(\D x^a + N^a \D r)(\D x^b + N^b \D r) \, ,
\label{ds2}
\end{equation}
where $h_{\mu\nu}\equiv g_{\mu\nu}-n_\mu n_\nu$ is the induced metric, while $N$ and $N^a$ are the lapse and the shift respectively. In this notation, the metric tensor and its inverse read
\begin{equation}
g_{\mu\nu}
	 = \begin{pmatrix}
				h_{ab}	&		N_a\\
			N_b				&		N^2+N^c N_c
		\end{pmatrix} ,
\qquad
g^{\mu\nu}
	 = \begin{pmatrix}
			h^{ab}+N^{-2}N^a N^b	&		-N^{-2}N^a \\
			-N^{-2}N^b	&		N^{-2}	
		\end{pmatrix} \, ,
\label{G}
\end{equation}
where the Latin indices $a,b,c\ldots$ are used for temporal and angular coordinates: $\lbrace a,b,c\ldots\rbrace=\lbrace t, \theta, \phi\rbrace$.
The unitary gauge is fixed by requiring that constant-$r$ hypersurfaces coincide with the uniform scalar field hypersurfaces, i.e. $\delta\Phi(x^a,r)\equiv0$. With this choice, $n_r = N$ and $n_a =0$. Therefore, $g_{r a}= h_{ra} = N_a$ and $h_{rr} = N^a N_a$.
By construction, the following orthogonality conditions hold:
\begin{equation}
h^\mu_\nu n^\nu =0 \, ,
\qquad
n^\mu \nabla_\nu n_\mu =0 \, .
\label{r4}
\end{equation}
Covariant derivatives acting on the $(2+1)$-dimensional hypersurface can be defined in the standard way as
\begin{equation}
D_a V_b = h^\mu_a h^\nu_b \nabla_\mu V_\nu \, ,
\label{Dcov}
\end{equation}
for some generic vector $V_\mu$.
Moreover, the extrinsic curvature can be constructed by projecting on the hypersurface as
\begin{equation}
K_{\mu\nu} = h^\alpha_\mu h^\beta_\nu \nabla_\alpha n_\beta = h^\alpha_\mu \nabla_\alpha n_\nu
=   \nabla_\mu n_\nu -  n^\alpha n_\mu \nabla_\alpha n_\nu \, .
\label{excurv}
\end{equation}
In particular, the covariant temporal and angular components of the extrinsic curvature can be conveniently written also as
\begin{equation}
K_{ab}=\nabla_a n_b = -N\Gamma_{ab}^r
	= \dfrac{1}{2N}\left( \partial_r h_{ab} - D_a N_b - D_a N_b \right) \, .
\label{Eij}
\end{equation}

\subsection{Gauss-Codazzi equation}

The $(2+1)$-dimensional Riemann tensor is defined in the standard way as
\begin{equation} 
-\hat{R}_{\alpha\beta\mu\nu}V^\alpha = D_\mu D_\nu V_\beta -D_\nu D_\mu V_\beta \, .
\end{equation}
After some lengthy computation, one can derive the Gauss-Codazzi equation
\begin{equation}
h^\tau_\mu h^\rho_\nu h^\sigma_\beta R_{\alpha\sigma\tau\rho}
= \hat{R}_{\alpha\beta\mu\nu}  + K_{\mu\beta}K_{\nu\alpha} - K_{\nu\beta} K_{\mu\alpha} \, .
\label{gauss3}
\end{equation}
Contracting both sides of \eqref{gauss3} with the full metric $g^{\mu\alpha}g^{\nu\beta}$ and after some manipulations, one finds
\begin{equation}
R = \hat{R} - K_{\mu\nu}K^{\mu\nu} + K^2 - 2\nabla_\mu(Kn^\mu-n^\nu\nabla_\nu n^\mu) \, ,
\label{Gauss-Codazzi}
\end{equation}
which relates the full Ricci scalar $R$ to the intrinsic curvature $\hat R$. As expected, formally the only difference with respect to the standard ADM decomposition based on a time foliation consists in some signs.

\subsection{Effective theory in unitary gauge}

In parallel with \cite{Cheung:2007st}, one can write the most general Lagrangian in unitary gauge by requiring invariance only under the residual (temporal and angular) diffeomorphisms. Therefore, besides the standard Riemann tensor, covariant derivatives and contractions thereof, one can make use of additional building blocks consisting in: explicit functions of $r$; operators with free $r$-upper indices, such as $g^{rr}$ and $R^{rr}$; the extrinsic curvature $K_{\mu\nu}$. Notice that, because of the Gauss-Codazzi relation \eqref{gauss3}, $(3+1)$-dimensional objects and their projected versions are not independent. Therefore, one can forget about the induced Riemann $\hat{R}_{\mu\nu\alpha\beta}$ and contractions thereof in the Lagrangian. Moreover, for the same reason, one can also avoid the use of the induced metric and $(3+1)$-dimensional covariant derivatives.

As a result, the most general action in the unitary gauge $\delta\Phi\equiv0$ takes on the form
\begin{equation}
S = \int\D^4 x\sqrt{-g} \, \mathcal{L}\left(g_{\mu\nu}, \epsilon^{\mu\nu\alpha\beta} ,
R_{\mu\nu\alpha\beta}, g^{rr} , K_{\mu\nu} ,\nabla_\mu ; r 
\right) \, .
\label{T}
\end{equation}

\subsubsection{Effective action for perturbations}

The result \eqref{T} represents the most general theory that is compatible with the residual symmetries after fixing the gauge $\delta\Phi\equiv0$. The logic underlying the construction of the effective theory for perturbations closely follows the one of \cite{Creminelli:2006xe,Cheung:2007st}, but the result turns out to differ considerably in a few aspects, as already discussed in Sec. \ref{Sec:EFT}. Indeed, the different degree of symmetry of the background \eqref{GMN-2} with respect to the case of FLRW cosmologies crucially affects both the number of independent operators in the EFT for perturbations and their transformation properties under residual (temporal and angular) diffeomorphisms.
In the following, we will explicitly construct the EFT \eqref{EFT-ss0} and make such differences manifest. 
Let us start with the tadpole Lagrangian. 

In full generality, up to linear order in perturbations, the effective theory contains the following tadpoles:
\begin{equation}
S_{\text{tadpole}} = \int\D^4 x\sqrt{-g}\left[ -\Lambda(r)
- f(r)g^{rr} + \kappa^{\mu\nu}(r)K_{\mu\nu} +
\xi^{\mu\nu\alpha\beta}(r) R_{\mu\nu\alpha\beta} \right] \, ,
\label{EFTss}
\end{equation}
one for each building block in \eqref{T}. $\Lambda(r)$, $f(r)$, $\kappa^{\mu\nu}(r)$ and $\xi^{\mu\nu\alpha\beta}(r)$ are arbitrary functions of the background metric and its derivatives.\footnote{The Einstein Hilbert term simply corresponds to taking $\frac{\Mpl^2}{2}\bar g^{\mu\alpha}\bar g^{\nu\beta}$ in $\xi^{\mu\nu\alpha\beta}(r)$.} If $\kappa^{\mu\nu}(r)$ and $\xi^{\mu\nu\alpha\beta}(r)$ were proportional the background metric only, then the last two tadpoles in \eqref{EFTss} would simply be $\kappa(r)K$ and $\xi(r)R$ and, getting rid of $K=\nabla_\mu n^\mu$ by an integration by parts, one would conclude that \eqref{EFTss} contains only $3$ free functions. By contrast, since in general $\partial_r \bar g_{ab} \, \cancel{\propto} \, \bar g_{ab}$, the matrices $\kappa^{\mu\nu}(r)$ and $\xi^{\mu\nu\alpha\beta}(r)$ have in principle many more independent entries corresponding to additional free functions in the theory \eqref{EFTss}. In the following, we are going to make this statement more quantitative, in particular we are going to show that eventually only $4$ are actually independent (i.e. $\kappa^{\mu\nu}(r)$ and $\xi^{\mu\nu\alpha\beta}(r)$ contain only $2$ free functions, the other $2$ being $\Lambda(r)$ and $f(r)$).

Let us start recalling that the orthogonality condition $K^{\mu\nu}n_\mu=0$ allows to use the induced metric to raise and lower indices in the tadpole $\kappa^{\mu\nu}(r)K_{\mu\nu}$, to be read therefore as $\kappa^{ab}(r)K_{ab}$.
Since the matrix $\kappa^{ab}(r)$ is a function of background quantities, it must carry the same symmetries of $\bar h_{ab}$, meaning that it has to be a diagonal matrix with only two independent entries: $\kappa_{ab}\equiv\text{diag}(\kappa_{tt},\kappa_{\theta\theta},\kappa_{\theta\theta}\sin^2\theta )$.
As a result, the tadpole $\kappa^{ab}(r)K_{ab}$ can be explicitly written as
\begin{equation}
\kappa^{ab}(r)K_{ab} = \kappa^{tt}(r)K_{tt} + \kappa^{\theta\theta}(r)\left(K_{\theta\theta}+\frac{K_{\phi\phi}}{\sin^2\theta}  \right) \, ,
\label{kappaterms}
\end{equation}
where $\kappa^{tt}(r)$ and $\kappa^{\theta\theta}(r)$ are the two free functions of $r$.
Furthermore, since the trace of the extrinsic curvature can always be recast in terms of $\Lambda(r)$ and $f(r)g^{rr}$ up to a total derivative by an integration by parts ($K=\nabla_\mu n^\mu$), we are free to add a term of the type $-\kappa^{\theta\theta}(r)c^2(r)h^{ab}K_{ab}$ to the Lagrangian \eqref{EFTss} in order to cancel the last term in \eqref{kappaterms}. This means that $\kappa^{ab}(r)K_{ab}$ contains actually only one free independent function.

Let us now focus on the last tadpole $\xi^{\mu\nu\alpha\beta}(r) R_{\mu\nu\alpha\beta}$. Again, since $\xi^{\mu\nu\alpha\beta}(r)$ is a background tensor, because of rotational invariance, the angular components are not independent from each other. In addition, taking into account the symmetry structure of the Riemann tensor, one infers that it contains in principle $8$ arbitrary functions. Let us start analyzing those corresponding to the coefficients of
\begin{equation}
{R^r}_{trt} \, , \qquad
({R^\theta}_{r\theta r}+{R^\phi}_{r\phi r}) \, , \qquad
({R^\theta}_{t\theta t}+{R^\phi}_{t\phi t}) \, , \qquad
{R^\theta}_{\phi\theta\phi} \, .
\label{elimable}
\end{equation}
First of all, notice that the term involving ${R^r}_{trt}$ can be always eliminated in favor of the other tadpoles. Indeed, consider the operator $K_{\mu\nu}R^{\alpha\mu\beta\nu}n_\alpha n_\beta$, which in unitary gauge reads $(g^{rr})^{-1}K_{\mu\nu}R^{r\mu r\nu}$. Using the definition of Riemann tensor and Eq. \eqref{excurv},
\begin{equation}
\begin{split}
& K_{\mu\nu}  R^{\alpha\mu\beta\nu} n_\alpha n_\beta  = -K_{\mu\nu}n_\beta\left( \nabla^\beta\nabla^\nu n^\mu -  \nabla^\nu\nabla^\beta n^\mu \right)
\\
&	= -K^{\mu\nu}n^\beta\left[ \nabla_\beta K_{\nu\mu} + \nabla_\beta\left( n^\alpha n_\nu\nabla_\alpha n_\mu \right)
-\nabla_\nu K_{\beta\mu} - \nabla_\nu\left( n^\alpha n_\beta\nabla_\alpha n_\mu \right)\right]
\\
&	= -K^{\mu\nu}\left[n^\beta \nabla_\beta K_{\nu\mu} + n^\alpha n^\beta \nabla_\alpha n_\mu \nabla_\beta n_\nu
-n^\beta \nabla_\nu K_{\beta\mu} - \nabla_\nu\left( n^\alpha\nabla_\alpha n_\mu \right)\right]
  \, .
\end{split} 
\label{analR}
\end{equation}
Therefore, up to integrations by parts, it can be re-written in terms of the tadpoles $f(r)g^{rr}$, $\Lambda(r)$ and $\kappa^{\mu\nu}(r)K_{\mu\nu}$. This means that we can forget about the first term in \eqref{elimable} in our counting of independent functions.
Moreover, taking the trace, the last three combinations in \eqref{elimable} can be all eliminated in favor of the Ricci tensor as\footnote{Since we are interested in terms that are linear in perturbations, we shall use the background metric to raise and lower indices.}, 
\begin{align}
R_{rr} & = {R^t}_{rtr}+ ({R^\theta}_{r\theta r}+{R^\phi}_{r\phi r}) \, , \\
R_{tt} & = {R^r}_{trt}+ ({R^\theta}_{t\theta t}+{R^\phi}_{t\phi t}) \, , \\
R_{\theta\theta}+\frac{R_{\phi\phi}}{\sin^{2}\theta}  & = -\frac{c^2}{a^2} ({R^\theta}_{t\theta t}+{R^\phi}_{t\phi t}) + 
c^2({R^\theta}_{r\theta r}+{R^\phi}_{r\phi r}) + 2{R^\phi}_{\theta\phi\theta}  \, .
\end{align}
Therefore, in full generality the tadpole $\xi^{\mu\nu\alpha\beta}(r) R_{\mu\nu\alpha\beta}$ can be always thought of as being a sum of the four remaining building blocks
\begin{equation}
R \, , \qquad
R_{rr} \, , \qquad
R_{tt} \, , \qquad
\left( R_{\theta\theta}+\frac{R_{\phi\phi}}{\sin^{2}\theta}\right) \, , 
\label{elimable2}
\end{equation}
with some arbitrary coefficients. Whether these are all independent or not is what we are going to show now. 

First, it is clear that the trace condition $R={R^t}_t+{R^r}_r+({R^\theta}_\theta+{R^\phi}_\phi)$ allows immediately to eliminate one of the last three terms in \eqref{elimable2}, say $R_{tt}$, with the only effect of redefining the coefficients of the others. 
Second, consider the identity
\begin{equation}
R_{\mu\nu}K^{\mu\nu} 	= R_{\mu\nu}\nabla^\mu n^\nu -\frac{1}{2} R_{\mu\nu}n_\alpha\nabla^\alpha(n^\mu n^\nu) \, .
\end{equation}
After simple integrations by parts, up to linear order in perturbations $R_{\mu\nu}K^{\mu\nu}\supseteq g^{rr}$, $ R$, $R^{rr}$, $ K$ only. Thus, one is always free to add to the Lagrangian the operator $R_{\mu\nu}K^{\mu\nu}$ with some proper coefficient in such a way to get rid of also $({R^\theta}_\theta+{R^\phi}_\phi)$ in \eqref{elimable2}, in analogy with the discussion around \eqref{analR}. Finally, the identity 
\begin{equation}
(g^{rr})^{-1}R^{rr} = R_{\mu\nu}n^\mu n^\nu = K^2 -K_{\mu\nu}K^{\mu\nu} + \nabla^\mu \left( n^\nu \nabla_\nu n_\mu - n_\mu K \right) \, ,
\end{equation}
which simply follows from the definition of the $(3+1)$-dimensional Riemann tensor, allows to re-express also $R^{rr}$ as a function of the other tadpole operators.

In conclusion, the action \eqref{EFTss} contains in general $4$ independent tadpoles:
\begin{equation}
S_{\text{tadpole}} = \int\D^4 x\sqrt{-g}\left[ \frac{1}{2}M^2_1(r) R -\Lambda(r) - f(r)g^{rr} - \alpha(r)\bar K_{\mu\nu} K^{\mu\nu} \right] \, ,
\label{tadLagr}
\end{equation}
where $M^2_1(r)$, $\Lambda(r)$, $f(r)$ and $\alpha(r)$ are the corresponding coefficients, some of which are to be fixed by Einstein equations.

\subsubsection{Quadratic action for perturbations}

In the previous Section, we have shown that the only independent operators that enter up to linear order in perturbations are those in \eqref{tadLagr}. The next step is all possible operators at quadratic order that are compatible with the symmetries of the system. In the spirit of an effective description, we will classify the operators in terms of the number of derivatives.

\paragraph{Zero-th order in derivatives.} At the zero-th order in derivatives, the only quadratic operator in perturbations is given by $M^4_2(r)(\delta g^{rr})^2$, where the coefficient $M^4_2(r)$ is in principle an arbitrary function of $r$ of dimensions $4$ in energy to be fixed experimentally.

\paragraph{First order in derivatives.} At the first order in derivatives, the only non-trivial operators we can add are of the form
\begin{equation}
f_{ab}(r) \delta g^{rr}\delta K^{ab} \, ,
\label{oo1}
\end{equation}
where $f_{ab}$ is an $r$-dependent matrix which must have the same symmetries of the background metric \eqref{GMN-2}. In other words, it is diagonal and with only two free entries:
\begin{equation}
f_{ab}(r)  \equiv \begin{pmatrix}
f_1(r) & \\
& f_2(r) \gamma_{ij}
\end{pmatrix} \, ,
\qquad
\gamma_{ij} \equiv \begin{pmatrix}
1 & \\
& \sin^2\theta
\end{pmatrix} \, ,
\qquad
\lbrace i,j\rbrace=\lbrace\theta,\phi\rbrace \, ,
\label{fh}
\end{equation}
for some arbitrary $f_1(r)$ and $f_2(r)$.
As a result, at first order in derivatives there are only two independent operators in the effective theories, that we choose to write as
\begin{equation}
M_3^3(r) \delta g^{rr }\delta K \, ,
\qquad
M_4^2(r) \bar K_{ab} \delta g^{rr }\delta K^{ab} \, .
\end{equation}

\paragraph{Second order in derivatives.} At the second order in the number of derivatives, we have in principle many more independent operators allowed by symmetries. Let us start with those involving the extrinsic curvature only, which schematically read
\begin{equation}
f_{abcd}(r) \delta K^{ab}\delta K^{cd} \, ,
\label{pert2d}
\end{equation}
where again $f_{abcd}(r)$ has the same isometries of the background. Therefore, it is clear that the only independent free functions in $f_{abcd}(r)$ are the coefficients of the operators
\begin{equation}
 (\delta K^{tt})^2 , \quad \gamma_{ij }\delta K^{tt}\delta K^{ij} , \quad
(\gamma_{ij}\delta K^{ij})^2 ,  \quad \gamma_{ij }\delta K^{ti}\delta K^{tj} , \quad
\gamma_{ij}\delta K^{jk}\gamma_{kl}\delta K^{li} \, .
\end{equation}
Let us now consider operators that involve the Riemann tensor. Analogous considerations to Eq. \eqref{analR} suggest that we can disregard operators of the type $R^{r\mu  r\nu}$. Then, we can focus only on those where the Riemann is projected onto the $(2+1)$-hypersurface, which can be written as
\begin{equation}
f_{abcd}(r) \delta \hat{R}^{abcd} \delta g^{rr} \, .
\label{pert2dintr}
\end{equation}
Because of the background isometries, only $2$ are independent:
\begin{equation}
\gamma_{ij }\delta\hat{R}^{titj} \, ,
\qquad
 \gamma_{ij } \gamma_{kl }\delta\hat{R}^{ikjl} \, .
\end{equation}
Finally, one could also think of adding operators involving derivatives of $\delta g^{rr}$: the combinations allowed by the residual symmetries in the EFT are $(\partial_\mu\delta g^{rr})^2$, $(\partial_r\delta g^{rr})^2$, $(\partial_r\delta g^{rr})\delta K$ and $(\partial_r\delta g^{rr})\bar K_{ab}\delta K^{ab}$. In general these will be associated with higher order equations of motion. Whether this fact is related to the presence of pathologies is beyond the scope of our work and will be discussed elsewhere.

In conclusion, at second order in perturbations there are $14$ independent operators admitted by symmetries, that we write as follows:
\begin{align}
& S^{(2)} =  \int\D^4x \, \sqrt{-g} \Big[
	 M_2^4(r)(\delta g^{rr})^2
	+M_3^3(r) \delta g^{rr }\delta K  + M_4^2(r) \bar K_{\mu\nu} \delta g^{rr }\delta K^{\mu\nu}
\nonumber\\
&	+ M_5^2(r)(\partial_r\delta g^{rr})^2
	+M_6^2(r) (\partial_r\delta g^{rr})\delta K  + M_7(r) \bar K_{\mu\nu} (\partial_r\delta g^{rr})\delta K^{\mu\nu}
	+ M_8^2(r)(\partial_a\delta g^{rr})^2
\nonumber\\
&	+M_9^2(r)(\delta K)^2 + M_{10}^2(r)\delta K_{\mu\nu}\delta K^{\mu\nu} 
+ M_{11}(r) \bar K_{\mu\nu}\delta K \delta K^{\mu\nu} 
+ M_{12}(r) \bar  K_{\mu\nu}\delta K^{\mu\rho} \delta {K^{\nu}}_\rho 
\nonumber\\
&	+ \lambda(r)\bar  K_{\mu\rho}{\bar{K}^{\rho}}_{ \nu} \delta K \delta K^{\mu\nu}
+ M_{13}^2(r) \delta g^{rr } \delta\hat{R} 
+ M_{14}(r) \bar K_{\mu\nu} \delta g^{rr }\delta\hat{R}^{\mu\nu} 
\Big] \, .
\label{EFT-ss}
\end{align}

\section{Infinitesimal variations}
\label{app:variations}

In this section we derive the variations of the geometric ingredients of the radially foliated manifold, collecting the results that have been used in the main text.

To begin with, the variation of the normal vector $n_\mu$ \eqref{ouv} under an infinitesimal transformation of the metric of the type $g_{\mu\nu}\rightarrow g_{\mu\nu}+\delta g_{\mu\nu}$ takes on the form
\begin{equation}
\delta n_\mu =  -\frac{1}{2}n_\mu n_\alpha n_\beta\delta g^{\alpha\beta} \, ,
\label{id1}
\end{equation}
leading also to
\begin{equation}
\delta n^\mu = n_\nu\delta g^{\mu\nu} -\frac{1}{2}n^\mu n_\alpha n_\beta\delta g^{\alpha\beta} \, .
\label{dnM}
\end{equation}
Together with the orthogonality condition $h^{\mu\nu}n_\mu=0$, Eq. \eqref{id1} can be used to derive the following identity:
\begin{equation}
n_\mu\delta h^{\mu\nu} = -h^{\mu\nu}\delta n_\mu = 0 \, .
\end{equation}
The variation $\delta K$ of the trace of the extrinsic curvature can be easily computed using the identity
\begin{equation}
\frac{1}{\sqrt{-g}} \partial_\mu \left( \sqrt{-g} \, X^\mu \right) = \nabla_\mu X^\mu 
\label{idgr}
\end{equation}
($X^\mu$ is a generic four-vector), which follows from the relation $\delta \sqrt{-g} = \frac{1}{2}\sqrt{-g} g^{\mu\nu}\delta g_{\mu\nu} $.
Using \eqref{idgr}, we can write
\begin{equation}
K = \nabla_\mu n^\mu = \frac{1}{\sqrt{-g}}\partial_\mu \left(\sqrt{-g} \, n^\mu \right) \, .
\end{equation}
Then, the variation $\delta K$ is easily computed as
\begin{equation}
\delta K 
	= \frac{1}{2} g_{\mu\nu}\delta g^{\mu\nu} K + \nabla_\mu\left[ 
	 n_\nu\delta g^{\mu\nu} -\frac{1}{2}n^\mu(n_\alpha n_\beta +g_{\alpha\beta})\delta g^{\alpha\beta} 
	 \right] \, ,
\label{varK}
\end{equation}
where we have used the result \eqref{dnM}.

Let us now focus on $\delta K_{\mu\nu}$. From the definition \eqref{excurv}, the variation of the extrinsic curvature reads
\begin{equation}
\begin{split}
\delta K_{\mu\nu} & = \nabla_\mu \delta n_\nu - \delta \Gamma^\rho_{\mu\nu} n_\rho 
-  \delta n^\rho n_\mu\nabla_\rho n_\nu
\\
&\qquad
-  n^\rho\delta n_\mu\nabla_\rho n_\nu
-  n^\rho n_\mu\nabla_\rho \delta n_\nu +  n^\rho n_\mu\delta\Gamma^\sigma_{\rho\nu}n_\sigma \, .
\end{split}
\end{equation}
Plugging in the variation of the Christoffel symbol,
\begin{equation}
\begin{split}
\delta \Gamma^\rho_{\mu\nu}
&	= \frac{1}{2}g^{\rho\sigma}\left(\nabla_\mu\delta g_{\nu\sigma} + \nabla_\nu\delta g_{\mu\sigma} - \nabla_\sigma\delta g_{\mu\nu} \right)
\\
&	= -\frac{1}{2}g^{\rho\sigma}\left(g_{\alpha\nu}g_{\beta\sigma}\nabla_\mu
	+ g_{\alpha\mu}g_{\beta\sigma}\nabla_\nu- g_{\alpha\mu}g_{\beta\nu}\nabla_\sigma \right)\delta g^{\alpha\beta}  \, ,
\end{split}
\label{varChr}
\end{equation}
it takes on the form
\begin{equation}
\begin{split}
\delta K_{\mu\nu} & = -\frac{1}{2}\nabla_\mu \left(n_\nu n_\alpha n_\beta\delta g^{\alpha\beta}\right) 
\\
&\qquad
	+ \frac{1}{2}\left(g_{\alpha\nu}n_\beta\nabla_\mu
	+ g_{\alpha\mu}n_\beta\nabla_\nu- g_{\alpha\mu}g_{\beta\nu}n^{\rho}\nabla_\rho \right)\delta g^{\alpha\beta}
\\
&\qquad
	- n_\beta n_\mu(\nabla_\alpha n_\nu)\delta g^{\alpha\beta} + n_\alpha n_\beta n_\mu n^\rho (\nabla_\rho n_\nu)\delta g^{\alpha\beta}
\\
&\qquad
	+ \frac{1}{2} n_\mu n^\rho \nabla_\rho \left( n_\nu n_\alpha n_\beta \delta g^{\alpha\beta} \right)
	- \frac{1}{2} n_\alpha n_\beta n_\mu\nabla_\nu\delta g^{\alpha\beta} \, ,
\end{split}
\label{dKMNv}
\end{equation}
and one can easily check that taking the trace the result \eqref{varK} is recovered.

Finally, we calculate $\delta \hat R$. Taking the variation of the trace of Eq. \eqref{gauss3}, 
\begin{equation}
\delta \hat{R} = 2h^{\mu\nu}R_{\mu\rho\nu\sigma}\delta h^{\rho\sigma} + h^{\mu\nu}h^{\rho\sigma}\delta R_{\mu\rho\nu\sigma} + 2K\delta K - 2K^{\mu\nu}\delta K_{\mu\nu}
  \, ,
\label{gauss4-22}
\end{equation}
where
\begin{equation}
\delta h^{\mu\nu} = \delta g^{\mu\nu} -  n_\lambda\left( n^\mu\delta g^{\nu\lambda} + n^\nu\delta g^{\mu\lambda} \right) + n^\mu n^\nu n_\alpha n_\beta\delta g^{\alpha\beta}
\label{dgvar}
\end{equation}
and
\begin{equation}
\delta R_{\mu\rho\nu\sigma} 
	= -R_{\beta\rho\nu\sigma} g_{\alpha\mu} \delta g^{\alpha\beta} 
+ g_{\mu\lambda} \left( \nabla_\nu\delta\Gamma^\lambda_{\rho\sigma}  - \nabla_\sigma\delta\Gamma^\lambda_{\rho\nu} 
\right) \, ,
\label{driemm}
\end{equation}
where $\delta\Gamma^\lambda_{\mu\nu}$ is written in Eq. \eqref{varChr}.
Therefore, after straightforward manipulations,
\begin{equation}
\begin{split}
\delta\hat{R}
&	=  \left( R_{\alpha\beta} - n^\mu n^\nu R_{\mu\alpha\nu\beta}
	- 3R_{\rho\alpha} n_\beta n^\rho
	+ 2R_{\mu\nu} n^\mu n^\nu n_\alpha n_\beta
	 \right) \delta g^{\alpha\beta}
\\
&\qquad
	+ h^{\nu}_\mu h^{\rho\sigma} \left( \nabla_\nu\delta\Gamma^\mu_{\rho\sigma}  - \nabla_\sigma\delta\Gamma^\mu_{\rho\nu} 
\right)
	+ 2K\delta K - 2K^{\mu\nu}\delta K_{\mu\nu}
  \, .
\end{split}
\label{gauss4-22b}
\end{equation}

\section{St\"uckelberg trick and decoupling limit}
\label{stueck}

In this appendix we collect the St\"uckelberg transformations that can be used to restore full gauge invariance in the theory \eqref{EFT-ss}. Without affecting the generality of the discussion, we will set here $b\equiv1$ in the background metric \eqref{GMN-2}.

Under a general transformation of coordinates, $x\rightarrow \tilde{x}(x)$, the metric changes as
\begin{equation}
\tilde{g}_{\mu\nu}(\tilde{x})
	 = \dfrac{\partial x^\alpha}{\partial \tilde{x}^\mu}
		\dfrac{\partial x^\beta}{\partial \tilde{x}^\nu}
		g_{\alpha\beta}(x) \, ,
\qquad
\tilde{g}^{\mu\nu}(\tilde{x})
	 = \dfrac{\partial \tilde{x}^\mu}{\partial x^\alpha}
		\dfrac{\partial \tilde{x}^\nu}{\partial x^\beta}
		g^{\alpha\beta}(x) \, .
\label{m2}
\end{equation}
Focusing in particular on transformations that leave time and angular coordinates invariant,
\begin{equation}
\begin{cases}
r\rightarrow \tilde{r} = r + \pi(x^a,r) \, , \\
x^a\rightarrow \tilde{x}^a = x^a \, ,
\end{cases}
\label{StcukDiff0}
\end{equation}
or, equivalently,
\begin{equation}
\tilde{x}^\mu(x) = x^\mu + \pi(x)\delta^\mu_r \, .
\label{xtilde}
\end{equation}
A generic scalar function of $r$ transforms as
\begin{equation}
F(r) \rightarrow F(r+\pi)= F(r)+F'(r)\pi + \dfrac{1}{2}F''(r)\pi^2+\ldots
\label{StuckForm-Scalar}
\end{equation}
while
\begin{align}
\partial_r 
	& \rightarrow  (1-\pi'+{\pi'}^2)\partial_r \, ,
\label{StuckDer1}
\\
\partial_a
	& \rightarrow  (-\partial_a\pi+\pi'\partial_a\pi)\partial_r + \partial_a \, .
\label{StuckDer2}
\end{align}
The transformation laws of the contravariant components of the metric take on the form
\begin{equation}
g^{rr} \rightarrow  g^{rr}(1+ 2\pi' + {\pi'}^2)
		+ 2g^{a r}\partial_a\pi + 2g^{a r}\pi'\partial_a\pi
		+ (\partial_a\pi)(\partial_b\pi)g^{ab} \, ,
\label{StuckForm-g00}
\end{equation}
\begin{equation}
g^{ra} \rightarrow  (1 + \pi')g^{ra} + (\partial_b\pi)g^{ab} \, ,
\label{StuckForm-g^0i}
\end{equation}
\begin{equation}
g^{ab} \rightarrow  g^{ab} \, .
\label{StuckForm-g^ij}
\end{equation}
On the other hand, the St\"uckelberg transformations for the covariant metric can be easily calculated by solving perturbatively in $r(\tilde{x})$ the equation $\tilde{r}(x) = r + \pi(x^a,r)$, from which one finds $r(\tilde{x}^a,\tilde{r}) = \tilde{r} - \pi(\tilde{x}^a,\tilde{r}-\pi) = \tilde{r} - \pi(\tilde{x}^a,\tilde{r})+\pi'\pi(\tilde{x}^a,\tilde{r})+\ldots$ and therefore
\begin{equation}
g_{ra} \rightarrow  g_{rr}\left[-\partial_a\pi 	+ 2\pi'\partial_a\pi + \ldots\right](x^a,r)
		+ g_{ra}\left[1-\pi' + {\pi'}^2+ \ldots\right](x^a,r) \, ,
\label{StuckForm-g_0i}
\end{equation}
\begin{multline}
g_{ab} \rightarrow  g_{rr}\left[\partial_a\pi\partial_b\pi + \ldots\right](x^a,r)
		+ g_{rb} \left[- \partial_a\pi + \pi'\partial_a\pi+\ldots\right](x^a,r) +
 \\
	 + g_{ra} \left[- \partial_b\pi + \pi'\partial_b\pi+\ldots\right](x^a,r)
		+ g_{ab} \, ,
\label{StuckForm-g_ij}
\end{multline}
\begin{equation}
g_{rr} \rightarrow  g_{rr}\left[1-2\pi' + 3{\pi'}^2+ \ldots\right](x^a,r) \, .
\label{StuckForm-g_00}
\end{equation}
These expressions highly simplify in the decoupling limit regime. Setting the metric to its background value \eqref{GMN-2}, they can be assembled as
\begin{equation}
g^{\mu\nu} \rightarrow  \begin{pmatrix}
	\bar{g}^{ab} & 	 (\partial_c\pi)\bar{g}^{ac}
	\\
	(\partial_c\pi)\bar{g}^{cb}
		& 	
	\bar{g}^{rr}(1+ 2\pi' + {\pi'}^2)
		+ (\partial_c\pi)(\partial_d\pi)\bar{g}^{cd} 
\end{pmatrix} \, ,
\label{StuckForm-g^munu}
\end{equation}
\begin{equation}
g_{\mu\nu} \rightarrow
	\begin{pmatrix}
	\bar{g}_{ab} + \bar{g}_{rr}\partial_a\pi\partial_b\pi
		& 	\bar{g}_{rr}(-\partial_a\pi 	+ 2\pi'\partial_a\pi)
	\\
	\bar{g}_{rr}(-\partial_b\pi 	+ 2\pi'\partial_b\pi)
		& 	\bar{g}_{rr}(1-2\pi' + 3{\pi'}^2)
	\end{pmatrix} \, .
\label{StuckForm-g_munu}
\end{equation}
Moreover, some useful equations are: 
\begin{align}
N_a = g_{ra}
	& \rightarrow -\partial_a\pi + 2\pi'\partial_a\pi + \mathcal{O}(\pi^3) \, ,
\label{StcuckSumm2} \\
N = \dfrac{1}{\sqrt{g^{rr}}}
	& \rightarrow  1- \pi' + {\pi'}^2
		-\frac{1}{2} (\partial_a\pi)(\partial_b\pi)\bar{g}^{ab} + \mathcal{O}(\pi^3) \, ,
\label{StcuckSumm3} \\
N^a =-N^2 g^{ra}
	& \rightarrow  -(1-2\pi')(\partial_b\pi)\bar{g}^{ab}+ \mathcal{O}(\pi^3) \, ,
\label{StcuckSumm4} \\
h^{ab} = g^{ab} - \dfrac{N^a N^b}{N^2}
	& \rightarrow \bar{g}^{ab}
	- (\partial_c\pi)(\partial_d\pi)\bar{g}^{ac}\bar{g}^{bd}
	+ \mathcal{O}(\pi^3) \, ,
\label{StcuckSumm5}
\end{align}
up to quadratic order in $\pi$. Furthermore, the $(2+1)$-dimensional Christoffel symbol
\begin{equation}
\hat{\Gamma}^c_{ab} =
	\dfrac{1}{2} h^{cd}\left( \partial_a h_{bd}
		+ \partial_b h_{ad}
		- \partial_d h_{ab} \right) 
\label{StuckChristoffel-g-kij}
\end{equation}
transforms as (up to first order)
\begin{equation}
\hat{\Gamma}^c_{ab} \rightarrow
	 \bar{\Gamma}^c_{ab} 
	 + \dfrac{1}{2} \bar{g}^{cd}
		\left(
		 -\bar{g}'_{bd}\partial_a\pi
		-\bar{g}'_{ad}\partial_b\pi
		+\bar{g}'_{ab}\partial_d\pi \right) + \mathcal{O}(\pi^2) \, ,
\label{StuckChristoffel-h-kij}
\end{equation}
where $\bar{\Gamma}^c_{ab}$ is defined as the background value of $\Gamma^c_{ab}$,
\begin{equation}
	 \bar{\Gamma}^c_{ab} \equiv
	  \dfrac{1}{2} \bar{g}^{cd}
		\left(
		\partial_a\bar{g}_{bd} + \partial_b\bar{g}_{ad} - \partial_d\bar{g}_{ab}
		  \right) \, .
\end{equation}
Together with \eqref{StcuckSumm2}-\eqref{StcuckSumm5}, Eq. \eqref{StuckChristoffel-h-kij} can be used to compute the transformation law of the extrinsic curvature, which up to linear order in $\pi$ reads
\begin{equation}
K_{ab}   \rightarrow 
	  \bar{K}_{ab}
		+ \bar D_a\bar D_b\pi 
	+ \mathcal{O}(\pi^2) \, ,
\label{tlawKij}
\end{equation}
where $\bar D_a$ is the projected covariant derivative computed on the background metric, while taking the trace
\begin{equation}
K     \rightarrow 
 \bar{K}
		+ \bar D_a\bar D^a\pi 
	+ \mathcal{O}(\pi^2) \, .
\label{tlawK}
\end{equation}
Therefore, the transformation laws for the perturbed quantities read
\begin{equation}
\delta K_{ab}   \rightarrow 
	  -\bar{K}'_{ab}\pi + \bar D_a\bar D_b\pi 
	+ \mathcal{O}(\pi^2) \, ,
	\qquad
	\delta K  \rightarrow 
	- \bar{K}'\pi
		+ \bar D_a\bar D^a\pi 
	+ \mathcal{O}(\pi^2) \, .
\label{tlawdKij}
\end{equation}
As already emphasized in the main text the perturbation $\delta K_{ab}$ for the extrinsic curvature does not transform as a tensor, as it is clear from the term $\bar{K}'_{ab}$ in \eqref{tlawdKij}.

\subsection{An explicit example: the cubic Galileon}
\label{app:declimcubic}

In this appendix we extend the discussion of Sec. \ref{sec:declim} of the decoupling limit in the EFT \eqref{EFT-ss0} by including operators with one extra derivative, i.e. operators of the form $(\delta g^{rr})^n\delta K$. Up to quadratic order in the number of fields, this yields
\begin{multline}
S =  \int\D^4x \, \sqrt{-g} \bigg[
\frac{\Mpl^2}{2} R -\Lambda(r) - f(r)g^{rr} - \alpha(r)\bar K_{\mu\nu} K^{\mu\nu}
\\
	+ M_2^4(r)(\delta g^{rr})^2
	+M_3^3(r) \delta g^{rr }\delta K  + M_4^2(r) \bar K_{\mu\nu} \delta g^{rr }\delta K^{\mu\nu}
 + \ldots
\bigg] \, .
\end{multline}
Let us focus for simplicity on the case of theories with $\alpha=M_4^2=0$, of which cubic Horndeski operators are genuine examples (see App. \ref{app:cubic galileon}).
Using \eqref{tlawdKij} yields the following quadratic action for the scalar mode in the decoupling limit:
\begin{multline}
S^{(2)}_\pi =  \int\D^4x \, ac^2\sin\theta \bigg\{
\left[\frac{1}{ac^2}\partial_r\left(ac^2 f' +ac^2 M_3^3\bar K' \right)  -\frac{1}{2}\left(\Lambda''+f'' \right) \right]\pi^2
\\
+\left[ f - \frac{a}{c^2}\partial_r\left(\frac{c^2}{a}M_3^3  \right) \right]\frac{\dot{\pi}^2}{a^2}
\\
-\left[f - \frac{1}{a}\partial_r\left(aM_3^3  \right) \right]
\left[\frac{(\partial_\theta\pi)^2}{c^2}+\frac{(\partial_\phi\pi)^2}{c^2\sin^2\theta}\right]
-\left(f - 4M_2^4 \right){\pi'}^2
\bigg\} \, ,
\label{EFT-ssG30bis}
\end{multline}
where the speeds of propagation are now modified according to
\begin{equation}
c_{\pi,r}^2 = \frac{f - 4M_2^4}{f - \frac{a}{c^2}\partial_r\left(\frac{c^2}{a}M_3^3  \right)} \, ,
\qquad
c_{\pi,\Omega}^2 = \frac{f - \frac{1}{a}\partial_r\left(aM_3^3  \right)}{f - \frac{a}{c^2}\partial_r\left(\frac{c^2}{a}M_3^3  \right)} \, .
\end{equation}
Notice that the operator $\delta g^{rr} \delta K$ is responsible for making the velocity along the angular direction non-unitary.
Furthermore, it is now instructive to consider the flat spacetime limit of \eqref{EFT-ssG30bis} and compare the result for instance with \cite{Nicolis:2004qq}.
To this end, it is convenient to make the field redefinition $\pi=\delta\Phi/\bar \Phi'+\ldots$, which holds at the leading order. Then, in the limit $a\rightarrow1$, $c\rightarrow r$, $\bar{K} \rightarrow \frac{2}{r}$, the action \eqref{EFT-ssG30bis} takes on the form
\begin{multline}
S^{(2)}_\varphi =  \int\D^4x \, ac^2\sin\theta \bigg\{
\bigg[\frac{1}{r^2}\partial_r\left(r^2 f' +r^2 M_3^3\bar K'\right)  -\frac{1}{2}\left(\Lambda''+f'' \right)
\\
- \frac{\bar \Phi'^2}{r^2}\partial_r\left[  \frac{r^2\bar \Phi''}{\bar \Phi'^3}\left(f - 4M^4 \right)
\right]
-\left(f - 4M^4 \right)\frac{\bar \Phi''^2}{\bar \Phi'^2}
 \bigg]\frac{\delta\Phi^2}{\bar \Phi'^2}
\\
+\left[ f - \frac{1}{r^2}\partial_r\left(r^2M_3^3  \right) \right]\frac{\dot{\delta\Phi}^2}{\bar \Phi'^2}
-\left[f - \partial_r\left(M_3^3  \right) \right]
\frac{(\partial_\Omega\delta\Phi)^2}{\bar \Phi'^2}
-\left(f - 4M^4 \right)\frac{{\delta\Phi'}^2}{\bar \Phi'^2}
\bigg\} \, .
\label{EFT-ssG30bis3}
\end{multline}
In order to make a comparison with \cite{Nicolis:2004qq}, we take the cubic galileon Lagrangian \cite{Nicolis:2008in} (see also App. \ref{app:cubic galileon})
\begin{equation}
\mathcal{L} = g_2(\partial\Phi)^2 + g_3(\partial\Phi)^2\square\Phi
= g_2 X + \frac{2g_3}{3}X^{3/2}K \, ,
\label{cubicexN}
\end{equation}
where $g_2$ and $g_3$ are generic coupling constants.
Therefore, expanding the metric in fluctuations in the unitary gauge $\delta\Phi=0$ and comparing with \eqref{EFT-ssG30bis} yields
\begin{equation}
M^4(r)  = \frac{g_3\bar\Phi'^2}{4}\left(\bar\Phi''+\bar\Phi'\bar{K} \right)   \, ,
\qquad
M_3^3(r)  = g_3\bar\Phi'^3 \, ,
\label{M3}
\end{equation} 
\begin{equation}
f    =
-g_2\bar\Phi'^2 + g_3\bar\Phi'^2 \left(\bar \Phi''  -\bar \Phi'\bar{K}\right) \, ,
\qquad
\Lambda   =
g_3\bar\Phi'^2 \left(\bar \Phi''  +\bar \Phi'\bar{K}\right)  \, .
\label{M3-2}
\end{equation}

Using the expressions \eqref{M3}-\eqref{M3-2} and setting $g_2=-3$, $g_3=-1$, it is straightforward to check that \eqref{EFT-ssG30bis3} coincides with the quadratic action of \cite{Nicolis:2004qq}. In particular, one can show that the mass of $\delta\Phi$ in \eqref{EFT-ssG30bis3} is zero on the background equations of motion, as it should be since the theory \eqref{cubicexN} we started with is shift invariant.

\section{Cubic and Quartic Horndeski in unitary gauge}

The construction of the effective theory \eqref{EFT-ss0} is based only on the breaking pattern of Poincaré down to time translations and spatial rotations. Additional symmetries ({\it e.g.} \cite{Finelli:2018upr}) or the requirement of not having any unwanted ghost like degree of freedom on top of the scalar and tensor modes (see {\it e.g.} \cite{Langlois:2017mxy} for a discussion in the context of FLRW backgrounds), will further constrain the $r$-dependent coefficients in \eqref{EFT-ss0}. In addition, one could also expect more constraints coming from causality and analyticity \cite{Baumann:2015nta}. Postponing the study of all these points to future work, in this appendix we confine ourselves to show which kinds of operators are generated in the specific class of Horndeski theories \cite{Horndeski:1974wa}, which besides of having second order equations of motion \cite{Deffayet:2011gz} are known to be protected against large quantum corrections \cite{Pirtskhalava:2015nla,Santoni:2018rrx}. As an example, we will focus on the cubic Horndeski and, as a prototype of theory that yields the additional $\alpha$-tadpole in \eqref{EFT-ss0}, we will discuss the quartic Horndeski.

Let us start from the definition \eqref{ouv} of the unit vector perpendicular to the hypersurface. Then, differentiating and using Eq. \eqref{excurv}, 
\begin{equation}
\nabla_\mu\nabla_\nu\Phi
	= \sqrt{X}\nabla_\mu n_\nu + \frac{n_\nu\nabla_\mu X}{2\sqrt{X}}
	=  \sqrt{X}\left(K_{\mu\nu}+n^\alpha n_\mu\nabla_\alpha n_\nu\right) + \frac{n_\nu\nabla_\mu X}{2\sqrt{X}} \, ,
\label{phinm}
\end{equation}
where $X \equiv \nabla_\mu\Phi\nabla^\mu\Phi$.
On the other hand,
\begin{multline}
\sqrt{X}n_\nu n^\alpha\nabla_\alpha n_\mu + \frac{1}{2X}\nabla^\alpha\Phi\nabla_\alpha Xn_\mu n_\nu
\\
= -\frac{1}{2X}n_\nu n^\alpha\nabla_\alpha X\nabla_\mu \Phi + n_\nu n^\alpha\nabla_\alpha\nabla_\mu\Phi
+ \frac{1}{2X}\nabla^\alpha\Phi\nabla_\alpha Xn_\mu n_\nu
\\
= \frac{1}{2\sqrt{X}}n_\nu\nabla_\mu\left(\nabla_\alpha\Phi\nabla^\alpha\Phi\right) \, .
\label{stepi}
\end{multline}
Eq. \eqref{stepi} can be used to re-write Eq. \eqref{phinm} as \cite{Gleyzes:2013ooa,Gleyzes:2014qga}
\begin{equation}
\nabla_\mu\nabla_\nu\Phi
=  \sqrt{X}\left(K_{\mu\nu}+n^\alpha n_\mu\nabla_\alpha n_\nu+n^\alpha n_\nu\nabla_\alpha n_\mu\right)
+  \frac{1}{2X}\nabla^\alpha \Phi\nabla_\alpha Xn_\mu n_\nu \, .
\label{phinm-2}
\end{equation}

\subsection{Cubic Horndeski}
\label{app:cubic galileon}

Let us focus on the cubic Horndeski Lagrangian
\begin{equation}
\mathcal{L}_{\text{H}3}
	= G_3(\Phi,X)\square\Phi  \, ,
\label{H3L}
\end{equation}
where $G_3$ is an arbitrary function of $X\equiv \nabla_\mu\Phi\nabla^\mu\Phi$ and $\Phi$.
Taking the trace of Eq. \eqref{phinm}, or equivalently Eq. \eqref{phinm-2},
\begin{equation}
\square\Phi = \sqrt{X}K + \frac{\nabla^\mu\Phi\nabla_\mu X}{2X} \, ,
\end{equation}
and defining $F_3$ such that \cite{Gleyzes:2013ooa}
\begin{equation}
G_3=F_3+2XF_{3X} \, ,
\label{diffeqFG}
\end{equation}
after an integration by parts,
the cubic Horndeski Lagrangian \eqref{H3L} can be written as follows:
\begin{equation}
\begin{split}
\mathcal{L}_{\text{H}3}
&	= -F_{3X}\nabla_\mu X\nabla^\mu\Phi
	-XF_{3\Phi}
	 + 2XF_{3X}\left(\sqrt{X}K + \frac{\nabla^\mu\Phi\nabla_\mu X}{2X} \right) 
	\\
&	= 2X^{3/2}F_{3X}K -XF_{3\Phi} \, .
\label{ca}
\end{split}
\end{equation}
Solving the differential equation \eqref{diffeqFG} yields 
\begin{equation}
F_3(X) = \frac{1}{\sqrt{X}}\int\D X\frac{G_3(X)}{2\sqrt{X}}
\end{equation}
and 
\begin{equation}
2X^{3/2}F_{3X} = \sqrt{X}G_{3}(X)-\int\D X\frac{G_3(X)}{2\sqrt{X}}
= \int\D X \, \sqrt{X} \, G_{3X}(X) \, .
\end{equation}

\subsection{Quartic Horndeski}

Now we are going to rewrite the quartic  Horndeski Lagrangian 
\begin{equation}
\mathcal{L}_{\text{H}4} = G_4(\Phi,X)R
-2G_{4X}(\Phi,X)\left[(\square\Phi)^2-(\nabla_\mu\nabla_\nu\Phi)^2\right]
\label{L4}
\end{equation}
in terms of the geometric quantities of the radially foliated spacetime. 
Using the second identity in \eqref{r4} and the orthogonality condition $K_{\mu\nu}n^\nu=0$, we can write
\begin{equation}
\nabla_\mu\nabla_\nu\Phi\nabla^\mu\nabla^\nu\Phi = 
X\left(K_{\mu\nu}K^{\mu\nu}+2n^\alpha n^\beta \nabla_\alpha n_\mu \nabla_\beta n^\mu\right)
+  \frac{1}{4X^2}(\nabla^\alpha \Phi\nabla_\alpha X)^2 \, .
\end{equation}
Then, following \cite{Gleyzes:2013ooa}, the Lagrangian \eqref{L4} takes on the form
\begin{multline}
\mathcal{L}_{\text{H}4} = G_4R
-2G_{4X}\bigg[\left(\sqrt{X}K + \frac{\nabla^\mu \Phi\nabla_\mu X}{2X}\right)^2
\\
-X\left(K_{\mu\nu}K^{\mu\nu}+2n^\alpha n^\beta \nabla_\alpha n_\mu \nabla_\beta n^\mu\right)
-  \frac{1}{4X^2}(\nabla^\alpha \Phi\nabla_\alpha X)^2\bigg] \, .
\label{L4-2}
\end{multline}
Now, using that
\begin{multline}
n^\alpha \nabla_\alpha n_\mu = \frac{\nabla^\alpha \Phi}{\sqrt{X}}\nabla_\alpha \frac{\nabla_\mu \Phi}{\sqrt{X}}
= \frac{\nabla^\alpha \Phi\nabla_\mu \nabla_\alpha \Phi}{X} - \frac{\nabla^\alpha \Phi\nabla_\mu \Phi\nabla_\alpha X}{2X^2}
\\
=\frac{\nabla_\mu X}{2X} - \frac{n^\alpha n_\mu\nabla_\alpha X}{2X}
= \frac{1}{2X}\left(g^\alpha_\mu-n^\alpha n_\mu\right)\nabla_\alpha X
\end{multline}
and that $n_\mu\nabla_\beta n^\mu=0$, Eq. \eqref{L4-2} becomes
\begin{equation}
\mathcal{L}_{\text{H}4} = G_4 R
-2XG_{4X}\left(K^2-K_{\mu\nu}K^{\mu\nu}\right)
-2G_{4X} \nabla_\mu X \left(Kn^\mu -n^\beta\nabla_\beta n^\mu \right) \, .
\label{L4-3}
\end{equation}
Furthermore, after some final straightforward manipulations,
\begin{multline}
-2G_{4X} \nabla_\mu X \left(Kn^\mu -n^\beta\nabla_\beta n^\mu \right)
= -2\left( \nabla_\mu G_{4} - \sqrt{X}G_{4\Phi}n_\mu \right)\left(Kn^\mu -n^\beta\nabla_\beta n^\mu \right)
\\
= 2 G_{4} \nabla_\mu  \left(Kn^\mu -n^\beta \nabla_\beta n^\mu  \right)
+2\sqrt{X}G_{4\Phi}K
\\
=  G_{4}  \left( \hat R - K_{\alpha\beta }K^{\alpha\beta} + K^2 - R\right)
+2\sqrt{X}G_{4\Phi}K \, ,
\end{multline}
where we have integrated by parts and used the Gauss-Codazzi \eqref{Gauss-Codazzi}, and substituting in \eqref{L4-3}, we find
\begin{equation}
\mathcal{L}_{\text{H}4}  = G_4\hat R
+\left(G_4-2XG_{4X}\right)\left(K^2-K_{\mu\nu}K^{\mu\nu}\right)
+2\sqrt{X}G_{4\Phi}K \, .
\label{L4-4}
\end{equation}

\section{Transformation to Schr\"odinger-like form}\label{schf}
In this Section we summarise the procedure which can be systematically followed in order to transform a second order differential equation 
\begin{equation}
	{\rm c}_1(r) Q''(r)+2{\rm c}_2(r) Q'(r)+\left[ {\rm c}_3(r) \omega^2 + {\rm c}_4(r)\right] Q(r)=0
\end{equation}
into a Schr\"odinger-like equation of the form
\begin{equation}\label{app-e1}
	{\cal Q}''(\tilde{r})+\left[\omega^2 + V(\tilde{r}) \right]{\cal Q}(\tilde{r})=0\, .
\end{equation}
Without violating the generality of the argument, we are going to set ${\rm c}_3(r)=1$. This can be always achieved multiplying \eqref{app-e1} by an overall factor ${\rm c}_3^{-1}(r)$ which effectively coincide with redefining $ {\rm c}_i (r)\rightarrow {\rm c}_i (r)/{\rm c}_3 (r)$. 
Then, by the following coordinate redefinition 
\begin{equation}
r \rightarrow \tilde{r}(r)= \int_{r_c} ^ r \frac{\D l}{\sqrt{{\rm c}_1(l) } }\ \ \ \ \ \text{s.t.}\ \ \ \ \ \frac{\D}{\D \tilde{r}} = \sqrt{{\rm c}_1(r)}\frac{\D}{\D r}  
\end{equation}
and variable redefinition
\begin{equation}
	{ Q}(r) \rightarrow {\cal Q}(\tilde{r}(r))= \exp\left [  \int_{r_c} ^ r \left ( \frac{{\rm c}_2(l)}{{\rm c}_1(l)}-\frac{{\rm c}_1 '(l)}{4 {\rm c}_1(l)} \right )\D l  \right ]{ Q}(r)
\end{equation}
one obtains  \eqref{app-e1}  with the potential:
\begin{equation}\label{rwpot-gen}
V(\tilde{r}(r)) =
\frac{{\rm c}_1''(r)}{4}+ \frac{{\rm c}_2(r) {\rm c}_1'(r)}{{\rm c}_1(r)}-\frac{3}{16}\frac{ {\rm c}_1'(r)^2}{
   {\rm c}_1(r)}-\frac{{\rm c}_2(r)^2}{{\rm c}_1(r)}-{\rm c}_2'(r)+{\rm c}_4(r)   \,.
\end{equation}

\section{The Regge-Wheeler equations} \label{App: Regge-Wheeler and Zerilli equations}

The Regge-Wheeler equations, in the form given by Zerilli~\cite{Zerilli:1971wd}, are:
\begin{equation}
\label{ZerilliC6a}
{\rm h_0}'' - \dot{\rm h_1}' - {2\over r} \dot{\rm h_1} + \left(
  {4G{\cal M}
  \over r^2} - {\ell (\ell + 1) \over r} \right) {{\rm h_0} \over r -
  2G{\cal M}} = 0
\end{equation}
\begin{equation}
\label{ZerilliC6b}
\ddot{\rm h_1} - \dot{\rm h_0}' + {2\over r} \dot{\rm h_0}
+ (\ell -1)(\ell+2)(r - 2G{\cal M}) { {\rm h_1} \over r^3} = 0
\end{equation}
\begin{equation}
\label{ZerilliC6c}
\left(1 - {2 G{\cal M} \over r}\right) {\rm h_1}' - \left( 1 -
  {2G{\cal M} \over
    r}\right)^{-1} \dot{\rm h_0} + {2G{\cal M} \over r^2} {\rm h_1} = 0
\end{equation}
They describe the GR dynamics of small perturbations around a Schwarzschild background.
Note that the Regge-Wheeler gauge has been chosen.

\section{Isolating the gravitational waves -- the large radius limit}
\label{sec:choosing gauge}

The GR equations governing the evolution of perturbations around a Schwarzschild black hole
were derived by \cite{Regge:1957td, Zerilli:1970se}. 
In this Appendix, we wish to describe how the equations can be understood in a heuristic way.
For simplicity, we focus on the odd perturbations.

In the odd sector, the only gauge transformation takes the form
$\tilde x_\mu = x_\mu + \xi_\mu$, with
\begin{equation}
\xi_\mu = (0, 0, \epsilon_i {}^j \partial_j \delta) \, ,
\end{equation}
under which 
\begin{eqnarray}
{\rm \tilde h_0} = {\rm h_0} + \dot\delta \quad , \quad
{\rm \tilde h_1} = {\rm h_1} + \delta' - {2c' \over c} \delta
\quad , \quad
{\rm \tilde h_2} = {\rm h_2} - 2\delta \, ,
\end{eqnarray}
where $\dot{(\phantom{x})} \equiv \partial_t(\phantom{x})$ and $(\phantom{x})' \equiv \partial_r(\phantom{x})$. 
The Regge-Wheeler gauge corresponds to choosing
${\rm \tilde h_2} = 0$:
\begin{eqnarray}
{\rm h_0}^{\rm RW} = {\rm h_0} + \dot\delta \quad , \quad
{\rm h_1}^{\rm RW} = {\rm h_1} + \delta' - {2c' \over c} \delta
\quad , \quad
0 = {\rm h_2}^{\rm RW} = {\rm h_2} - 2\delta \, ,
\end{eqnarray}
To isolate the gravitational waves,
we find it more transparent to keep ${\rm h_2}$ around, {\it i.e.} to set $\delta = {\rm h_2}/2$ in the expressions for ${\rm
  h_0}^{\rm RW}$ and ${\rm h_1}^{\rm RW}$. 
Further simplification is obtained by taking the large $r$ limit,
which includes $\partial_r {\,\rm fluc.} \gg {\,\rm fluc.}/r$ -- with the understanding that
$\partial_r$ pulls out a factor of the momentum, and likewise
for $\partial_t$. 
For instance $\delta' \gg c' \delta /c$. 

We substitute the above expressions for ${\rm
  h_0}^{\rm RW}$ and ${\rm h_1}^{\rm RW}$ in terms of ${\rm h_0, h_1}$ and ${\rm h_2}$ into the standard
Regge-Wheeler equations~\cite{Regge:1957td, Zerilli:1971wd}, which for completeness are summarized in Appendix \ref{App: Regge-Wheeler and Zerilli equations}. After taking the large $r$ limit, these equations reduce to
\begin{subequations}
\begin{eqnarray}
 &{\rm h_0}'' - \dot {\rm h_1}' \simeq 0& \qquad {\rm from \,\,
  Eq.\,\,}(\ref{ZerilliC6a})
 \\
 &\ddot {\rm h_1} - \dot {\rm h_0}' \simeq 0& \qquad {\rm from \,\, Eq. \,\,}
(\ref{ZerilliC6b}) \\
 &(-\dot {\rm h_0} + {\rm h_1}') + {1\over 2} (-\ddot {\rm h_2} + {\rm
  h_2}'') \simeq 0& \qquad {\rm from \,\, Eq.\,\,} (\ref{ZerilliC6c}) \, .
\end{eqnarray}
\end{subequations}
These expressions make it manifest that ${\rm h_2}$ obeys a wave equation in the large $r$
limit. Interestingly, the wave equation lives in the Regge-Wheeler 
equation with the lowest number of derivatives (no more than one
derivative on ${\rm h_1}$ or ${\rm h_0}$). 
This large $r$ limit of the Regge-Wheeler equations is useful for
seeing where the gravitational waves live, but not helpful for
deducing the quasi-normal spectrum. The dynamics of the
modes at small $r$ is important for determining the latter.

To make progress with the finite $r$ form of the Regge-Wheeler
equations, we need to identify the variable that contains
the gravitational wave degree of freedom. Both ${\rm h_0}^{\rm RW}$
and ${\rm h_1}^{\rm RW}$ contains ${\rm h_2}$, which is what we are ultimately interested
in. However, because ${\rm h_0}$ transforms by a time
derivative of the gauge parameter ($\dot\delta$), it has no conjugate
momentum. 
Thus, ${\rm h_1}$ is the more promising quantity to focus
on in terms of obtaining an equation with the desired second
derivative (in time and radius) structure. 
Our goal therefore reduces to finding a second order equation of motion for
${\rm h_1}^{\rm RW}$ out of the Regge-Wheeler equations. 

This procedure can in turn be approached in two different ways,
recalling that ${\rm h_1}^{\rm RW} =  {\rm h_1} + {\rm h_2}'/2 -  {\rm
  h_2} c'/c$. Since ${\rm h_2}$ is the gravitational wave
of interest, one could choose the ${\rm h_1} = 0$ gauge such that
the equation for ${\rm h_1}^{\rm RW}$ becomes purely an equation for
${\rm h_2}$. A more sophisticated viewpoint is to note
that the combination ${\rm h_1} + {\rm h_2}'/2 - {\rm
  h_2} c'/c$ is gauge invariant (terms involving $\delta$
cancel under a gauge transformation). This is the gauge invariant
combination that contains the gravitational wave degree of freedom;
obtaining a single equation governing its evolution is precisely
what we want. Contrast this with the other gauge invariant combination
involving ${\rm h_0}$ and ${\rm h_1}$: that removes the gravitational
wave degree of freedom and is therefore not what we want to focus on. 
(Another useful combination is $h_0 + \dot{h}_2/2$.)

Now, there are three Regge-Wheeler equations.
Because the gravitational wave of interest lives in Eq. (\ref{ZerilliC6c}), this provides a natural starting point which gives an
expression for $\dot {\rm h_0}^{\rm RW}$ (adopting the standard
Schwarzschild form for $a, b, c$ in the background metric):
\begin{eqnarray}
\dot {\rm h_0}^{\rm RW} = (1- 2G{\cal M}/r) \partial_r ( [1 - 2G{\cal M}/r] {\rm
  h_1}^{\rm RW} ) \, .
\end{eqnarray}
This
then motivates the use of Eq. (\ref{ZerilliC6b}) because it
depends on $\dot {\rm h_0}^{\rm RW}$ and its derivative,
but not ${\rm h_0}^{\rm RW}$. Thus, substituting the above into
Eq. (\ref{ZerilliC6b}):
\begin{eqnarray}
\label{h1RWeq}
&& \ddot {\rm h_1}^{\rm RW} - \partial_r \left( \left[1 - {2G{\cal M} \over
  r}\right] \partial_r \left( \left[1 - {2G {\cal M} \over r}\right] {\rm h_1}^{\rm RW} \right) \right)
  \nonumber \\
&& + {2\over r} \left(1 - {2G{\cal M} \over r}\right) \partial_r
   \left( \left[1 - {2G{\cal M} \over r}\right] {\rm
  h_1}^{\rm RW} \right) + {(\ell -1) (\ell +2) \over r^2} \left(1 -
   {2G{\cal M} \over r}\right) {\rm
  h_1}^{\rm RW} = 0 \, .\nonumber \\
{}
\end{eqnarray}
Eq. (\ref{h1RWeq}) lends itself to further simplification:
introducing the tortoise coordinate $\tilde{r}$ defined by
\begin{equation}
\partial_{\tilde{r}} \equiv (1 - 2 G{\cal M}/r) \partial_r \, ,
\label{rstar}
\end{equation}
and
multiplying ${\rm h_1}^{\rm RW}$ by a suitable function of $r$ to
remove the first derivative term (see Appendix \ref{schf}). 
Thus one obtains the standard
Schr\"odinger-like equation for determining the spectrum of odd quasi-normal modes
of a Schwarzschild black hole in GR (see {\it e.g.}~\cite{Nollert:1999ji}):
\begin{equation}
\label{RWequation}
\left[ \frac{d^2}{d \tilde{r}^2} + \omega^2 - 
\left(1 - {2G{\cal M} \over r}\right)
\left( {\ell(\ell+1)\over r^2} - {6 G{\cal M} \over r^3}\right)
\right] \left[ \left(1 - {2G{\cal M} \over r}\right)
{{\rm h_1} \over r} \right] = 0 \, .
\end{equation}

The even sector could be treated in a similar way, though it is
considerably more complex.

\section{Bianchi identities}
\label{bianchiIdentities}

The Bianchi identities tell us that not all equations of motion are
independent. In manipulating the equations that govern black hole perturbations
which are often quite complicated, it is useful to know how the equations of motion
are related. We give the relations here. 

Under a gauge transformation $\tilde x^\mu = x^\mu -
\xi^\mu$, the metric transforms according to
\begin{equation}
\delta g_{\mu\nu} = \xi^\gamma \partial_\gamma g_{\mu\nu} +
g_{\gamma\mu} \partial_\nu \xi^\gamma + g_{\gamma\nu} \partial_\mu
\xi^\gamma \, ,
\end{equation}
and the scalar field transforms according to
\begin{equation}
\delta\Phi = \xi^\gamma \partial_\gamma \Phi \, .
\end{equation}

Using the fact that $\xi^\mu$ is an arbitrary function of space and
time, it can be shown, using arguments along the lines of those for
proving Noether's second theorem, that the equations of motion are related by:
\begin{equation}
0 = {\delta S \over \delta \Phi} \partial_\gamma \Phi
+ {\delta S \over \delta g_{\mu\nu}} \partial_\gamma g_{\mu\nu}
- 2 \partial_\mu \left( {\delta S \over \delta g_{\mu\nu}} g_{\gamma\nu}
  \right) \, ,
\end{equation}
where $\gamma = 0, 1, 2, 3$ for four identities.
The form of the identities are fully general. 
When the metric and scalar are separated into background and
fluctuations, and the background equations of motion are enforced,
the Bianchi identities yield the following relations to first order in
perturbations:
\begin{equation}
0 = {\delta S \over \delta (\delta\Phi)} \partial_\gamma \bar\Phi
+  {\delta S \over \delta (\delta g_{\mu\nu})} \partial_\gamma \bar g_{\mu\nu}
- 2 \partial_\mu \left( {\delta S \over \delta (\delta g_{\mu\nu})}
  \bar g_{\gamma\nu}
  \right) \, ,
\end{equation}
Note that in the above formulation, it is important that gauge-fixing
is performed after the equations of motion are written down.
For instance, suppose one chooses a gauge in which the spatial part of
the metric is diagonal; it is important the equations of motion that
followed from varying the off-diagonal parts of the spatial metric are also used.

\bibliographystyle{utphys}
\bibliography{EFTofSSB}

\end{document}